\documentclass[useAMS,usenatbib,usegraphicx]{mn2e_mod1}
\newcommand*{\preprint}{}
\usepackage[breaklinks,colorlinks,citecolor=black,linkcolor=black,urlcolor=black]{hyperref}

\newcommand*{\mysub}[2]{\ensuremath{#1_{\mathrm{#2}}}}
\newcommand*{\unit}[1]{\ensuremath{\mathrm{\, #1}}}

\newcommand*{\phmin}{\hspace{1.9ex}} 

\newcommand*{\Omegam}{\mysub{\Omega}{m}}
\newcommand*{\Omegab}{\mysub{\Omega}{b}}
\newcommand*{\Omegal}{\mysub{\Omega}{\Lambda}}
\newcommand*{\Omegade}{\mysub{\Omega}{DE}}

\newcommand*{\fgas}{\mysub{f}{gas}}

\newcommand*{\Mgas}{\mysub{M}{gas}}
\newcommand*{\Mlens}{\mysub{M}{lens}}
\newcommand*{\mgas}{\mysub{m}{gas}}
\newcommand*{\mlens}{\mysub{m}{lens}}

\newcommand*{\LCDM}{\ensuremath{\Lambda}CDM}
\newcommand*{\rhocr}{\mysub{\rho}{cr}}

\newcommand*{\dA}{D}

\newcommand*{\atr}{\mysub{a}{tr}}
\newcommand*{\wet}{\mysub{w}{et}}
\newcommand*{\Mnu}{\ensuremath{\sum m_\nu}}
\newcommand*{\Neff}{\mysub{N}{eff}}
\newcommand*{\betas}{\mysub{\beta}{s}}
\newcommand*{\gammat}{\mysub{\gamma}{t}}
\newcommand*{\moment}{\ensuremath{\mathcal{M}}}
\newcommand*{\fnl}{\mysub{f}{NL}}
\newcommand*{\ns}{\mysub{n}{s}}
\newcommand*{\deltac}{\mysub{\delta}{c}}

\newcommand*{\second}{\unit{s}}

\newcommand*{\erg}{\unit{erg}}
\newcommand*{\cm}{\unit{cm}}
\newcommand*{\km}{\unit{km}}

\newcommand*{\Mpc}{\unit{Mpc}}

\newcommand*{\keV}{\unit{keV}}
\newcommand*{\eV}{\unit{eV}}
\newcommand*{\Msun}{\ensuremath{\, M_{\odot}}}

\newcommand*{\E}[1]{\ensuremath{\times 10^{#1}}}
\newcommand*{\expectation}[1]{\ensuremath{\left\langle #1 \right\rangle}}
\newcommand*{\ltsim}{\ {\raise-.75ex\hbox{$\buildrel<\over\sim$}}\ }
\newcommand*{\gtsim}{\ {\raise-.75ex\hbox{$\buildrel>\over\sim$}}\ }
\newcommand*{\proptosim}{\ {\raise-.75ex\hbox{$\buildrel\propto\over\sim$}}\ }

\newcommand*{\like}{\ensuremath{\mathcal{L}}}

\newcommand*{\secref}{Section}
\newcommand*{\appref}{Appendix}
\newcommand*{\eqnref}{Equation}
\newcommand*{\figref}{Figure}
\newcommand*{\tabref}{Table}

\newcommand*{\Chandra}{{\it Chandra}}
\newcommand*{\Planck}{{\it Planck}}

\newcommand*{\Nobs}{\mysub{N}{det}}
\newcommand*{\nobs}{\mysub{\tilde{n}}{det}}

\defcitealias{Navarro9611107}{NFW}
\newcommand*{\NFW}{\citetalias{Navarro9611107}}
\defcitealias{von-der-Linden1208.0597}{W}
\defcitealias{Kelly1208.0602}{t}
\defcitealias{Applegate1208.0605}{G}
\newcommand*{\wtg}{\citetalias{von-der-Linden1208.0597}\citetalias{Kelly1208.0602}\citetalias{Applegate1208.0605}}
\defcitealias{Mantz0909.3098}{M10a}
\newcommand*{\mare}{\citetalias{Mantz0909.3098}}
\defcitealias{Mantz0909.3099}{M10b}
\newcommand*{\maerd}{\citetalias{Mantz0909.3099}}
\newcommand*{\emten}{\mare{},\hyperlink{MtenBht}{b}}
\defcitealias{Mantz1402.6212}{M14}
\newcommand*{\mamrakls}{\citetalias{Mantz1402.6212}}

\newcommand*{\figscale}{1.0}

\renewcommand*{\preprint}{IGC-14/6-1}

\begin{document}

\title[Weighing the Giants IV: Cosmology and Neutrino Mass]{Weighing the Giants IV: Cosmology and Neutrino Mass}

\author[A. B. Mantz et al.]{Adam B. Mantz,$^{1,2}$\thanks{Corresponding author e-mail: \href{mailto:amantz@kicp.uchicago.edu}{\tt amantz@kicp.uchicago.edu}} {} 
  Anja von der Linden,$^{3,4,5}$
  Steven W. Allen,$^{3,4,6}$ \newauthor
  Douglas E. Applegate,$^7$
  Patrick L. Kelly,$^8$
  R. Glenn Morris,$^{3,6}$
  David A. Rapetti,$^5$ \newauthor
  Robert W. Schmidt,$^9$
  Saroj Adhikari,$^{10}$
  Mark T. Allen,$^{3,4}$
  Patricia R. Burchat,$^{3,4}$ \newauthor
  David L. Burke,$^{3,6}$
  Matteo Cataneo,$^5$
  David Donovan,$^{11}$
  Harald Ebeling,$^{11}$ \newauthor
  Sarah Shandera,$^{10}$
  Adam Wright$^{3,4,6}$\medskip\\
  $^1$Department of Astronomy and Astrophysics, University of Chicago, 5640 South Ellis Avenue, Chicago, IL 60637, USA\\
  $^2$Kavli Institute for Cosmological Physics, University of Chicago, 5640 South Ellis Avenue, Chicago, IL 60637, USA\\
  $^3$Kavli Institute for Particle Astrophysics and Cosmology, Stanford University, 452 Lomita Mall, Stanford, CA 94305, USA\\
  $^4$Department of Physics, Stanford University, 382 Via Pueblo Mall, Stanford, CA 94305, USA\\
  $^5$Dark Cosmology Centre, Niels Bohr Institute, University of Copenhagen, Juliane Maries Vej 30, 2100 Copenhagen, Denmark\\
  $^6$SLAC National Accelerator Laboratory, 2575 Sand Hill Road, Menlo Park, CA  94025, USA\\
  $^7$Argelander-Institute for Astronomy, Auf dem H\"ugel 71, D-53121 Bonn, Germany\\
  $^8$Department of Astronomy, University of California, Berkeley, CA 94720, USA\\
  $^9$Astronomisches Rechen-Institut, Zentrum f\"ur Astronomie der Universit\"at Heidelberg, M\"onchhofstrasse 12-14, D-69120 Heidelberg, Germany\\
  $^{10}$Institute for Gravitation and the Cosmos, Pennsylvania State University, University Park, PA 16802, USA\\
  $^{11}$Institute for Astronomy, 2680 Woodlawn Drive, Honolulu, HI 96822, USA
}
\date{Submitted 16 July 2014. Accepted 7 October 2014.}

\pagerange{\pageref{firstpage}--\pageref{lastpage}} \pubyear{2014}
\maketitle
\label{firstpage}

\begin{abstract}
  We employ robust weak gravitational lensing measurements to improve cosmological constraints from measurements of the galaxy cluster mass function and its evolution, using X-ray selected clusters detected in the ROSAT All-Sky Survey. Our lensing analysis constrains the absolute mass scale of such clusters at the 8 per cent level, including both statistical and systematic uncertainties. Combining it with the survey data and X-ray follow-up observations, we find a tight constraint on a combination of the mean matter density and late-time normalization of the matter power spectrum, $\sigma_8(\Omegam/0.3)^{0.17}=0.81\pm0.03$, with marginalized, one-dimensional constraints of $\Omegam=0.26\pm0.03$ and $\sigma_8=0.83\pm0.04$. For these two parameters, this represents a factor of two improvement in precision with respect to previous work, primarily due to the reduced systematic uncertainty in the absolute mass calibration provided by the lensing analysis. Our new results are in good agreement with constraints from cosmic microwave background (CMB) data, both  WMAP and \Planck{} (plus WMAP polarization), under the assumption of a flat \LCDM{} cosmology with minimal neutrino mass. Consequently, we find no evidence for non-minimal neutrino mass from the combination of cluster data with CMB, supernova and baryon acoustic oscillation measurements, regardless of which all-sky CMB data set is used (and independent of the recent claimed detection of B-modes on degree scales). We also present improved constraints on models of dark energy (both constant and evolving), modifications of gravity, and primordial non-Gaussianity. Assuming flatness, the constraints for a constant dark energy equation of state from the cluster data alone are at the 15 per cent level, improving to $\sim 6$ per cent when the cluster data are combined with other leading probes.
\end{abstract}


\clearpage
\section{Introduction} \label{sec:intro}

Great strides have been made in recent years in the use of galaxy cluster surveys as probes of the halo mass function, and thereby of cosmology and fundamental physics (for a review, see \citealt*{Allen1103.4829}). Cluster surveys covering the entire extragalactic sky, or a significant fraction of it, now exist at X-ray \citep{Truemper1993Sci...260.1769T, Ebeling1998MNRAS.301..881E, Ebeling1004.4683, Bohringer0405546}, optical/IR (e.g.\ \citealt{Koester0701265, Rykoff1303.3562}) and millimeter \citep{Reichardt1203.5775, Hasselfield1301.0816, Planck1303.5089} wavelengths, and a number of independent groups have published cosmological constraints in broad agreement with one another based on these data (e.g.\ \citealt{Eke9802350, Donahue9907333, Henry0002365, Henry0404142, Borgani2001ApJ...561...13B, Reiprich0111285, Seljak0111362, Viana0111394, Allen0208394, Pierpaoli0210567, Schuecker0208251, Vikhlinin0212075, Vikhlinin0812.2720, Voevodkin0305549, Dahle0608480, Mantz0709.4294, Mantz0909.3098,  Henry0809.3832, Rozo0902.3702, Sehgal1010.1025, Benson1112.5435, Planck1303.5080}).

These cluster survey data have provided highly competitive constraints on dark energy and modifications of gravity (e.g.\ \citealt{Vikhlinin0812.2720, Mantz0909.3098, Schmidt0908.2457,  Rapetti1205.4679}), as well as measurements of the late-time normalization of the matter power spectrum ($\sigma_8$, defined by \eqnref~\ref{eq:sigma2def}, below). Constraints on $\sigma_8$ are a key complement to measurements of the amplitude of the power spectrum at high redshift from the CMB in many cosmological models of interest, particularly those where the dark energy equation of state or neutrino masses are free parameters. Since cosmological data currently provide our best limits on the species-summed neutrino mass (\citealt*{Mantz0911.1788}; \citealt{Reid0910.0008}), improving constraints on $\sigma_8$ is a priority.

Previous constraints on $\sigma_8$ from clusters have been systematically limited due to fundamental uncertainties regarding the absolute calibration of cluster mass measurements (for a discussion, see \citealt{von-der-Linden1402.2670}). The most widespread observational techniques used to estimate masses, based on X-ray data or optical spectroscopy, assume that the measured thermal/kinetic energies accurately reflect the underlying gravitating mass, and are thus subject to a theoretically uncertain bias. Recently, measurements of the gravitational lensing of background galaxies due to clusters have emerged as a potential avenue for providing a more accurate absolute mass calibration, since weak-lensing mass measurements are expected to be nearly unbiased when the analysis is restricted to the appropriate radial range (e.g.\ \citealt{Becker1011.1681}) and systematic effects in the shear measurements and photometric redshifts can be accounted for \citep{Applegate1208.0605}. Thanks to the availability of wide field-of-view imagers with superb image quality, such as SuprimeCam at the Subaru telescope and MegaCam/MegaPrime at the Canada-France-Hawaii Telescope (CFHT), unbiased weak lensing measurements for large samples of clusters are now within reach.

The {\it Weighing the Giants} project was conceived in order to provide just such an accurate and precise calibration of cluster masses for studies of cosmology, and for the closely related analysis of cluster scaling relations. The project involves 51 massive clusters that have previously been used in cosmological studies (\citealt{Allen0706.0033, Mantz0909.3098, Mantz0909.3099}, hereafter \emten{}). Details of the lensing data and their analysis appear in Papers~I--III of this series \citep{von-der-Linden1208.0597, Kelly1208.0602, Applegate1208.0605}, which we collectively refer to as \wtg{} below. The \wtg{} lensing analysis has already been used to calibrate mass estimates based on X-ray observations that assume hydrostatic equilibrium (\citealt{von-der-Linden1402.2670}; Applegate et~al., in prep.), particularly in the context of the cosmological constraints available from gas mass fraction (\fgas{}) measurements in relaxed clusters \citep{Mantz1402.6212}. Here we apply the lensing data to cosmological tests based on the cluster mass function (also referred to as cluster counts), specifically by incorporating the \wtg{} data into the \emten{} analysis of X-ray cluster survey and follow-up data. A companion paper (WtG\,V, in prep.) explores the astrophysical consequences of our mass calibration for cluster scaling relations, which are necessarily constrained simultaneously with cosmological parameters in our analysis.

Given both the widespread expectation that the ``correct'' answers for cosmological parameters will be consistent with those determined from CMB data for a spatially flat, cosmological-constant model, and the potential of galaxy cluster surveys to provide high-precision cosmological constraints, minimizing the possibility of observer bias is paramount in such work. The \wtg{} lensing analysis employed a procedure whereby those working on it were blind in all comparisons to independent mass estimates, in particular (but not limited to) those from X-ray observations and from lensing results in the literature, until the lensing analysis was finalized (see \citealt{Applegate1208.0605} for a full discussion). This entire lensing analysis was completed before the cosmological analysis presented here had begun. Although we did not explicitly blind cosmological parameter results in this work, the constraints reported here are simply those that follow from incorporating the \wtg{} lensing data into an already mature analysis pipeline (\mare{}), which is a simple and straightforward addition (\secref~\ref{sec:like}).

This paper is organized as follows. \secref{}~\ref{sec:data} describes our cluster data and the external cosmological probes with which we combine them, while \secref{}~\ref{sec:model} outlines the analysis procedure and the models fitted to the data. Our results are presented in \secref~\ref{sec:cosmores}. \secref~\ref{sec:discussion} considers the importance of the lensing and X-ray follow-up data to the analysis, and the potential gains from obtaining an expanded lensing data set and combining surveys at different wavelengths. We conclude in \secref~\ref{sec:conclusions}. Best-fitting parameter values reported here always correspond to modes of the marginalized posterior distributions, and uncertainties correspond to 68.3 per cent confidence maximum-likelihood intervals, unless otherwise specified. We make occasional use of a reference cosmological model, which has Hubble parameter $h=H_0/100\km\second^{-1}\Mpc^{-1}=0.7$, mean matter density in units of the critical density $\Omegam=0.3$, and cosmological constant energy density $\Omegal=0.7$. We use the standard definition of cluster masses and characteristic radii in terms of a spherical overdensity, $\Delta$, with respect to the critical density at the cluster's redshift: $M_\Delta = (4\pi/3) \Delta \rhocr(z) r_\Delta^3$.

\section{Data} \label{sec:data}

The data set employed here consists of three X-ray flux-limited samples of clusters (i.e.\ redshifts and fluxes, along with the associated selection functions), as well as deeper follow-up X-ray data and/or high-quality optical imaging for a subset of the detected clusters. As in \emten{}, the cluster samples used here are based on the BCS \citep{Ebeling1998MNRAS.301..881E}, REFLEX \citep{Bohringer0405546}, and Bright MACS \citep{Ebeling1004.4683} catalogs, themselves compiled from the ROSAT All-Sky Survey (RASS; \citealt{Truemper1993Sci...260.1769T}). In the cases of BCS and REFLEX (covering redshifts $z<0.3$), we use only clusters with 0.1--2.4\keV{} luminosities $>2.5\E{44}\erg\second^{-1}$ (as estimated for our reference cosmology) to eliminate low-mass clusters and groups; this cut has no impact on the Bright MACS sample ($0.3<z<0.5$). We depart slightly from \emten{} by using a higher flux limit of $5\E{-12}\erg\second^{-1}\cm^{-2}$ in the 0.1--2.4\keV{} band when selecting clusters from the BCS in order to avoid incompleteness that affects the lowest fluxes in BCS at all redshifts (see \citealt{Ebeling1998MNRAS.301..881E}).\footnote{For the REFLEX and Bright MACS catalogs, we respectively use flux limits of $3$ and $2\E{-12}\erg\second^{-1}\cm^{-2}$, as in \emten{}.} We have also expanded the allowance for overall incompleteness/impurity for Bright MACS to $\pm10$ per cent from the $\pm5$ per cent previously assumed for MACS and other surveys in \mare{}, reflecting the greater challenges affecting the MACS survey construction. Finally, we have removed Abell~2318, RX\,J0250.2$-$2129 and RX\,J1050.6$-$2405 from the data set, as these appear consistent with their X-ray emission being dominated by active galactic nuclei (AGN) rather than the intracluster medium (A.~Edge, private communication; other deletions from the published catalogs are listed in \maerd{}). However, these changes to the data set are not significant enough to affect any of our cosmological results, as we have verified by explicitly comparing constraints using the old and new samples. The new sample contains a total of 224 clusters.

X-ray luminosities and gas masses were derived from ROSAT and/or \Chandra{} data for 94 clusters in \maerd{}. We employ these measurements again in the present work, in addition to the survey data, to improve constraints on the cluster scaling relations and refine the mass information available for individual clusters (see \secref{}s~\ref{sec:scalingmod} and \ref{sec:followup}).\footnote{Since the analysis of \maerd{}, the model for the contaminant affecting the \Chandra{} ACIS detectors (including its time dependence) has been modified slightly. An overall bias in gas masses or luminosities from follow-up observations would have no effect on the cosmological analysis in this work, since gas mass is used only as an empirically calibrated mass proxy, and luminosities from follow-up data are cross-calibrated to the ROSAT survey luminosities (see \mare{}). Nevertheless, we note that directly comparing luminosities and gas mass profiles for 59 clusters in common between the \maerd{} and \citet{Mantz1402.6212} generations of analysis (not all of which were published in each paper), shows agreement at the per cent level.}

The new data that are central to this work are the measurements of weak gravitational lensing for 50 massive clusters,\footnote{While the full \wtg{} analysis employs 51 clusters, we omit Abell~370 from this work, since it has fundamentally different selection properties from our data set (i.e.\ it is not X-ray selected).} which are used to calibrate the absolute cluster mass scale. These data and their analysis are described in \wtg{}. Specifically, we use the shear profiles derived from the simpler ``color-cut'' method of that work, which are available for the entire data set, rather than those from the ``$p(z)$'' method, which are available for just over half of the sample.\footnote{The more robust $p(z)$ masses have been used to characterize the bias and scatter of the color-cut method \citep{Applegate1208.0605}, and this information is fed into the analysis presented here (specifically it factors into the width of the lensing-to-true mass normalization; see \secref~\ref{sec:scalingmod}). The larger number of clusters for which we can do a color-cut analysis makes this cross-calibration approach preferable to relying exclusively on $p(z)$ clusters.} Of the 50 \wtg{} clusters, 27 belong to the flux-limited sample identified above, and are straightforward to incorporate into the likelihood function for  cosmology and scaling relations described in \mare{} and reviewed in \secref~\ref{sec:like}. The remaining 23 cannot be used to constrain the X-ray luminosity--mass relation because, even though they are X-ray selected, we do not have a robustly quantified selection function for them with which to account for selection biases. However, they can still be used to calibrate the relation linking gas and total mass, to the extent that the correlation of intrinsic scatters in luminosity and gas mass at fixed total mass is small (e.g.\ \citealt{Allen1103.4829}). We have verified empirically that including these additional lensing data in this way (see \secref~\ref{sec:like}) does not bias our cosmological results.

In addition to the measurements of redshift, X-ray luminosity, gas mass and total mass (integrated over radii $\ltsim r_{500}$), we take advantage of the cosmological information available from X-ray measurements of the gas mass fraction, $\fgas$, at $\sim r_{2500}$ for relaxed clusters (\citealt{Mantz1402.6212}, hereafter \mamrakls{}).\footnote{We use the term \fgas{} generically to refer to the \mamrakls{} data set in this paper, or $\fgas($0.8--1.2$\,r_{2500})$ when necessary for clarity. The integrated gas mass fraction that is constrained at radii $\sim r_{500}$ from the X-ray and lensing follow-up observations that form part of the cluster counts data set will be referred to as $\fgas(r_{500})$.} More precisely, these \fgas{} measurements are made in a spherical shell spanning 0.8--1.2\,$r_{2500}$, where theoretical and observational uncertainties due to various astrophysical effects (e.g.\ AGN feedback, gas cooling and clumping, etc.) are minimized and where X-ray spectroscopy permits precise total mass estimates. These data provide additional constraints on dark energy parameters and, when combined with external priors on the cosmic mean baryon density ($\Omegab$), produce tight constraints on $\Omegam$. These $\fgas($0.8--1.2$\,r_{2500})$ data do not constrain $\sigma_8$, although their constraint on \Omegam{} is useful for breaking the degeneracy between the two parameters in cluster counts data.

Our baseline cluster analysis uses all the data described above, the RASS cluster catalogs, mass proxies from X-ray follow-up data, lensing data and \fgas{} measurements (but see \secref~\ref{sec:s8}), and also incorporates Gaussian priors on the Hubble parameter ($h=0.738\pm0.024$; \citealt{Riess1103.2976}) and the cosmic baryon density ($100\,\Omegab h^2=2.202\pm0.045$; \citealt{Cooke1308.3240}). (Note that these external priors are not required or used when the cluster data are combined with CMB data.) In \secref~\ref{sec:cosmores}, we present results from these cluster data, and compare and combine our results with those from independent cosmological probes. Specifically, we use all-sky CMB data from the {\it Wilkinson Microwave Anisotropy Probe} (WMAP 9-year release; \citealt{Bennett1212.5225, Hinshaw1212.5226}) and the \Planck{} satellite (1-year release, including WMAP polarization data, called \Planck{}+WP below; \citealt{Planck1303.5075}), as well as high-multipole data from the Atacama Cosmology Telescope (ACT; \citealt{Das1301.1037}) and the South Pole Telescope (SPT; \citealt{Keisler1105.3182, Reichardt1111.0932, Story1210.7231}). We also include the Union 2.1 compilation of type Ia supernovae \citep{Suzuki1105.3470} and baryon acoustic oscillation (BAO) data from the combination of results from the 6-degree Field Galaxy Survey (6dF; $z=0.106$; \citealt{Beutler1106.3366}) and the Sloan Digital Sky Survey (SDSS, $z=0.35$ and $0.57$; \citealt{Padmanabhan1202.0090, Anderson1303.4666}). Technical details of our use of these non-cluster data can be found in \mamrakls{}.

\section{Model and Analysis Methods} \label{sec:model}

\mare{} provide a detailed description of the analysis procedure for the cluster survey and X-ray follow-up data, including models for the cosmological background, halo abundance and measurement process employed in this work. Here we review the most relevant aspects of the analysis and describe the additions necessary to include the new gravitational lensing data. For details of the analysis of the \fgas{} data, see \mamrakls{}.

\subsection{Cosmological Model} \label{sec:cosmomodel}

As in \mare{} and \mamrakls{}, we consider cosmological models with a Friedmann-Robertson-Walker metric, containing radiation, baryons, neutrinos, cold dark matter (CDM), and dark energy. For the cluster data, the key parameters describing the average universe are the Hubble parameter ($h$), the cosmic densities of baryons (\Omegab), neutrinos (parametrized by their species-summed mass, \Mnu{}), matter (in total, \Omegam{}) and dark energy (\Omegade, or \Omegal{} in the case of a cosmological constant), and the global curvature density ($\Omega_k$). We adopt an evolving parametrization of the dark energy equation of state \citep{Rapetti0409574},
\begin{equation} \label{eq:wdef}
  w = w_0 + w_a \left( \frac{z}{z+\mysub{z}{tr}} \right) = w_0 + w_a \left( \frac{a^{-1}-1}{a^{-1}+\atr^{-1}-2} \right),
\end{equation}
where $a=(1+z)^{-1}$ is the scale factor. In this model, $w$ takes the value $w_0$ at the present day and $\wet=w_0+w_a$ in the high-redshift limit (i.e.\ at ``early times''), with the timing of the transition between the two determined by \atr{}. \eqnref~\ref{eq:wdef} contains as special cases the cosmological constant model (\LCDM{}; $w_0=-1$ and $w_a=0$), constant-$w$ models ($w_a=0$), and the simpler evolving-$w$ model adopted by \citet{Chevallier0009008} and \citet{Linder0208512} ($\atr=0.5$). \citet{Allen0706.0033} and \mamrakls{} provide more details regarding calculations using this model. Note that, as in \mare{}, we propagate the effect of dark energy density and velocity perturbations (when $w\neq-1$) on linear scales when evaluating the matter power spectrum.

The variance of the linearly evolved density field, smoothed by a spherical top-hat window of comoving radius $R$, enclosing mass $M=4\pi \rho R^3 / 3$, is
\begin{equation}
  \label{eq:sigma2def}
  \sigma^2(R,z) = \frac{1}{2\pi^2} \int_0^\infty k^2 P(k,z) |W_R(k)|^2 dk,
\end{equation}
where $P(k,z)$ is the linear power spectrum evolved to redshift $z$ and $W_R(k)$ is the Fourier transform of the window function. The matter power spectrum is parametrized by an amplitude, conventionally $\sigma_8 = \sigma(R=8h^{-1}\Mpc,z=0)$, and the scalar spectral index, $\ns$. We express the halo mass function, the expected number density as a function of redshift and mass, in the standard way:
\begin{equation}
  \label{eq:massfunction}
  \expectation{\frac{dn(M,z)}{dM}} = \frac{\rho}{M} \frac{d\ln \sigma^{-1}}{dM} f(\sigma,z).
\end{equation}
As in \mare{}, we use the \citet{Tinker0803.2706} parametrization of $f(\sigma,z)$, including its explicit redshift dependence. To account for systematic uncertainties in the mass function, including for models other than \LCDM{}, the effects of baryons, etc., we marginalize over priors at the 10 per cent level both in the baseline function, $f(\sigma,z=0)$, and in the redshift dependent terms from \citet[][see details in \mare{}]{Tinker0803.2706}.

In \eqnref~\ref{eq:massfunction}, as well as in the correspondence of mass and scale (i.e.\ $M\propto\rho R^3$) entering $W_R(k)$, $\rho$ refers to the sum of baryon and CDM densities, i.e.\ matter \emph{not} including neutrinos. Similarly, neutrinos are not included in the power spectrum used in \eqnref~\ref{eq:sigma2def}. \citet{Costanzi1311.1514} have shown that this choice results in the \citet{Tinker0803.2706} fitting formula providing a more accurate approximation to the mass function in $N$-body simulations with massive neutrinos than the analogous calculations including the neutrino density everywhere (see also \citealt{LoVerde1405.4858}). For our baseline model with $\Mnu=0.056\eV$ (the minimum value allowed by neutrino oscillation data), this distinction is completely negligible, but it has a small impact on our constraints (tightening them) at values $\Mnu\gtsim0.3\eV$, consistent with estimates of the magnitude of the effect by \citet{Costanzi1311.1514} and the level of systematic uncertainty adopted in our analysis.

Note that \secref{}s~\ref{sec:growthindex} and \ref{sec:nongauss} introduce modifications to the evolution of the power spectrum and the mass function in order to investigate departures from General Relativity (GR) and non-Gaussianities in the primordial perturbation field. These are outlined in the respective sections.

\subsection{Cluster Scaling Relations} \label{sec:scalingmod}

Connecting the predicted mass function to a flux-limited survey requires a scaling relation -- a stochastic function consisting of a mean relation and a model for intrinsic scatter -- linking mass and X-ray luminosity. Additional observables that have a smaller intrinsic scatter at fixed mass (i.e.\ better mass proxies, namely gas mass and temperature in the case of X-ray follow-up observations) can improve cosmological constraints by refining the information available for individual clusters (e.g.\ \citealt*{Wu0907.2690}; see also \secref~\ref{sec:followup}). It is therefore advantageous to define joint scaling relations, describing the trends and joint scatter of several observables as a function of mass, as we do below. Due to the ubiquity of selection biases in cosmological samples and the steepness of the mass function, accurate constraints on scaling relations (and cosmology) can only be obtained from a simultaneous cosmology+scaling relation analysis that properly accounts for the influence of the mass function and the survey selection function on the observed data (see \secref~\ref{sec:like}, \emten{} and \citealt{Allen1103.4829}).

Our model for the cluster scaling relations is that of \emten{}, expanded to include the new weak lensing observations. We describe the scaling of each observable cluster property with mass as a power law, and the joint intrinsic scatter as a multi-dimensional log-normal distribution. For this purpose, we define the logarithmic total mass within $r_{500}$ as\footnote{To simplify interpretation of the intrinsic scatter terms, we use natural logarithms in the scaling relation model, a change of notation with respect to \emten{}.}
\begin{eqnarray}
  \label{eq:scalingmass}
  m &=& \ln\left(\frac{E(z)M_{500}}{10^{15}\Msun}\right),
\end{eqnarray}
with $E(z)=H(z)/H_0$. The corresponding definitions for observables -- luminosity (0.1--2.4\,keV{} band), center-excised temperature, gas mass and lensing mass -- are
\begin{eqnarray}
  \label{eq:MLTdefs}
  \ell &=& \ln\left(\frac{L_{500}}{E(z)10^{44}\erg\second^{-1}}\right),  \\
  t &=& \ln\left(\frac{kT_{500}}{\keV}\right), \nonumber\\
  \mgas &=& \ln\left(\frac{E(z)\Mgas{}_{,500}}{10^{15}\Msun}\right), \nonumber\\
  \mlens &=& \ln\left(\frac{E(z)\Mlens{}_{,500}}{10^{15}\Msun}\right). \nonumber
\end{eqnarray}
The quantities in \eqnref{}~\ref{eq:MLTdefs} represent intrinsic properties of a given cluster, as distinct from measured values (to which they are related by a model for measurement scatter); along with $m$, they are free parameters of the model.\footnote{Note that, while $m$ represents true mass, the quantities in \eqnref~\ref{eq:MLTdefs} need not be identically the true luminosity, average temperature, etc.\ for a cluster (although they do correspond to the measured quantities generally described as such). For example, asphericity might result in a departure of \mgas{} from the true gas mass within $r_{500}$, an effect that contributes to the intrinsic scatter of the \mgas{}--$m$ relation. Similarly, \mlens{} refers to the spherical mass that would be reconstructed from an ideal shear profile (i.e.\ without statistical error), which is in general different from the true mass due to projected structure.} With these definitions, power-law scaling relations become linear relations between $\bmath{y} \equiv (\ell,t,\mgas,\mlens)$ and $m$. For a given cluster, the expectation value of $\bmath{y}$ is $\bmath{\beta_0} + \bmath{\beta_1} m$, and we assume a multivariate Gaussian intrinsic scatter in $\bmath{y}$ at fixed $m$; i.e.
\begin{equation} \label{eq:scaling}
  P(\bmath{y}|m) \propto |\mathbf{\Sigma}|^{-1/2} \exp\left( -\frac{1}{2} \bmath{\eta}^\mathrm{t} \mathbf{\Sigma}^{-1} \bmath{\eta} \right) ,
\end{equation}
where $\mathbf{\Sigma}$ is a covariance matrix and $\bmath{\eta}=\bmath{y}-(\bmath{\beta_0}+\bmath{\beta_1}m)$. The normalizations ($\bmath{\beta_0}$), slopes ($\bmath{\beta_1}$) and diagonal elements of $\mathbf{\Sigma}$ are in general free parameters that we allow the data to fit (though see below). Following \maerd{}, we also fit the off-diagonal covariance between $\ell$ and $t$ (which turns out to be consistent with zero; \maerd{}). For simplicity, and because there is no particular expectation for a non-zero covariance, we fix the off-diagonal covariance terms involving \mlens{} and \mgas{} to zero (see discussion in \appref~\ref{sec:scaling}).

For the \mlens--$m$ relation, we assume a slope of unity and place priors on the normalization and intrinsic scatter. Specifically, we adopt a Gaussian prior on the normalization, $\beta_{0,\mlens} = 0.99 \pm 0.07$, encoding the expected bias (and its uncertainty) of weak lensing masses due to triaxiality, line-of-sight structure, the assumption of a \citet*[][hereafter \NFW{}]{Navarro9611107} mass profile, systematic biases affecting shear measurements, photometric redshift errors, and the statistical uncertainty accrued in cross-calibrating $p(z)$ (5-filter) and color-cut (3-filter) lensing data. (Full details can be found in \citealt{Applegate1208.0605}.) We constrain the scatter between \mlens{} and $m$ with a wide Gaussian prior, $20\pm10$ per cent, where the central value is motivated by the simulations of \citet{Becker1011.1681}.\footnote{Comparing the scatter in two mass bins, both lower in mass than the clusters in our lensing sample, these simulations imply that the intrinsic scatter decreases as a function of mass. We have tested whether a power-law dependence of the scatter on mass would change our results, marginalizing over indices in the range $\pm0.35$, and find that this has a negligible effect on our cosmological constraints. This is due to the small range in mass covered by our lensing data, and the fact that, when X-ray mass proxy information is also included in the analysis, the data are able to directly constrain the intrinsic scatter at the pivot mass of the lensing sample (\appref~\ref{sec:scaling}). Note that the width of our prior on the intrinsic scatter, significantly greater than the uncertainties reported by \citet{Becker1011.1681}, partly reflects differences between their analysis and ours, such as our use of a fixed \NFW{} concentration parameter (\secref~\ref{sec:like}).}

The \mgas--$m$ relation deserves some additional consideration, since the value and evolution of its normalization, $\beta_{0,\mgas} = \ln \fgas(r_{500})$, carry additional cosmological information (\citealt{Sasaki9611033, Pen9610090, Allen0205007, Allen0405340, Allen0706.0033, Allen1103.4829, Ettori0211335, Ettori0904.2740, Battaglia1209.4082, Planelles1209.5058}; \mamrakls). In principle, this information could be used in tandem with the more precise $\fgas($0.8--1.2\,$r_{2500})$ measurements of \mamrakls{}, given a suitable model for their covariance. In practice, the low precision of our mass constraints at $r_{500}$ for individual clusters (due to the scatter in $\mlens|m$) significantly limits the information available from the \mgas--$m$ relation. In addition, the measurement correlation between the two \fgas{} values is negligible, since the total masses are estimated independently from different data (lensing vs.\ X-ray) and the gas mass measured in the 0.8--1.2\,$r_{2500}$ shell is a small fraction of that integrated within $r_{500}$. We therefore simplify the analysis by keeping the model for $\fgas($0.8--1.2\,$r_{2500})$, used for the \mamrakls{} data, independent of the parameters of the \mgas--$m$ relation. In addition to allowing the normalization, mass dependence and intrinsic scatter of the \mgas{}--$m$ relation to vary, we marginalize over a $\pm5$ per cent uniform prior on the evolution of the normalization, of the form $\fgas(r_{500},z) = \fgas(r_{500},z=0)(1+\alpha_f z)$. This form, and the prior itself, are identical to those used to describe the evolution in $\fgas($0.8--1.2\,$r_{2500})$ in \mamrakls{}, but $\alpha_f$ is varied independently of the corresponding parameter at $r_{2500}$. We constrain the intrinsic scatter in $\mgas|m$ with a uniform prior spanning 0.0--0.10, where 0.10 corresponds to the high end of the confidence interval for the fractional intrinsic scatter of $\fgas(r_{500})$, measured from the \mamrakls{} data (Mantz et~al., in preparation).

\subsection{Likelihood Function} \label{sec:like}
The complete likelihood of the X-ray and lensing data set takes the same form as in \mare{},
\begin{equation}
  \label{eq:growth_like_propto}
  \like \propto e^{-\expectation{\Nobs}} \prod_{i=1}^{\Nobs} \expectation{\nobs{}_{,i}}.
\end{equation}
Here $\expectation{\Nobs}$ is the expected number of cluster detections in the survey data for a given set of model parameters, accounting for the selection function. The product runs over the \Nobs{} detected clusters, and accounts for their redshifts, survey fluxes and any follow-up measurements. Following \mare{}, we use an abbreviated notation where $\bmath{x}$ stands for the true values of $z$ and $m$; $\bmath{y}$ stands for the intrinsic values of $\ell$, $t$, $\mgas$ and $\mlens$ (as above); and $\bmath{\hat{y}}$ stands for the measured values of $\bmath{y}$, plus the X-ray survey flux, $\hat{F}$. Similarly, $\bmath{\hat{x}}$ indicates measured values of $\bmath{x}$, although in practice we model any mass estimates as response variables of the scaling relations (i.e.\ components of $\bmath{\hat{y}}$). The per-cluster likelihood term can then be expressed as
\begin{eqnarray} \label{eq:bayes_Pobs}
  \expectation{\nobs{}_{,i}} & = & \int d\bmath{x} \int d\bmath{y} \expectation{\frac{dN}{d\bmath{x}}} \, P(\bmath{y}|\bmath{x}) \, P(\bmath{\hat{x}_i},\bmath{\hat{y}_i}|\bmath{x},\bmath{y}) \nonumber\\
  & & \times P(I|\bmath{x},\bmath{y},\bmath{\hat{x}_i},\bmath{\hat{y}_i}).
\end{eqnarray}
Here, $\expectation{dN/d\bmath{x}} = \expectation{d^2N/dzdm}$ can be calculated from the mass function and cosmic expansion history,
\begin{eqnarray}
  \expectation{\frac{d^2N}{dzdm}} = M \, \frac{dV}{dz} \expectation{\frac{dn(M,z)}{dM}},
\end{eqnarray}
where $V$ is the comoving volume as a function of redshift. The likelihood associated with the scaling relations is simply the function $P(\bmath{y}|\bmath{x})$ given in \eqnref~\ref{eq:scaling}. The remaining factors are respectively the likelihoods associated with the measurements, $P(\bmath{\hat{x}_i},\bmath{\hat{y}_i}|\bmath{x},\bmath{y})$, and selection function (the probability to be $I$ncluded in the data set), $P(I|\bmath{x},\bmath{y},\bmath{\hat{x}_i},\bmath{\hat{y}_i})$, for a particular cluster. These are written in a general form in \eqnref~\ref{eq:bayes_Pobs} and can be simplified for our purposes, as we detail below.

In the case of a cluster with a precisely determined redshift  (i.e.\ measured spectroscopically, which is the case for all our clusters), the integral $d\bmath{x}=dz\,dm$ can be replaced by an integral over mass only ($dm$) at fixed $z$.\footnote{This is equivalent to factoring the term associated with the redshift measurement, $P(\hat{z}_i|z)$, out of $P(\bmath{\hat{x}_i},\bmath{\hat{y}_i}|\bmath{x},\bmath{y})$, and approximating it as a delta function.} For a given parent cluster sample, our selection function is simply a function of redshift and detected X-ray survey flux; hence, the final term reduces to $P(I|z,\hat{F})$, a function that is tabulated for each of the BCS, REFLEX and Bright MACS samples \citep{Ebeling1998MNRAS.301..881E, Ebeling1004.4683, Bohringer0405546}. Note that, as in \mare{}, we marginalize over separate allowances for the overall completeness/purity of each cluster sample. The measurement term can be factored into survey, X-ray follow-up and lensing parts, since these three observations are independent; to be explicit,
\begin{eqnarray}
  P(\bmath{\hat{x}_i},\bmath{\hat{y}_i}|\bmath{x},\bmath{y}) & = & P(\hat{F}|z,\ell,t) \, P(\hat{\ell},\hat{t},\hat{m}_\mathrm{gas}|z,m,\ell,t,\mgas) \nonumber\\
  & & \times P(\hat{m}_\mathrm{lens}|z,\mlens).
\end{eqnarray}
The X-ray measurement models we employ are identical to those in \mare{}, and we refer the interested reader there for full details. In brief, the survey flux model straightforwardly follows from the intrinsic cluster luminosity, temperature and redshift, with the appropriate K-correction, and accounts for Poisson scaling of the measurement uncertainties with true flux. The model for X-ray follow-up measurements of mass proxies accounts not only for the straightforward statistical uncertainties in each measurement and their covariance (due to being measured from the same data), but also for their aperture dependence (i.e.\ the difference between the aperture used in the measurement and the true value of $r_{500}$ according to $m$ and the cosmological model). 

To evaluate the likelihood associated with the lensing data for a cluster, we compare the shear profile measured by \wtg{}\footnote{Hence, the term $\hat{m}_\mathrm{lens}$ in our equations should be interpreted as shorthand for the measured shear profile of a cluster.} (specifically, using the color-cut method) to the shear profile predicted from an \NFW{} profile with mass given by $\mlens$ and concentration parameter $c=4$ (consistent with the mean concentration measured in \wtg{} and the mean population concentration in $N$-body simulations; \citealt{Neto0706.2919}). The profiles are measured in annuli about the X-ray center in the radial range 750\,kpc to 3\,Mpc (in our reference cosmological model),\footnote{This radial range is chosen to minimize sensitivity to the assumed concentration, avoid high values of shear and cluster galaxy contamination in cluster centers, and reduce the effect of possible mis-centering, as discussed in detail by \citet{Applegate1208.0605}.} where the annuli are chosen to contain approximately equal numbers of galaxies (at least 300). We write
\begin{equation}
 \ln P(\hat{m}_\mathrm{lens}|z,\mlens) = -\frac{1}{2} \sum_j \left[\frac{\hat{g}_j - g_j(z,\mlens, c=4)}{\sigma_{g,j}}\right]^2,
\end{equation}
where $\hat{g}_j$ is the azimuthally averaged tangential shear measured in annulus $j$, and $\sigma_{g,j}$ is its uncertainty, determined by bootstrapping the galaxy population in each annulus.\footnote{As described in \wtg{}, corrections for shear calibration are applied on a per-galaxy basis, whereas corrections for contamination by cluster member galaxies are applied to the average shear measured in each annulus.} The predicted shear at projected radius $\theta_j$ is evaluated as
\begin{equation} \label{eq:shear}
  g_j(z,\mlens,c) = \frac{\expectation{\betas} \gammat{}_{,\infty}(\theta_j; \mlens,c)}{1 - \frac{\expectation{\betas^2}}{\expectation{\betas}} \kappa_\infty(\theta_j; \mlens,c)},
\end{equation}
where $\gammat{}_{,\infty}$ and $\kappa_\infty$ are respectively the tangential shear and convergence of a source at infinite redshift due to a lens at redshift $z$ with an \NFW{} mass distribution given by \mlens{} and $c$ \citep{Wright2000ApJ...534...34W}. \betas{} encodes the dependence on the redshift of the cluster and the lensed sources,
\begin{equation}
 \betas = \frac{D_\mathrm{LS} D_{\mathrm{O}\infty}}{D_\mathrm{OS} D_{\mathrm{L}\infty}},
\end{equation}
where the terms on the right hand side are variously the angular diameter distances separating the lens (L), source (S), observer (O), and a fictitious source at infinite redshift ($\infty$). Note that these terms introduce a cosmology dependence to the predicted shear. The averages of \betas{} and $\betas^2$ that appear in \eqnref~\ref{eq:shear} are evaluated using the distribution of galaxy redshifts in the COSMOS field, after replicating the same catalog selection cuts applied to each cluster field, such as the removal of the cluster red sequence.  More details can be found in \citet{Applegate1208.0605}.

\section{Cosmological Results} \label{sec:cosmores}

Our results are produced using {\sc cosmomc}\footnote{\url{http://cosmologist.info/cosmomc/}} (\citealt{Lewis0205436}; October 2013 version), appropriately modified to evaluate the likelihoods of the \fgas{}\footnote{\url{http://www.slac.stanford.edu/~amantz/work/fgas14/}} and cluster counts data. Cosmological calculations were performed using the {\sc camb}\footnote{\url{http://www.camb.info/}} package of \citet*{Lewis9911177}, suitably modified to implement the evolving-$w$ model of \citet{Rapetti0409574}, including the corresponding dark energy density perturbations (see also \mamrakls{}).

When analyzing cluster data alone, we incorporate Gaussian priors on the Hubble parameter, $h=0.738\pm0.024$ \citep{Riess1103.2976}, and mean baryon density, $100\,\Omegab h^2=2.202\pm0.045$ \citep{Cooke1308.3240}; we additionally fix the scalar spectral index of density perturbations to $\ns=0.95$ in this case.\footnote{Since the cluster data probe the amplitude of the power spectrum over a very limited range of scales, there is a degeneracy between \ns{} and $\sigma_8$ constraints from clusters alone. However, varying \ns{} within the range allowed by CMB data ($\Delta\ns\sim0.03$) would result in a sub-per-cent shift in our clusters-only value of $\sigma_8$ (\mare{}).} When CMB data are included in the fit, these three parameters are allowed to vary freely, along with the optical depth to reionization. With the exception of \secref{}s~\ref{sec:nu} and \ref{sec:growthindex}, we assume a minimal value of the species-summed neutrino mass, $\Mnu=0.056\eV$,\footnote{For this mass, our results are not sensitive to the distinction between, e.g., models with a single massive neutrino species and those with three degenerate neutrinos.} and the standard effective number of relativistic species, $\Neff=3.046$.

In \secref~\ref{sec:s8}, we begin by discussing our constraints on \Omegam{} and $\sigma_8$, two parameters on which clusters with accurately calibrated masses can provide powerful and largely model-independent constraints, and compare these with results from independent work. \secref~\ref{sec:nu} examines the implications of these results for cosmological constraints on neutrino masses, which depend sensitively on the accuracy of $\sigma_8$ measurements. Our constraints on dark energy parameters are presented in \secref~\ref{sec:demodels}. \secref{}s~\ref{sec:growthindex} and \ref{sec:nongauss} respectively investigate constraints on departures from GR and non-Gaussianities in the initial perturbation field.

\subsection{Cluster Constraints on \Omegam{} and $\sigma_8$} \label{sec:s8}

Within the standard class of cosmological models, constraints on \Omegam{} and $\sigma_8$ from cluster counts data at low redshifts are largely independent of the dark energy model assumed (e.g.\ \citealt{Vikhlinin0812.2720}, \mare). Constraints on these two parameters are typically degenerate, although data that probe the shape of the mass function or (more pertinently for this study) the growth of structure with time can break the degeneracy. Alternatively, or in addition, the gas mass fraction for relaxed clusters can be used to break the degeneracy by independently constraining \Omegam{}. Throughout this section, we use the \fgas{} data of \mamrakls{} in conjunction with the cluster counts and follow-up data (henceforth referring to their combination simply as ``clusters''); \secref~\ref{sec:followup} discusses the role of these individual components in more detail.

In the context of combining multiple cosmological probes, \Omegam{} is generally tightly constrained in any case. For this reason cluster-counts results are often reported in the form of a power law corresponding to the minor axis of the \Omegam--$\sigma_8$ confidence region, i.e.\ on the combination $\sigma_8\,\Omegam^\alpha$, where $\alpha$ encodes the slope of the degeneracy. From the cluster data, we find $\sigma_8(\Omegam/0.3)^{0.17} = 0.81\pm0.03$;\footnote{Our approach to choosing the exponent of this expression is to minimize the correlation between $\ln(\sigma_8\Omegam^{-\alpha})$ and $\ln(\sigma_8\Omegam^{1/\alpha})$ in the Markov chains from our analysis. Strictly speaking, the resulting value, $\alpha\sim0.17$, does not describe the minor axis of the confidence region (this would correspond to a slightly steeper value, $\alpha\sim0.23$), but rather generates the curves describing the best-fitting value and uncertainty of $\sigma_8$ as a function of \Omegam{}. Note that these values of $\alpha$ do not correspond simply to effective redshift and mass limits of the data set (e.g.\ \citealt{Weinberg1201.2434}) because we perform this analysis after marginalizing over systematic uncertainties, which limit the constraints on both \Omegam{} and $\sigma_8$. In addition, the cosmological dependences that enter into the measurement of cluster masses from real data, whether from X-ray or lensing observations, generally preclude such a simple interpretation.} the one-dimensional, marginalized constraints are $\Omegam=0.26\pm0.03$ and $\sigma_8=0.83\pm0.04$. These results are identical for \LCDM{} (with and without curvature) and flat constant-$w$ models. Even for models with free neutrino mass (\secref~\ref{sec:nu}), as well as for flat evolving-$w$ models (\secref~\ref{sec:demodels}), the cluster constraint on the width of the \Omegam--$\sigma_8$ ellipse remains equivalent to our result for the flat \LCDM{} case (although the slope of the degeneracy changes slightly). As expected, we find that the cluster constraints on $\sigma_8$ are limited by the precision of our overall mass calibration, parametrized by our prior on the normalization of the \mlens--$m$ relation (\secref~\ref{sec:scalingmod}). The a posteriori correlation of these parameters is such that a 10 per cent shift in the mass calibration implies a nearly 20 per cent shift in $\sigma_8$ at fixed $\Omegam$.

\begin{figure}
  \centering
  \includegraphics[scale=\figscale]{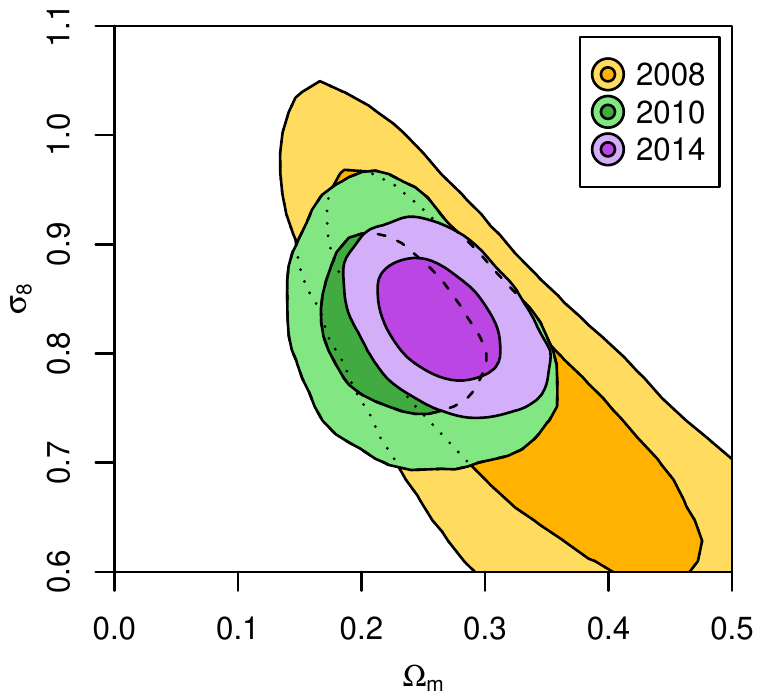}
  \caption{
    Constraints on \Omegam{} and $\sigma_8$ from this work (purple shading) and earlier works by these authors (yellow and green shading; \citealt{Mantz0709.4294}, \mare{}), accounting for systematic uncertainties. Dark and light shading respectively indicate the 68.3 and 95.4 per cent confidence regions. The underlying cluster survey data set is nearly identical across all three generations of results, but the approaches to calibrating cluster masses and the associated scaling relations have incorporated progressively better control of systematic uncertainties, leading to significantly tighter and more robust constraints. Contemporaneous priors on $h$ and $\Omegab h^2$ are included in each case (the improvement in these priors has negligible effect compared to the mass calibration). These results are essentially identical for flat and non-flat \LCDM{} models, and flat constant-$w$ dark energy models. In evolving-$w$ models and models with the neutrino mass free, the shape of the confidence region changes slightly, but its width ($\sigma_8$ at fixed \Omegam) remains the same.
  }
  \label{fig:oms8oldnew}
\end{figure}

\figref~\ref{fig:oms8oldnew} shows the joint constraints on $\sigma_8$ and \Omegam{} from clusters in the present analysis (purple shading)  along with previous results from these authors, namely \citet[][yellow banana]{Mantz0709.4294} and \mare{} (green shading), to emphasize the extent to which systematic uncertainties in mass calibration have decreased over time. In the first case, \citet{Mantz0709.4294} directly used hydrostatic mass estimates from the X-ray analysis of \citet{Reiprich0111285}, regardless of the clusters' dynamical states, marginalizing over generous allowances for the bias and scatter of these estimates with respect to the true masses (20 per cent uncertainties in each). \mare{} instead employed gas mass as a proxy for total mass, calibrating this relation using a hydrostatic X-ray analysis of relatively relaxed clusters by \citet{Allen0706.0033}, and marginalizing over systematic allowances for non-thermal support and instrument calibration at the $\sim15$ per cent level. As discussed above, the present work is calibrated to a gravitational lensing data set, providing $\sim8$ per cent precision on the mass calibration (\wtg). With only minor differences, these three results rest on the same underlying X-ray cluster catalogs. However, given their very different mass calibration strategies, the level of agreement between them, particularly considering the blind nature of the \wtg{} analysis, is encouraging.

Comparing our current results with \mare{}, we note that in both cases the \Omegam{} constraint is largely dictated by \fgas{} data. The inclusion of a robust mass calibration in both the cluster counts and \fgas{} analyses has led to significant improvement in the constraints on both parameters shown here,\footnote{Improvements in the modeling of the gas depletion in clusters also contribute to the improved \Omegam{} constraint from \fgas{} data; see \mamrakls{}.} nearly a factor of two reduction in the area of the 95.4 per cent confidence region.

\begin{figure*}
  \centering
  \includegraphics[scale=\figscale]{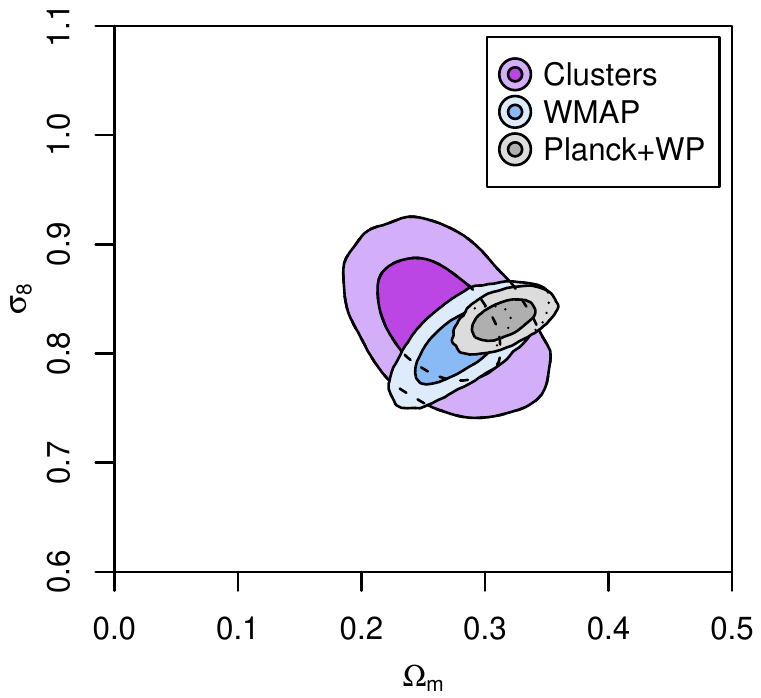}
  \hspace{1cm}
  \includegraphics[scale=\figscale]{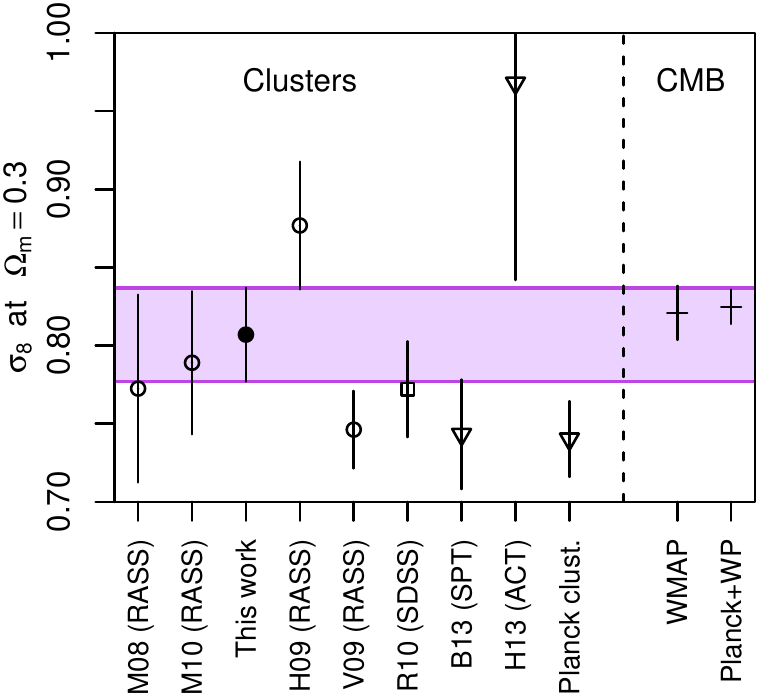}
  \caption{
    Left: Constraints from our cluster data (with standard priors on $h$ and $\Omegab h^2$) are compared with results from WMAP and \Planck+WP CMB data, assuming a flat \LCDM{} cosmology with minimal neutrino mass (assuming the normal mass hierarchy). Dark and light shading respectively indicate the 68.3 and 95.4 per cent confidence regions, including systematic uncertainties. The three sets of constraints are mutually consistent.
    Right: A number of marginalized constraints on $\sigma_8$ from the literature are compared at a common, concordance value of $\Omegam=0.3$. Results from clusters are shown by circles (X-ray surveys), squares (optical surveys) or triangles (SZ surveys), with crosses showing CMB constraints. The error bars include each author's estimate of the systematic uncertainties whenever possible (see text for details and references). The shaded, horizontal band reflects the 68.3 per cent confidence interval for our new result (filled circle), $\sigma_8(\Omegam/0.3)^{0.17}=0.81\pm0.03$. The mass calibration provided by the (blinded) \wtg{} lensing analysis (reflected by our $\sigma_8$ value) is in agreement with previous work by the same group (\citealt{Mantz0709.4294}; \mare{}), within the quoted statistical plus systematic uncertainties, and provides good agreement in $\sigma_8$ with CMB measurements, but is offset from some other cluster analyses.
  }
  \label{fig:oms8cmb}
  \label{fig:oms8comp}
\end{figure*}
 
The left panel of \figref~\ref{fig:oms8cmb} compares the new cluster constraints to results from WMAP (blue shading) and \Planck+WP (gray shading) CMB data for flat \LCDM{} models. Our results are consistent with either CMB data set. In particular, there is no tension between our cluster constraints and the 1-year \Planck+WP CMB results, in contrast to the \Planck{} analysis of cluster counts based on their own Sunyaev-Zel'dovich (SZ) effect cluster detections (\citealt{Planck1303.5080}; see also \citealt{von-der-Linden1402.2670}). 

Our result for $\sigma_8$ is compared to those of a selection of other galaxy cluster studies in the right panel of \figref~\ref{fig:oms8comp}. Given the parameter degeneracy, it is instructive to compare $\sigma_8$ constraints for a fixed, canonical value of $\Omegam$, in this case 0.3.\footnote{This choice is motivated by the tight constraints on $\Omegam\approx0.3$ obtained from the combination of available cosmological data, essentially independent of the model assumed (see \tabref~\ref{tab:death}).} In most cases (exceptions are noted below), the uncertainties on $\sigma_8$ at fixed \Omegam{} are reported to be limited by the absolute cluster mass calibration.\footnote{Where the authors provide an estimate of systematic uncertainty in their results, we include it in the figure, even if their ``baseline'' reported results include only statistical errors (e.g.\ \citealt{Vikhlinin0812.2720, Planck1303.5080}). When systematic allowances are included in the baseline results, but more conservative allowances are also considered (e.g.\ in a discussion section), we show the baseline results (e.g.\ \citealt{Rozo0902.3702, Benson1112.5435}). The exception is \citet{Henry0809.3832}, who do not explicitly account for systematic uncertainty in the mass calibration (the precision of constraints on the scaling relations limits the $\sigma_8$ measurement in this case).} The values and error bars in the figure thus primarily reflect the mass calibration used in each study and the adopted uncertainty  in that calibration, rather than, e.g., differences in the analysis methods used. Note that the present work (shaded region) is the first to self-consistently incorporate a mass calibration from weak lensing mass estimates, including a rigorous quantification of all systematic uncertainties.\footnote{A common practice has been to estimate rough systematic uncertainties by comparing hydrostatic mass estimates to lensing estimates, using a small number of clusters and assuming a fixed cosmology. Note that the X-ray/lensing mass ratio inferred from data in reality does depend on cosmological parameters (Applegate et~al., in prep.). Note also that in several cases hydrostatic mass calibrations have been implemented as priors on scaling relations, rather than by  directly incorporating mass estimates and simultaneously fitting the cosmology and scaling relation models. This approach makes it virtually impossible to properly account for all covariances among the parameters.}

Also shown in the right panel of \figref~\ref{fig:oms8comp} are the constraints from CMB anisotropy power spectra measured from WMAP and \Planck+WP data. For comparison to the similar figure in \citet[][their Figure~10]{Planck1303.5080}, note that here we show constraints on $\sigma_8$ at a fixed value of $\Omegam=0.3$, \emph{not} constraints on the combination $\sigma_8(\Omegam/0.27)^{0.3}$. The difference is negligible for cluster data but significant in the case of CMB data, for which that particular power-law approximately corresponds to the long axis of the parameter degeneracy (i.e.\ the least constrained direction). Given that the combination of available cosmological probes currently constrains \Omegam{} to be $\approx0.3$ to a precision of $\sim0.015$, essentially independent of the cosmological model assumed (see \tabref~\ref{tab:death}), taking a fixed value of \Omegam{} is arguably a more sensible choice for evaluating the tension, or lack thereof, among the cluster and CMB results.

The first three points shown in the figure are those of \citet{Mantz0709.4294}, \mare{} and this work, discussed above. Turning to the other results shown based on X-ray selected clusters, \citet{Henry0809.3832} analyzed a subset of the HIFLUGCS sample of \citet{Reiprich0111285}, calibrating the mass scaling relations by jointly fitting early weak lensing measurements, \Chandra{} X-ray hydrostatic masses and simulated clusters, with the results dominated by the X-ray mass estimates and simulations. The claimed precision on the mass calibration from this procedure is $<4$ per cent with no additional systematic uncertainty accounted for; the reported $\sigma_8$ constraints are instead limited by uncertainty in the slope and scatter of the mass--temperature relation. The analysis of \citet{Vikhlinin0812.2720} employs a combination of low-redshift RASS clusters and clusters at $0.35<z<0.9$ from the 400 square degree ROSAT catalog \citep{Burenin0610739}. Their mass calibration was based on hydrostatic estimates from \Chandra{} data, with a systematic uncertainty of $\sim9$ per cent estimated by comparing to the lensing data available at the time \citep{Hoekstra0705.0358, Zhang0802.0770}, assuming a fixed cosmology.

The results of \citet{Rozo0902.3702} are based on the optically selected MaxBCG catalog, derived from SDSS. Their mass calibration is from a stacked lensing analysis of the SDSS data, with a 6 per cent systematic allowance assigned based on the level of agreement between two analyses of the lensing data (systematics common to both analyses are not accounted for\footnote{Moreover, the 6 per cent agreement between the two lensing studies was reached only after correcting one of the methods by 18 per cent. While the motivation for the correction is ultimately well justified, it is difficult to completely dismiss the possibility of confirmation bias in such a case.}). Due to the steepness of the mass function and the nature of stacked analysis, this adopted calibration uncertainty applies most directly to the low-richness end of the cluster sample. The $\sigma_8$ constraints from this analysis are most dependent on the masses of high-richness clusters, where the statistical uncertainty is greater. The error budget for $\sigma_8$ is predominantly determined by this statistical component (see \citealt{Rozo0902.3702}).

The SZ cluster constraints shown include those from the SPT, ACT and \Planck{} cluster surveys. The SPT analysis of \citet{Benson1112.5435} ultimately uses the same X-ray mass calibration as that of \citet{Vikhlinin0812.2720}, and indeed the agreement between the two results is very close; the SPT constraints are slightly less tight due to an allowance for evolution in the mass calibration between the low-redshift calibration sample of \citet{Vikhlinin0812.2720} and the typical redshifts of SPT clusters. The ACT results shown in \figref~\ref{fig:oms8comp} use a mass calibration derived from galaxy velocity dispersion measurements, with an adopted systematic uncertainty of 15 per cent in mass \citep{Hasselfield1301.0816, Sifon1201.0991}. In this case, the particularly large uncertainties in $\sigma_8$ are most likely dominated by the small size of the data set used to constrain the SZ scaling relation (7 clusters with dynamical masses) rather than the 15 per cent prior on the mass calibration itself. The \Planck{} cluster results are marginalized over a uniform prior of $^{+20}_{-10}$ per cent in the mass calibration (for comparison to the Gaussian priors used elsewhere, this has a standard deviation of $\sim8.7$ per cent). For their main analysis, which does not account for this systematic uncertainty, the error bar is approximately half as large. The mass calibration in this case is tied to hydrostatic estimates based on XMM-{\it Newton} X-ray data.

Since the systematic uncertainties associated with the cluster mass scale have been only rough estimates in previous works, the right panel of \figref~\ref{fig:oms8comp} is in some sense more illustrative than informative. We would argue that earlier analyses based on X-ray masses for relaxed clusters should have included systematic uncertainties in their mass calibrations no smaller than the $\sim15$ per cent allowance included in \mare{}, and thus have comparable uncertainty in $\sigma_8$, and those that used hydrostatic masses for even unrelaxed clusters should include even larger uncertainties. Note that this does not necessarily imply better agreement among the cluster results themselves, given the considerable overlap in the clusters used for these hydrostatic mass calibrations (generally X-ray bright ROSAT clusters at redshifts $<0.3$). This only underscores the utility of mass estimates that are independent of X-ray detector calibrations and cluster dynamical state.

Apart from this work, the other major result based on a weak lensing calibration, and the only other result (excepting \citealt{Hasselfield1301.0816}) not ultimately based on X-ray hydrostatic mass estimates, is that of \citet{Rozo0902.3702}. While their adopted 6 per cent uncertainty in the mass calibration is arguably likely to be underestimated, it is interesting that their $\sigma_8$ measurement is the closest to ours of all the independent cluster results considered here.

We note that our value of $\sigma_8$ is marginally larger than some recent results from ground- and space-based cosmic shear and weak lensing tomography. For example, \citet{Kilbinger1212.3338} find $\sigma_8=0.74\pm0.03$ (again at fixed $\Omegam=0.3$) from a 2-dimensional cosmic shear analysis of CFHTLenS data.\footnote{\citet{Heymans1303.1808} obtained a nearly identical constraint, $0.74_{-0.04}^{+0.03}$, from a tomographic analysis of the CFHTLenS data.} The tomographic lensing analysis of HST COSMOS data by \citet{Schrabback0911.0053} yields a value of $0.75\pm0.08$, which is nominally lower than our constraints, but consistent within the uncertainties. 

\subsection{Constraints on Neutrino Mass} \label{sec:nu}

For a given amplitude of the matter power spectrum at the surface of last scattering, the predicted amplitude at low redshifts depends on the species-summed mass of neutrinos, \Mnu{}, with larger values of \Mnu{} corresponding to smaller values of $\sigma_8$ (for a review, see \citealt{Lesgourgues0603494}). Exploiting this degeneracy, constraints on $\sigma_8$ from clusters can be used in conjunction with CMB data (and other cosmological probes) to place limits on \Mnu{} that are considerably stronger than current laboratory experiments (e.g.\ \citealt{Allen1103.4829}). However, accurate constraints can only be obtained to the extent that the cluster and CMB measurements of the power spectrum amplitude are unbiased. Over the years, the combination of different data sets has led to gradually tightening upper limits on \Mnu{}, including occasional claims of a preference for $\Mnu>0$ (e.g.\ \citealt{Allen0306386, Tegmark0310723, Tereno0810.0555, Vikhlinin0812.2720, Mantz0911.1788, Reid0910.0008, Thomas0911.5291, Riemer-Sorensen1112.4940, Benson1112.5435, Burenin1301.4791, Reichardt1203.5775, Planck1303.5080, Beutler1403.4599, Dvorkin1403.8049}).

\begin{table}
  \begin{center}
    \caption{
      Posterior modes and 95.4 per cent confidence upper limits on \Mnu{} (in eV) from the combination of cluster, CMB, supernova and BAO data. The combined data sets include either WMAP (Comb$_\mathrm{WM}$) or \Planck+WP (Comb$_{Pl}$) all-sky CMB data; ACT and SPT CMB data are included in both cases. In addition to models with $r$ free, we show results employing a prior, $r=0.20^{+0.07}_{-0.05}$, based on results from the \citet{BICEP2-Collaboration1403.3985}.
    }
    \label{tab:nu}
    \vspace{1ex}
    \begin{tabular}{lcccc}
      \hline
      Model &\multicolumn{2}{c}{Comb$_\mathrm{WM}$} & \multicolumn{2}{c}{Comb$_{Pl}$} \\
      & Mode & 95.4\% lim. & Mode & 95.4\% lim. \\
      \hline
      flat \LCDM{} &  0.11 & 0.33 & 0.00 & 0.22 \\
      $\Omega_k$ free & 0.02 & 0.41 & 0.00 & 0.29 \\
      $w_0$ free & 0.00 & 0.46 &  0.08 & 0.38 \\
      $\Neff$ free & 0.05 & 0.31 & 0.06 & 0.29 \\
      $r$ free & 0.08 & 0.32 & 0.00 & 0.24 \\
      $r$ prior & 0.19 & 0.41 & 0.02 & 0.25 \\
      \hline
    \end{tabular}
  \end{center}
\end{table}

\begin{figure*}
  \centering
  \includegraphics[scale=\figscale]{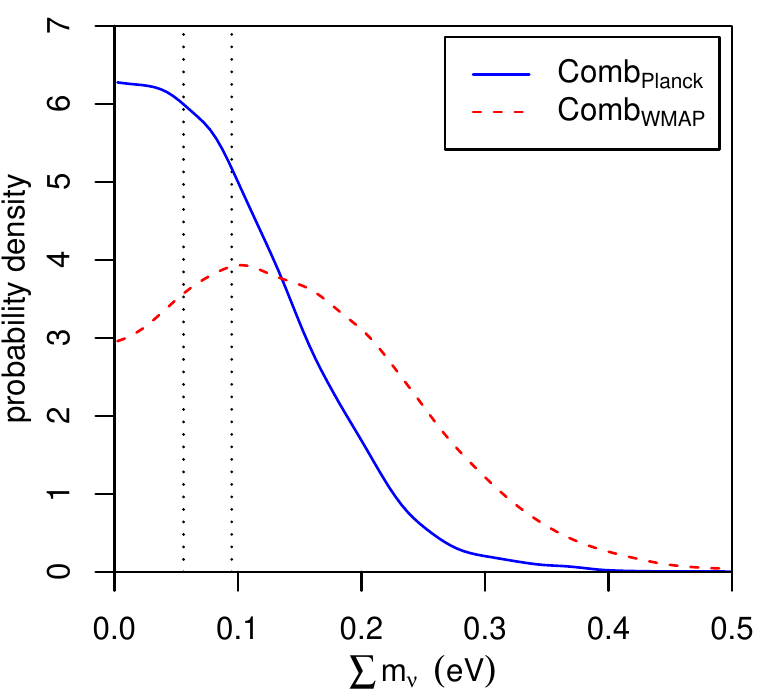}
  \hspace{1cm}
  \includegraphics[scale=\figscale]{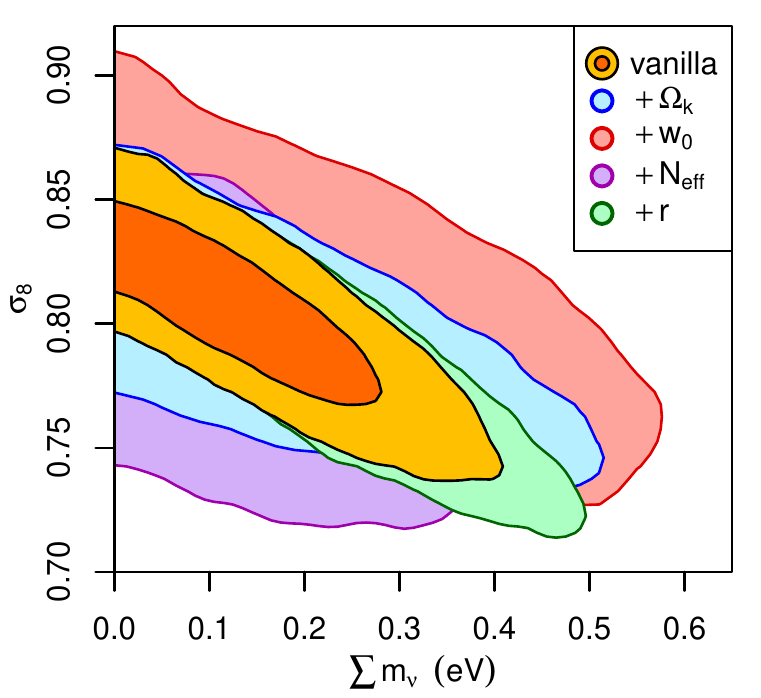}
  \caption{
    Left: Marginalized posterior distributions for \Mnu{}, assuming a flat \LCDM{} background and standard value of \Neff{}, from the combination of cluster, CMB, supernova and BAO data. The combined data include either \Planck+WP (solid line) or WMAP (dashed line) all-sky CMB data; ACT and SPT CMB data are included in both cases. Vertical, dotted lines indicate the minimum values of \Mnu{} implied by flavor oscillation measurements for the normal and inverted hierarchies (0.056 and 0.095\eV{}, respectively). Both data combinations are consistent with $\Mnu=0$.
    Right: For the combination using WMAP CMB data (see \appref~\ref{sec:planckfigs} for the equivalent \Planck+WP figure), we show 68.3 and 95.4 per cent confidence regions on $\Mnu$ and $\sigma_8$ for the flat \LCDM+\Mnu{} (``vanilla'') model (yellow/orange shading). The other regions correspond to 95.4 confidence (only) for models with an additional degree of freedom: spatial curvature (blue), the dark energy equation of state (red), the effective number of relativistic species (purple), or the amplitude of primordial tensor perturbations (green). In the latter case, we include the prior $r=0.20^{+0.07}_{-0.05}$, based on measurements by the \citet{BICEP2-Collaboration1403.3985}. The constraints when not including this additional prior are somewhat tighter (\tabref~\ref{tab:nu}).
  }
  \label{fig:mnu}
  \label{fig:mnus8}
\end{figure*}
 
The \citet{Planck1303.5080} recently published constraints on $\sigma_8$ from their SZ-detected clusters which, when combined with \Planck+WP CMB data and BAO, imply a $>2\sigma$ rejection of $\Mnu=0$. However, \citet{von-der-Linden1402.2670} have shown that the mass calibration used by the \Planck{} team is biased low compared with the \wtg{} lensing measurements. Our analysis of X-ray selected clusters, using the \wtg{} mass calibration, produces a $\sigma_8$ value that is consistent with both \Planck+WP and WMAP CMB measurements when assuming a minimal neutrino mass. We therefore expect that the combination of our data with either CMB data set will be fully consistent with minimal neutrino mass.

To quantify this, we first consider the simple case of a flat \LCDM{} model with the standard effective number of relativistic species ($\Neff=3.046$) and with \Mnu{} as a free parameter. Throughout this section, we model the three standard model neutrinos as being degenerate in mass. The posterior distributions for \Mnu{}, marginalizing over a flat \LCDM{} background, are shown in the left panel of \figref~\ref{fig:mnu} for the combinations of cluster, CMB, supernova and BAO data, where either WMAP (dashed line) or \Planck+WP (solid line) data are included in the combination (ACT and SPT CMB data are always included). Both results are consistent with the minimum summed neutrino mass implied by flavor oscillation data for either the normal or inverted hierarchies (vertical, dotted lines), or indeed with $\Mnu=0$. For the data combination including WMAP, the 68.3 and 95.4 per cent confidence intervals are $\Mnu=0.11\pm0.10\eV$  and $0.11_{-0.11}^{+0.22}\eV$;  the  corresponding limits for the combination including \Planck+WP data are $\Mnu=0_{-0.00}^{+0.12}\eV$ and $0_{-0.00}^{+0.22}\eV$. Using our gravitational lensing cluster mass calibration, there is thus no evidence for non-minimal or even non-zero neutrino mass in the best current cosmological data.

\tabref~\ref{tab:nu} shows the posterior modes and 95.4 per cent upper limits on \Mnu{} when additional cosmological parameters are free to vary: either global curvature ($\Omega_k$), the dark energy equation of state (constant-$w$), the effective number of relativistic species (\Neff{}), or the amplitude of the primordial tensor perturbation spectrum ($r$, the tensor-to-scalar ratio). In the latter case, we list constraints with $r$ completely free as well as results including the recent constraint from BICEP2, $r=0.20^{+0.07}_{-0.05}$, as an additional prior,\footnote{To be precise, we adopt a Gaussian prior of $-1.55\pm0.28$ on $\ln(r)$, which provides a good approximation to the posterior distribution for $r$ presented by the \citet{BICEP2-Collaboration1403.3985}. Forgoing the prior while leaving $r$ free actually tightens the limits on \Mnu{} somewhat, because the remaining data prefer a smaller value of $r$ than BICEP2 measures (e.g.\ \citealt{Story1210.7231}; see also e.g.\ \citealt{Flauger1405.7351, Mortonson1405.5857} and references therein).} fixing the tensor spectral index to zero as in the BICEP2 analysis \citep{BICEP2-Collaboration1403.3985}. \figref~\ref{fig:mnus8} shows the 95.4 per cent confidence regions for each case, from the full combination of data (including WMAP CMB data; see \appref~\ref{sec:planckfigs} for the equivalent \Planck+WP figure). The constraints are weaker in the more general models, particularly when $w$ is allowed to vary. Even in this case, however, there remains a degeneracy between \Mnu{} and $\sigma_8$. We comment on the prospects for improving neutrino mass limits further through tighter $\sigma_8$ measurements in \secref~\ref{sec:newlens}.

Of all the scenarios that we consider, the only ones that show even a marginal preference ($\gtsim1\sigma$) for non-zero neutrino mass are the basic flat \LCDM{}+\Mnu{} model ($1.1\sigma$ significance) and the model including tensor modes and a BICEP2 prior ($1.5\sigma$ significance), both when using WMAP CMB data. Keeping in mind that the tightest limits on \Mnu{} come from the combination of a cluster $\sigma_8$ measurement with CMB data, our null result stands in stark contrast to works that have adopted lower cluster mass calibrations (i.e.\ smaller values of $\sigma_8$) and subsequently claimed detections of neutrino mass from cosmological data (e.g., recently, \citealt{Burenin1301.4791, Planck1303.5080, Beutler1403.4599, Dvorkin1403.8049}).

\begin{figure*}
  \centering
  \includegraphics[scale=\figscale]{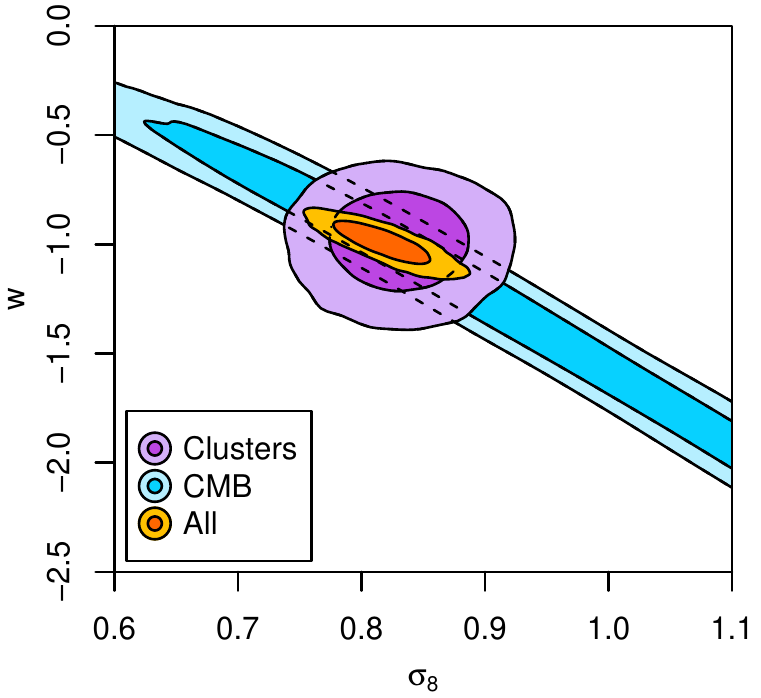}
  \hspace{1cm}
  \includegraphics[scale=\figscale]{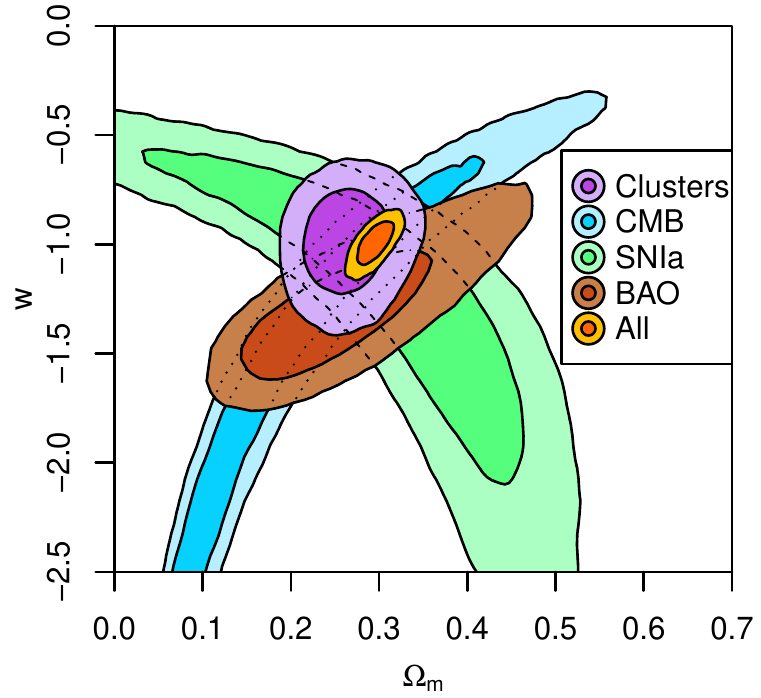}
  \caption{
    Constraints on constant-$w$ dark energy models with minimal neutrino mass from our cluster data (with standard priors on $h$ and $\Omegab h^2$) are compared with results from CMB (WMAP, ACT and SPT), supernova and BAO (also including priors on $h$ and $\Omegab h^2$) data, and their combination. The priors on $h$ and $\Omegab h^2$ are not included in the combined constraints. Dark and light shading respectively indicate the 68.3 and 95.4 per cent confidence regions, accounting for systematic uncertainties.
  }
  \label{fig:s8w}
  \label{fig:omw}
\end{figure*}

\subsection{Constraints on Dark Energy Models} \label{sec:demodels}

\begin{figure}
 \centering
 \includegraphics[scale=\figscale]{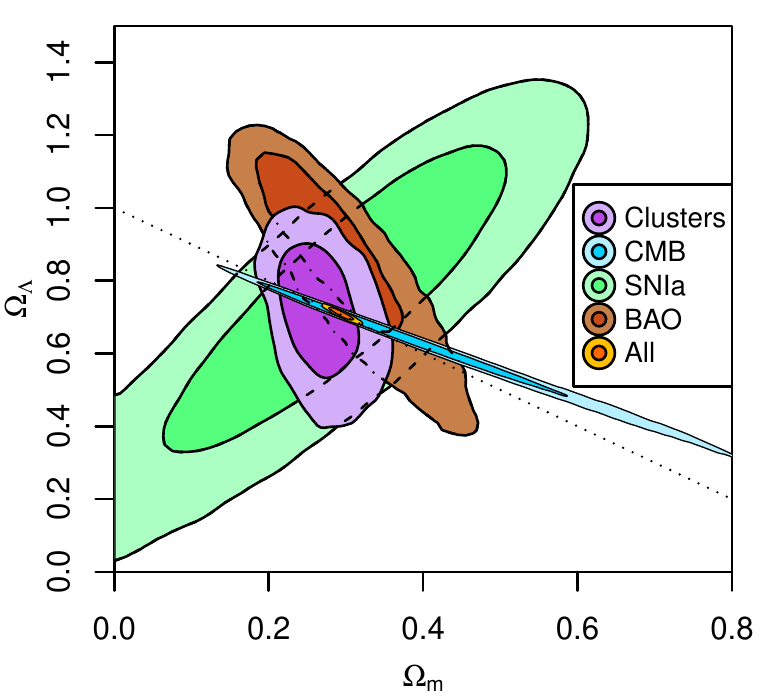}
 \caption{
   Constraints on \LCDM{} models (including curvature) with minimal neutrino mass from our cluster data (with standard priors on $h$ and $\Omegab h^2$) are compared with results from CMB (WMAP, ACT and SPT), supernova and BAO (also including priors on $h$ and $\Omegab h^2$) data, and their combination. The priors on $h$ and $\Omegab h^2$ are not included in the combined constraints. Dark and light shading respectively indicate the 68.3 and 95.4 per cent confidence regions, accounting for systematic uncertainties. The dotted line denotes spatially flat models.
 }
 \label{fig:lcdm}
\end{figure}

 We next investigate the constraints on dark energy models afforded by the cluster data alone, and their combination with other cosmological probes. The results appear in \tabref~\ref{tab:death}. For spatially flat, constant-$w$ models, our cluster data alone provide identical constraints on \Omegam{} and $\sigma_8$ to the flat \LCDM{} case, and additionally constrain the equation of state: $w=-0.98\pm0.15$. Note that the precision of the $w$ constraint is identical to what was obtained from the combination of cluster counts and \fgas{} data by \mare{} (i.e.\ without lensing data) as we would expect; the addition of weak lensing data for 50 clusters significantly enhances constraints on $\sigma_8$, but (due to the relatively low precision of lensing masses for individual clusters) has not tightened constraints on the redshift-dependent signal that determines $w$. Even so, the constraints on $w$ are impressive and competitive with the best other cosmological probes (below), as well as independent results from X-ray (\citealt{Vikhlinin0812.2720}; see also \citealt{Burenin1202.2889}) and SZ-selected clusters \citep{Benson1112.5435}.

The left panel of \figref~\ref{fig:s8w} shows the constraints in the $\sigma_8$--$w$ plane from clusters alone, CMB data alone, and the combination of clusters, CMB, supernova and BAO data for constant-$w$ models. (For figures in this section, ``CMB'' refers to the combination of WMAP data with ACT and SPT power spectra. Figures obtained using \Planck+WP data instead of WMAP are qualitatively and quantitatively similar, and appear in \appref~\ref{sec:planckfigs} for completeness; see also \tabref~\ref{tab:death}.) The joint constraints on \Omegam{} and $w$ from the various data sets are shown in the right panel of \figref~\ref{fig:omw}. From the combination of data, we obtain $w=-0.99\pm0.06$ ($-1.03\pm0.06$ for the combination using \Planck+WP data).

\begin{table*}
  \begin{center}
    \caption{
      Marginalized (one-dimensional) best-fitting values and 68.3 per cent maximum-likelihood confidence intervals for the parameters of various dark energy models, including systematic uncertainties. The parametrization of the equation of state is defined in \secref~\ref{sec:cosmomodel}. The ``Clusters'' data incorporates X-ray survey data, X-ray follow-up observations (providing mass proxies in general and \fgas{} measurements for relaxed clusters), and weak lensing data (\wtg). The ``Comb$_\mathrm{WM}$'' combination of data refers to the union of our cluster data set with CMB power spectra from WMAP \citep{Hinshaw1212.5226}, ACT \citep{Das1301.1037} and SPT \citep{Keisler1105.3182, Reichardt1111.0932, Story1210.7231}, the Union 2.1 compilation of type Ia supernovae \citep{Suzuki1105.3470}, and baryon acoustic oscillation measurements at $z=0.106$ \citep{Beutler1106.3366}, $z=0.35$ \citep{Padmanabhan1202.0090} and $z=0.57$ \citep{Anderson1303.4666}. ``Comb$_{Pl}$'' is identical, with the exception that 1-year \Planck{} data (plus WMAP polarization; \citealt{Planck1303.5075}) are used in place of the complete 9-year WMAP data. The clusters-only constraints incorporate standard priors on $h$ and $\Omegab h^2$ (\secref~\ref{sec:data}; \citealt{Riess1103.2976, Cooke1308.3240}). Parameters listed with no error bars for a given model are fixed. Parameters with no value listed are not relevant, given the other parameters that are fixed in that model. For the models in which $w_a$ is a free parameter (bottom section of table), there is no sensitivity to the transition time parameterized by \atr{}; therefore, \atr{} is either fixed (to 0.5) or is marginalized over the range 0.5 to 0.95 (indicated by the ``---'' symbol in the \atr{} column). The last column indicates in which figure, if any, the corresponding results are displayed.
    }
    \label{tab:death}
    \vspace{1ex}
    \begin{tabular}{cccccccccc}
      \hline\vspace{-2.5ex}\\
      Data & $\sigma_8$ & $\Omegam$ & $\Omegade$ & \phmin$10^3\Omega_k$ & \phmin$w_0$ & \phmin$w_a$ & \phmin$\wet$ & $\atr$ & Fig.\\
      \hline\vspace{-2ex}\\
      Clusters & $0.830\pm0.035$ & $0.259\pm0.030$ & & \phmin0 & $-1$ & \phmin0 & & & \ref{fig:oms8oldnew}\vspace{3.0ex}\\
      Clusters & $0.830\pm0.035$ & $0.261\pm0.032$ & $0.728\pm0.115$ & \phmin$8\pm110$ & $-1$ & \phmin0 & & & \ref{fig:lcdm}\vspace{1ex}\\
      Comb$_\mathrm{WM}$ & $0.814\pm0.019$ & $0.294\pm0.010$ & $0.709\pm0.011$ & $-3\pm4$ & $-1$ & \phmin0 & & & \ref{fig:lcdm}\vspace{1ex}\\
      Comb$_{Pl}$ & $0.823\pm0.013$ & $0.302\pm0.009$ & $0.698\pm0.009$ & \phmin$0\pm4$ & $-1$ & \phmin0 & & & \ref{fig:planckfigs}\vspace{3.0ex}\\
      Clusters & $0.831\pm0.036$ & $0.261\pm0.031$ & & \phmin0 & $-0.98\pm0.15$ & \phmin0 & & & \ref{fig:s8w}\vspace{1ex}\\
      Comb$_\mathrm{WM}$ & $0.819\pm0.026$ & $0.295\pm0.013$ & & \phmin0 & $-0.99\pm0.06$ & \phmin0 & & & \ref{fig:s8w}\vspace{1ex}\\
      Comb$_{Pl}$ & $0.833\pm0.021$ & $0.297\pm0.013$ & & \phmin0 & $-1.03\pm0.06$ & \phmin0 & & & \ref{fig:planckfigs}\vspace{3.0ex}\\
      Comb$_\mathrm{WM}$ & $0.818\pm0.023$ & $0.289\pm0.014$ & $0.715\pm0.016$ & $-5\pm5$ & $-1.03\pm0.07$ & \phmin0 & & &\vspace{1ex}\\
      Comb$_{Pl}$ & $0.836\pm0.021$ & $0.292\pm0.014$ & $0.713\pm0.015$ & $-4\pm4$ & $-1.08\pm0.07$ & \phmin0 & & &\vspace{3.0ex}\\
      Clusters & $0.829\pm0.036$ & $0.261\pm0.026$ & & \phmin0 & $-0.69^{+0.32}_{-0.36}$ & $-1.6^{+1.9}_{-1.3}$ & $-2.3^{+1.6}_{-1.0}$ & 0.5 & \ref{fig:clwevol}a\vspace{1ex}\\
      Comb$_\mathrm{WM}$ & $0.816\pm0.027$ & $0.292\pm0.015$ & & \phmin0 & $-1.04^{+0.13}_{-0.18}$ & \phmin$0.3^{+0.4}_{-0.6}$ & $-0.8^{+0.3}_{-0.4}$ & 0.5 & \ref{fig:clwevol}a\vspace{1ex}\\
      Comb$_{Pl}$ & $0.835\pm0.021$ & $0.298\pm0.015$ & & \phmin0 & $-0.96^{+0.15}_{-0.18}$ & $-0.3^{+0.6}_{-0.5}$ & $-1.2^{+0.4}_{-0.4}$ & 0.5 & \ref{fig:planckfigs}\vspace{3.0ex}\\
      Clusters & $0.827\pm0.036$ & $0.262\pm0.023$ & & \phmin0 & $-0.71^{+0.62}_{-0.42}$ & $-1.0^{+1.5}_{-1.4}$ & $-1.4^{+0.8}_{-1.1}$ & --- &\vspace{1ex}\\
      Comb$_\mathrm{WM}$ & $0.818\pm0.025$ & $0.291\pm0.015$ &  & \phmin0 & $-1.09^{+0.23}_{-0.22}$ & \phmin$0.2^{+0.5}_{-0.5}$ & $-0.9^{+0.2}_{-0.2}$ & --- &\vspace{1ex}\\
      Comb$_{Pl}$ & $0.834\pm0.02$1 & $0.300\pm0.015$ &  & \phmin0 & $-0.97^{+0.24}_{-0.20}$ & $-0.2^{+0.5}_{-0.5}$ & $-1.1^{+0.2}_{-0.3}$ & --- &\vspace{3.0ex}\\
      Comb$_\mathrm{WM}$ & $0.822\pm0.026$ & $0.294\pm0.015$ & $0.713\pm0.016$ & $-8\pm6$ & $-0.93^{+0.24}_{-0.20}$ & $-0.4^{+1.0}_{-1.1}$ & $-1.3^{+0.8}_{-0.9}$ & 0.5 & \ref{fig:wevol}b\vspace{1ex}\\
      Comb$_{Pl}$ & $0.840\pm0.022$ & $0.302\pm0.015$ & $0.705\pm0.016$ & $-8\pm5$ & $-0.87^{+0.26}_{-0.20}$ & $-0.8^{+0.9}_{-1.4}$ & $-1.6^{+0.7}_{-1.1}$ & 0.5  & \ref{fig:planckfigs}\vspace{3.0ex}\\
      Comb$_\mathrm{WM}$ & $0.822\pm0.025$ & $0.295\pm0.016$ & $0.712\pm0.016$ & $-7\pm7$ & $-0.97^{+0.40}_{-0.22}$ & $-0.1^{+0.6}_{-1.2}$ & $-1.1^{+0.5}_{-0.7}$ & --- & \ref{fig:wevol}b\vspace{1ex}\\
      Comb$_{Pl}$ & $0.838\pm0.021$ & $0.304\pm0.016$ & $0.703\pm0.016$ & $-7\pm5$ & $-0.71^{+0.24}_{-0.36}$ & $-1.1^{+1.1}_{-0.7}$ & $-1.3^{+0.3}_{-0.8}$ & --- & \ref{fig:planckfigs}\vspace{1ex}\\
      \hline
    \end{tabular}
  \end{center}
\end{table*}

\figref~\ref{fig:lcdm} presents the analogous results for \LCDM{} models including global curvature. Here the cluster data constrain the dark energy density to be $\Omegal=0.73\pm0.12$, a 60 per cent improvement relative to the constraints from the \fgas{} data alone (\mamrakls). The combination of all data strongly prefers spatial flatness, with $10^3\Omega_k=-3\pm4$ and $0\pm4$ from the combinations using WMAP and \Planck+WP data, respectively.

\begin{figure*}
  \centering
  \includegraphics[scale=\figscale]{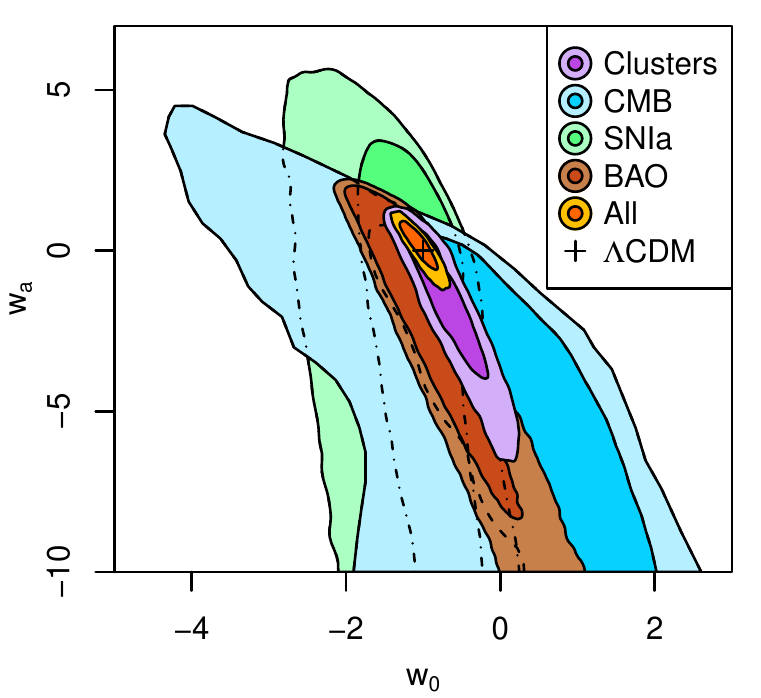}
  \hspace{1cm}
  \includegraphics[scale=\figscale]{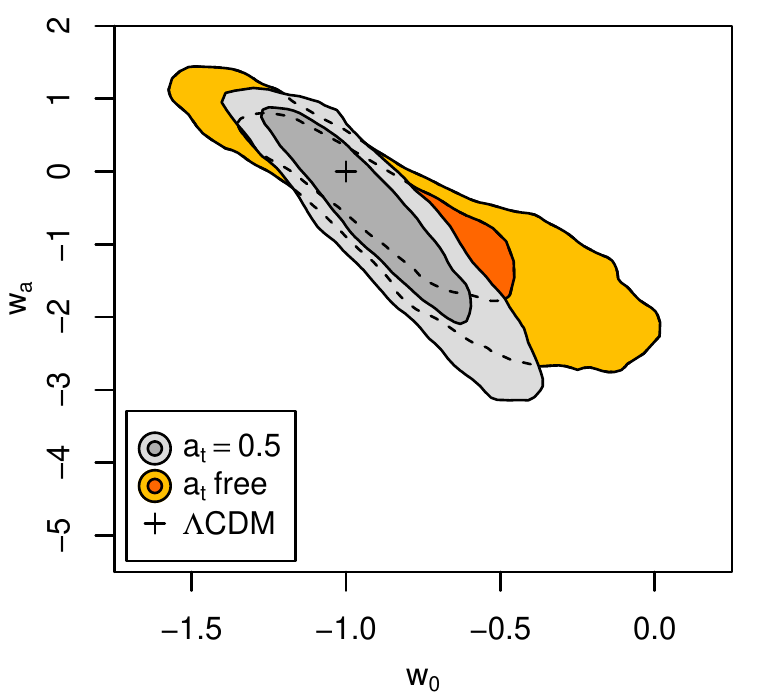}
  \caption{
    Left: Constraints on evolving-$w$ dark energy models with minimal neutrino mass and without global curvature from our cluster data (with standard priors on $h$ and $\Omegab h^2$) are compared with results from CMB (WMAP, ACT and SPT), supernova and BAO (also including priors on $h$ and $\Omegab h^2$) data, and their combination. The priors on $h$ and $\Omegab h^2$ are not included in the combined constraints. Dark and light shading respectively indicate the 68.3 and 95.4 per cent confidence regions, accounting for systematic uncertainties. The cross indicates the \LCDM{} model ($w_0=-1$, $w_a=0$).
    Right: Constraints on evolving-$w$ models with global curvature as a free parameter from the combination of cluster, CMB, supernova and BAO data. For the model with $\atr$ free, this parameter is marginalized over the range $0.5<\atr<0.95$ (see \eqnref~\ref{eq:wdef}). In all cases, we find consistency with the standard cosmological-constant model.
  }
  \label{fig:clwevol}
  \label{fig:wevol}
\end{figure*}

Turning to models with an evolving equation of state, we first consider the simplest case without spatial curvature. With this assumption, the cluster data alone are able to constrain the $w_0$ and $w_a$ parameters of the evolving dark energy model (see \eqnref~\ref{eq:wdef}), even when \atr{} is free (marginalized over $0.5<\atr<0.95$). Individual constraints from cluster, CMB, supernova and BAO data are shown in the left panel of \figref~\ref{fig:clwevol}, along with constraints from the combination of data, when \atr{} is fixed to 0.5. Regardless of which all-sky CMB data set is used and whether or not \atr{} is fixed, we find consistency with the cosmological-constant model.

\tabref~\ref{tab:death} also shows constraints for models including both free curvature ($\Omega_k$) and an evolving equation of state. In all cases, the cluster data, and the combinations of cluster and other leading data sets, remain consistent with spatial flatness and a cosmological constant (see the right panel of \figref~\ref{fig:wevol} for models including free curvature). Comparing to \mamrakls{}, who use identical \fgas{}, CMB, supernova and BAO data but not cluster counts, we generally find improvement in the constraints on $w_0$, and less so for $\Omega_k$ and $w_a$. In the most general model we consider, the constraint on $w_0$ shrinks from $-0.99\pm0.34$ to $-0.97^{+0.40}_{-0.22}$ for the combination using WMAP CMB data (from $-0.75\pm0.34$ to $-0.71^{+0.24}_{-0.36}$ for the combination using \Planck+WP data).

\subsection{Constraints on Modifications of Gravity} \label{sec:growthindex}

While dark energy (in the form of a cosmological constant) has been a mainstay of the standard cosmological model since the discovery that the expansion of the Universe is accelerating, other explanations for acceleration are possible. In particular, various modifications to GR in the large-scale/weak-field limit have been proposed (for recent
reviews see, e.g., \citealt{Frieman0803.0982, Clifton1106.2476, Joyce1407.0059}). Being sensitive to the action of gravity in this regime, the growth of cosmic structure has the potential to distinguish between dark energy and modified gravity theories that predict identical expansion histories.

\begin{figure*}
  \centering
  \includegraphics[scale=\figscale]{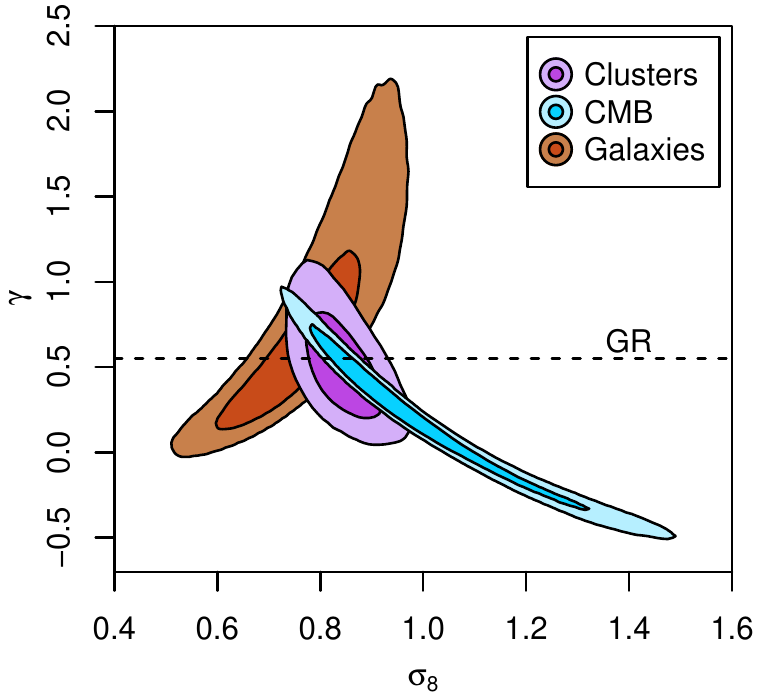}
  \hspace{1cm}
  \includegraphics[scale=\figscale]{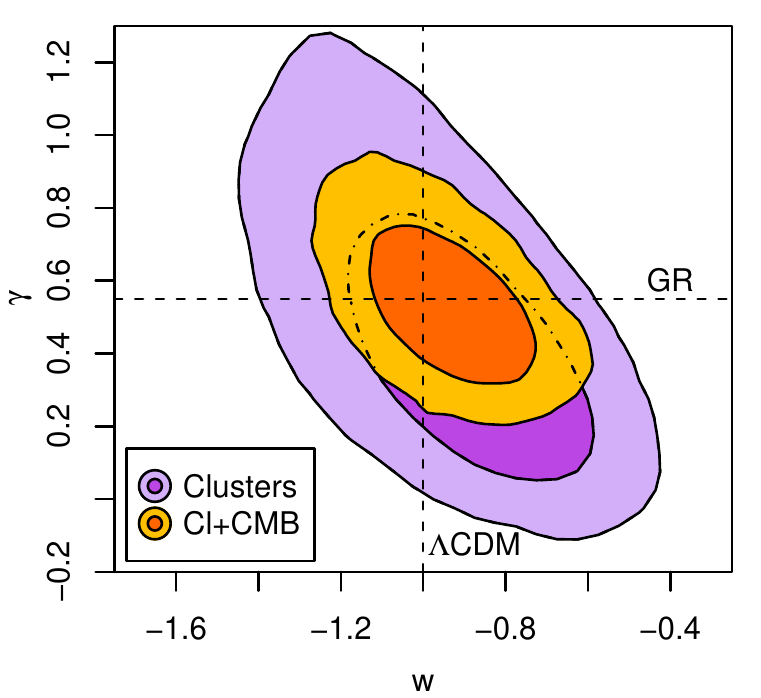}
  \caption{
    Constraints on models where the growth index of cosmic structure formation, $\gamma$, is a free parameter. Dark and light shading respectively indicate the 68.3 and 95.4 per cent confidence regions, accounting for systematic uncertainties. Left: Constraints from clusters, the CMB, and galaxy survey data individually, marginalizing over the standard flat \LCDM{} parametrization of the cosmic expansion history. Note that the treatment of the galaxy survey data uses a multivariate Gaussian approximation to constraints from RSD and the AP effect (see also \citealt{ Rapetti1205.4679}). GR corresponds approximately to $\gamma=0.55$ (dashed line). Right: Constraints from clusters and the combination of clusters and the CMB for models where $w$ is allowed to be free in the parametrization of the expansion history (this parameter does not directly affect the growth history in this model). Here the horizontal and vertical dashed lines respectively correspond to the standard models for the growth of cosmic structure (GR) and the expansion of the Universe (\LCDM{}). In these figures, `CMB' refers to the combination of ACT, SPT and WMAP data; see \appref~\ref{sec:planckfigs} for the corresponding figures using \Planck+WP instead of WMAP data.
  }
  \label{fig:gamma}
\end{figure*}

A simple and entirely phenomenological approach involves modifying the growth rate of density perturbations at late times, when the growth is approximately scale-independent. We adopt the simple parametrization in terms of the growth index, $\gamma$ (e.g.\ \citealt{Linder0507263}),
\begin{equation} \label{eq:growthindex}
  f(a) = \frac{d\ln\,\delta}{d\ln\,a} = \Omegam(a)^\gamma,
\end{equation}
where $\delta$ is the linear density contrast in synchronous gauge (at any scale), and where $\gamma=0.55$ approximately corresponds to GR for a wide range of expansion histories compatible with current data \citep{Polarski0710.1510}. Note that constraints on the growth index serve only as a useful consistency check of GR, rather than directly testing GR against alternative models of gravity. Constraints on $\gamma$ from earlier versions of our cluster analysis (in conjunction with contemporaneous cosmological data) are presented by \citet{Rapetti0812.2259, Rapetti0911.1787, Rapetti1205.4679}. Independent constraints from other data sets have been obtained by, e.g., \citet{Nesseris0710.1092}, \citet{di-Porto0707.2686}, \citet{Samushia1206.5309, Samushia1312.4899} and \citet{Beutler1403.4599}.

\begin{table}
  \begin{center}
    \caption{
      Marginalized best-fitting values and 68.3 per cent maximum-likelihood confidence intervals for the growth index ($\gamma$), $\sigma_8$, and $w$ from clusters (Cl), the CMB and galaxy survey data (gal). Here $\gamma$ determines the late-time growth of cosmic structure, and $w$ should be interpreted purely as a modification to the \LCDM{} expansion model (but not directly to the growth). Subscripts `WM' and `$Pl$' denote the use of WMAP or \Planck{}+WP data in combination with ACT and SPT. Note: $^a$the combinations with galaxy survey data should be treated with caution due to the caveats noted in the text.
    }
    \label{tab:gamma}
    \vspace{1ex}
    \begin{tabular}{lccc}
      \hline
      Data & $\gamma$ & $\sigma_8$ & $w$\\
      \hline
      Cl & $0.48\pm0.19$ & $0.833\pm0.048$ & $-1$\\
      Cl+CMB$_\mathrm{WM}$ & $0.56\pm0.13$ & $0.824\pm0.037$ & $-1$\\
      Cl+CMB$_\mathrm{WM}$+gal$^a$ & $0.66\pm0.06$  & $0.802\pm0.016$ & $-1$\\
      Cl+CMB$_{Pl}$ & $0.58\pm0.12$ & $0.824\pm0.037$ & $-1$\\
      Cl+CMB$_{Pl}$+gal & $0.67\pm0.06$  & $0.799\pm0.015$ & $-1$\vspace{1ex}\\
      Cl & $0.39\pm0.24$ & $0.850\pm0.055$ & $-0.90\pm0.19$\\
      Cl+CMB$_\mathrm{WM}$ & $0.52\pm0.14$ & $0.817\pm0.040$ & $-0.94\pm0.13$\\
      Cl+CMB$_\mathrm{WM}$+gal & $0.60\pm0.08$  & $0.792\pm0.020$ & $-0.91\pm0.08$\\
      Cl+CMB$_{Pl}$ & $0.57\pm0.14$ & $0.828\pm0.040$ & $-1.01\pm0.13$\\
      Cl+CMB$_{Pl}$+gal & $0.63\pm0.07$  & $0.799\pm0.015$ & $-0.96\pm0.07$\\
    \hline
    \end{tabular}
  \end{center}
\end{table}
 
We follow \citet{Rapetti1205.4679}, investigating the constraints on $\gamma$ from our cluster data, the integrated Sachs-Wolfe (ISW) effect on the CMB,\footnote{Cosmic growth also leaves an imprint at high multipoles through CMB lensing, but currently the CMB constraints on $\gamma$ primarily come from the ISW effect.} and measurements of redshift-space distortions (RSD) and the Alcock-Paczynski (AP) effect from galaxy survey data. In practice, we use {\sc camb} to calculate and tabulate $P(k,z)$ assuming GR, then modify these values from $z=30$ (well into the matter-dominated regime, where $f\rightarrow1$ independent of $\gamma$) onward to be consistent with the growth given by \eqnref~\ref{eq:growthindex}. This modified power spectrum is then integrated when evaluating the cluster mass function (\eqnref{}s~\ref{eq:sigma2def}--\ref{eq:massfunction}). For details of the calculation of the ISW effect in this model, see \appref~\ref{sec:isw}; as in earlier sections, we use CMB data from ACT, SPT, and either \Planck+WP or WMAP. The galaxy survey data include results from 6dF \citep{Beutler1204.4725}, SDSS \citep{Reid1203.6641} and the WiggleZ Dark Energy Survey \citep{Blake1108.2637}. Their likelihood is approximated by a multivariate Gaussian, encoding measurements of $f\sigma_8(z)$ and $F(z)=(1+z)\dA(z)H(z)/c$ at several redshifts, assuming zero neutrino mass; here $\dA$ is the angular diameter distance, and $c$ is the speed of light. For consistency, we fix $\Mnu=0$ in this section for all data sets, rather than using the baseline value of $0.056\eV$ employed elsewhere in this paper. Due to the approximate nature of the galaxy survey likelihood used here, compared with the analysis of cluster and CMB data, we urge caution in interpreting the results that combine all three data sets. However, the level of precision that is in principle available from this combination (\tabref~\ref{tab:gamma}) motivates a more complete analysis of the galaxy survey data, i.e.\ accounting for all parameter covariances, in future work.

The left panel of \figref~\ref{fig:gamma} shows the constraints on $\gamma$ and $\sigma_8$ from clusters, the CMB and galaxy survey data individually. In addition to the parameters shown, we marginalize over the standard set of free parameters of the flat \LCDM{} model. In the case of CMB or galaxy survey data alone, there are strong but complementary degeneracies (as discussed by \citealt{Rapetti1205.4679}), whereas the cluster data (with standard priors) constrain the entire model; the marginalized constraints from clusters are $\gamma=0.48\pm0.19$ and $\sigma_8=0.83\pm0.05$.

All three data sets shown are individually consistent with $\gamma=0.55$. Their combination has a marginal ($<2\sigma$) preference for higher values of $\gamma$ (\tabref~\ref{tab:gamma}), though this should be viewed with caution in light of the caveats mentioned above (see also \citealt{Beutler1312.4611}). The combination of clusters and the CMB (without galaxy survey data) is fully consistent with GR.

In the right panel of \figref~\ref{fig:gamma}, we present constraints on models when additional freedom is introduced into the model for the cosmic expansion in the form of the $w$ parameter. In this model, $w$ should not be interpreted as the dark energy equation of state, but simply as a phenomenological departure from the cosmic expansion model given by \LCDM{}, in the same way that $\gamma$ parametrizes departures of the growth history from that given by GR. (In particular, dark energy perturbations associated with values of $w$ different from $-1$ are not included in the growth equations, which instead depend on $\gamma$ through \eqnref~\ref{eq:growthindex}.) The figure shows constraints from clusters alone, and the combination of cluster and CMB data. Here again, the clusters and clusters+CMB data are fully consistent with the standard $w=-1$, $\gamma=0.55$ model, although the full combination, including the galaxy survey data, exhibits mild ($<2\sigma$) tension (\tabref~\ref{tab:gamma}).

\subsection{Constraints on Non-Gaussianity} \label{sec:nongauss}

\begin{figure*}
  \centering
  \includegraphics[scale=\figscale]{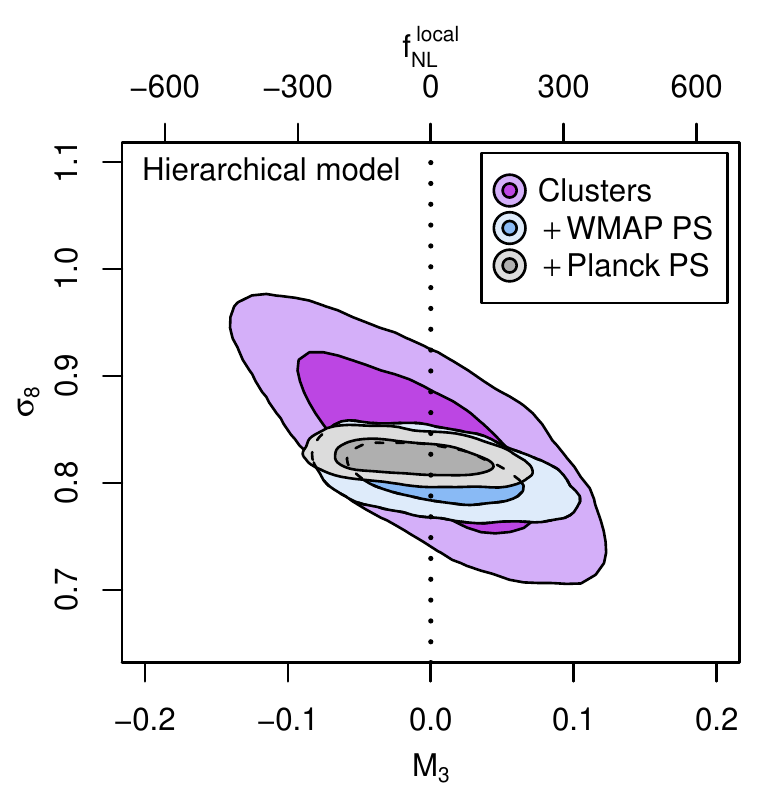}
  \hspace{1cm}
  \includegraphics[scale=\figscale]{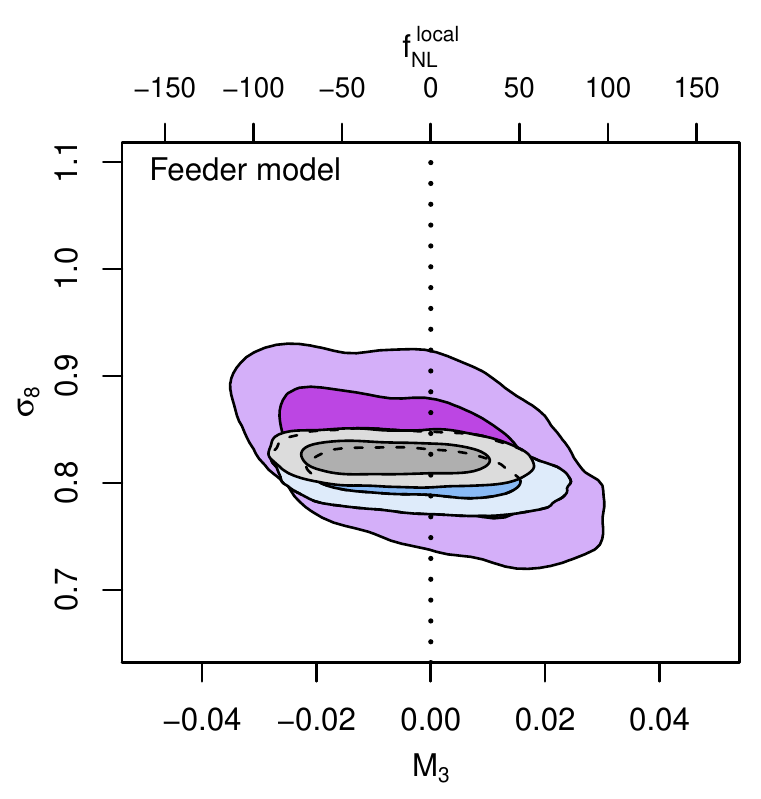}
  \caption{
    Constraints on hierarchical-type and feeder-type inflation models, in which the level of primordial non-Gaussianity is parameterized by $\moment_3$ (see text), from clusters and the combination of cluster and CMB data. (Note that feeder models generate more non-Gaussianity for a given value of $\moment_3$, hence the difference in scale between the two panels.) Dark and light shading respectively indicate the 68.3 and 95.4 per cent confidence regions, accounting for systematic uncertainties. When combining cluster and CMB data, we use only the CMB power spectra (PS) here (not bi- or trispectra). 
  }
  \label{fig:nongauss}
\end{figure*}

In the standard cosmological model, the primordial density perturbations sourced by inflation are assumed to be Gaussian, in which case their statistical properties are completely described by the power spectrum (i.e.\ two-point correlation function). However, many viable inflation models produce non-Gaussianity, which results in non-vanishing higher-order correlations (see, e.g., \citealt{Bartolo0406398}). CMB and galaxy survey studies of non-Gaussianity typically focus on constraining the amplitude of the bispectrum (three-point function), parametrized by \fnl{}, for a given ``triangle'' template configuration of momentum vectors (e.g.\ \citealt{Bennett1212.5225, Planck1303.5084}).

\begin{table*}
  \begin{center}
    \caption{
      Best-fitting values and 68.3 per cent confidence intervals for $\sigma_8$ and the non-Gaussian parameter $\moment_3$ from the cluster data set and its combination with CMB data for hierarchical-type (H) and feeder-type (F) inflation models. Note that we use only the CMB power spectra here (not bi- or trispectra). Hence, the CMB data refine our results by improving the constraints on $\sigma_8$ (and, to a lesser extent, other cosmological parameters), but do not directly constrain the non-Gaussian model. We also list the equivalent constraints on the level of non-Gaussianity in the bispectrum, \fnl{}, for the three canonical triangle configurations (local, equilateral and orthogonal; see \citealt{Shandera1304.1216} and \citealt{Adhikari1402.2336} for details of this conversion).
    }
    \label{tab:ng}
    \vspace{1ex}
    \begin{tabular}{clcrrrr}
      \hline\vspace{-2ex}\\
      Model & \multicolumn{1}{c}{Data} & $\sigma_8$ & $10^3\moment_3$ & $\fnl^\mathrm{local}$ & $\fnl^\mathrm{equil}$ & $\fnl^\mathrm{orthog}$\vspace{0.5ex}\\
      \hline\vspace{-1.5ex}\\
      H & Clusters & $0.835\pm0.053$ & $8^{+40}_{-65}$ & $24^{+129}_{-210}$ & $88^{+471}_{-765}$ & $-123^{+1066}_{-656}$\vspace{1ex}\\
      H & Clusters+WMAP & $0.808\pm0.019$ & $-1^{+42}_{-36}$ & $-3^{+135}_{-116}$ & $-12^{+494}_{-424}$ & $16^{+590}_{-689}$\vspace{1ex}\\
      H & Clusters+\Planck+WP & $0.823\pm0.011$ & $-29^{+46}_{-24}$ & $-94^{+148}_{-77}$ & $-341^{+541}_{-282}$ & $475^{+393}_{-754}$\vspace{1ex}\\
      F & Clusters & $0.830\pm0.041$ & $4^{+8}_{-21}$ & $11^{+26}_{-68}$ & $41^{+94}_{-247}$ & $-57^{+344}_{-131}$\vspace{1ex}\\
      F & Clusters+WMAP & $0.810\pm0.017$ & $-7^{+16}_{-9}$ & $-23^{+50}_{-27}$ & $-85^{+182}_{-100}$ & $119^{+139}_{-254}$\vspace{1ex}\\
      F & Clusters+\Planck+WP & $0.823\pm0.011$ & $-15^{+19}_{-4}$ & $-48^{+60}_{-11}$ & $-174^{+218}_{-41}$ & $242^{+57}_{-303}$\vspace{1ex}\\
      \hline
   \end{tabular}
  \end{center}
\end{table*}

For clusters, non-Gaussianity manifests itself in an enhancement or suppression of the mass function at the highest masses, and respectively a corresponding suppression or enhancement at low masses, relative to the Gaussian case. Importantly, the cluster signal is influenced by the entire series of $n$-point correlation functions \citep{LoVerde0711.4126, Shandera1211.7361}, and therefore has the potential to distinguish competing models of inflation that have identical bispectra but a different scaling of higher-order moments (e.g.\ \citealt{Barnaby1109.2985}).

\citet{Shandera1304.1216} present constraints on two such inflation models, referred to as hierarchical-type (single-field inflation) and feeder-type (including interactions with a spectator field), based on the \emten{} data set. In this work, the free parameter describing the overall level of non-Gaussianity is the dimensionless third moment of the density perturbation field, smoothed on scales of $8h^{-1}\Mpc$, $\moment_3$; the two models above differ in the scaling of higher-order moments relative to $\moment_3$, and in the form of the modified, non-Gaussian mass function. In particular, the feeder scaling generates greater non-Gaussianity overall for a given value of $\moment_3$ than the hierarchical scaling.

More recently, \citet{Adhikari1402.2336} have performed $N$-body simulations of structure formation from non-Gaussian initial conditions. Their results for non-Gaussian mass functions broadly vindicate the analytic approach of \citet{Shandera1304.1216}, but motivate several refinements of the model, detailed in \citet{Adhikari1402.2336}, which we adopt here. We do not recapitulate these refinements here, but note that their net effect is to reduce the modification to the mass function for a given value of $\moment_3$ compared with the \citet{Shandera1304.1216} model, for both hierarchical- and feeder-type scalings. Consequently, our constraints on non-Gaussianity are weaker than those reported by \citet{Shandera1304.1216}, despite our addition of lensing data to the data set used in that work.\footnote{Empirically, and on a very technical note, we find that the most significant change to the model is due to the lower value of $\deltac$, a parameter whose value was assumed by \citet{Shandera1304.1216}, but which was fit to simulations by \citet{Adhikari1402.2336}; this directly impacts the non-Gaussian modification of the mass function, which depends on the ratio $\mysub{\nu}{c}=\deltac/\sigma(M)$. In detail, the \citet{Adhikari1402.2336} results are not precisely applicable to our analysis because the spherical overdensity they adopt to construct the mass function is different from the overdensity we use. However, a partial re-analysis of the simulation data indicates that the particular choice of overdensity has a small effect compared with the overall update due to \deltac{}, and that, if anything, the appropriate \deltac{} for our mass function may be slightly larger than the Adhikari  value. We therefore adopt the \citet{Adhikari1402.2336} prescription for the non-Gaussian mass function here, while noting that our new constraints may err on the conservative side. Future work in this area will benefit from more simulations, covering a more extensive selection of models, and investigating the dependence of the results on the halo finder employed.} Apart from primordial non-Gaussianity, we adopt a standard flat \LCDM{} model in this section.

Joint constraints on $\moment_3$ and $\sigma_8$ from our cluster analysis are shown in \figref~\ref{fig:nongauss}. As noted by \citet{Shandera1304.1216}, these two parameters are degenerate, particularly for hierarchical scaling. Improved constraints can therefore be obtained by incorporating additional data, namely the CMB power spectrum, to better constrain $\sigma_8$. Note that we do not use the CMB bispectrum or trispectrum to constrain $\moment_3$ here; the improvement in the figure comes entirely from breaking degeneracies between $\moment_3$ and other model parameters. \tabref~\ref{tab:ng} lists the constraints on these parameters for both non-Gaussian models, as well as the equivalent constraints on the amplitude of the bispectrum (\fnl{}) for the canonical local, equilateral and orthogonal momentum-space configurations (see \citealt{Shandera1304.1216} and \citealt{Adhikari1402.2336} for details of this conversion). In all cases, our results are consistent with Gaussianity.

In addition to \citet{Shandera1304.1216}, previous constraints on non-Gaussianity have been obtained by, e.g., \citet{Williamson1101.1290} and \citet{ Benson1112.5435} from the SPT cluster sample and \citet{Mana1303.0287} based on the MaxBCG sample. A direct comparison of the constraints is not completely straightforward, since these authors model the effects of non-Gaussianity on the mass function differently, but broadly speaking all these cluster constraints are consistent (see discussion in \citealt{Shandera1304.1216}), and all are consistent with Gaussianity. In the long term, combining the redshift coverage at high masses of X-ray and SZ surveys with the large mass range (and spatial clustering; e.g.\ \citealt{Mana1303.0287}) probed by optical surveys has the potential to significantly tighten cluster constraints on non-Gaussianity.

\section{Discussion} \label{sec:discussion}

\subsection{The Role of Follow-up Data} \label{sec:followup}

Although a cosmological test can be carried out using only cluster survey data, given a survey of sufficient area and depth (in both mass and redshift), this approach requires relatively strong assumptions regarding the form and scatter of the scaling relations. A straightforward benefit of incorporating additional measurements of masses or mass proxies for even a subset of discovered clusters is that these aspects of the model can be constrained rather than assumed, expanding the scope of cosmological models that can be investigated (e.g.\ \citealt{Majumdar0305341}). In the context of the Dark Energy Survey (DES), \citet{Wu0907.2690} have shown that significant gains in dark energy constraints can be obtained by incorporating X-ray or SZ mass proxy information, for example.

The present work uses three forms of follow-up data (in addition to spectroscopic redshift measurements): weak gravitational lensing observations, X-ray measurements of mass proxies (X-ray luminosity, temperature and gas mass within $r_{500}$), and X-ray measurements of \fgas{} at $r_{2500}$ for relaxed clusters. To a large extent, the X-ray \fgas{} analysis can be considered independent (\secref~\ref{sec:scalingmod}), providing additional constraints on \Omegam{} and dark energy parameters. As for the former two types of data, their complementarity  goes beyond the fact that X-ray observations are currently more numerous than lensing observations for the clusters in our data set. Namely, as we have emphasized, weak lensing provides nearly unbiased masses on average, but with a significant, irreducible intrinsic scatter on a cluster-by-cluster basis. In contrast, some X-ray (and SZ) mass proxies have a much smaller intrinsic scatter with mass, but the normalization of their scaling relations must be calibrated. The combination of the two types of observations thus provides a robust constraint on the cluster mass scale (from lensing), as well as more precise constraints on the slope and intrinsic scatter of scaling relations (and potentially on the shape of the mass function) than lensing alone can provide.

\begin{figure}
  \centering
  \includegraphics[scale=\figscale]{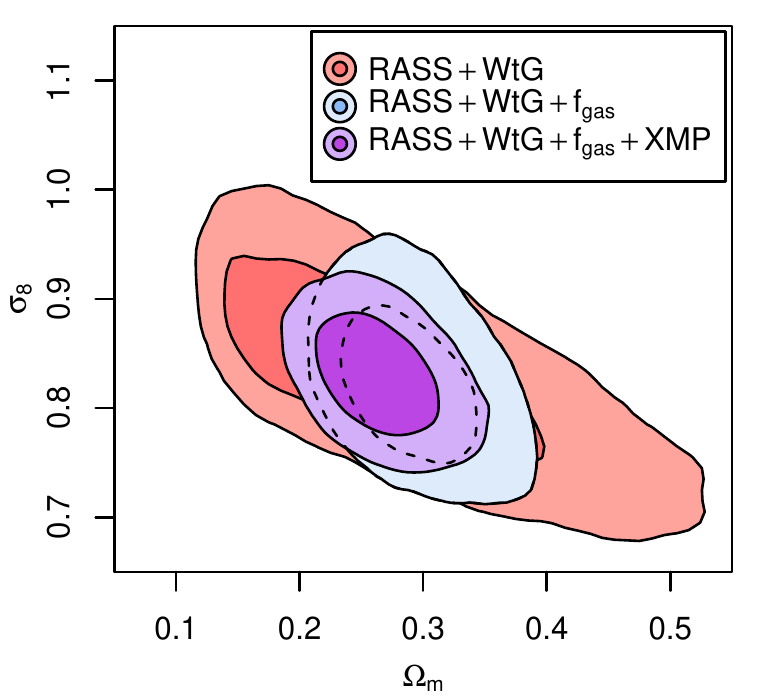}
  \caption{
    Constraints on \Omegam{} and $\sigma_8$ from subsets of our data. Dark and light shading respectively indicate the 68.3 and 95.4 per cent confidence regions, including systematic uncertainties and using standard priors on $h$ and $\Omegab h^2$. Red shading corresponds to the use of only the list of survey fluxes and redshifts for detected clusters (RASS) and the \wtg{} lensing observations. Blue shading adds X-ray measurements of \fgas{} for relaxed clusters, which constrain \Omegam{} but not $\sigma_8$, and purple regions also include X-ray mass proxies ($kT$ and \Mgas{}) from X-ray follow-up data.
  }
  \label{fig:followup}
\end{figure}
 
A cost/benefit analysis of these types of data in the spirit of \citet{Wu0907.2690} is beyond the scope of this paper. However, it is straightforward to ask how each contributes to our current results. \secref~\ref{sec:s8} has shown the importance of the lensing data for tightening constraints on $\sigma_8$ as well as \Omegam{} (see also \mamrakls{}) by straightforwardly comparing with the \mare{} results. \figref~\ref{fig:followup} shows how constraints on these two parameters respond to the addition of follow-up X-ray data, given a lensing mass calibration to start with. Red shading in the figure shows the constraints available from only the combination of the survey detections plus redshifts (RASS) and WtG lensing data. The classic $\sigma_8$--\Omegam{} degeneracy is apparent, but the redshift leverage of the data (which span $0<z<0.5$) is sufficient to break it. The width of the confidence region in this case is constrained to be $\sigma_8(\Omegam/0.3)^{0.17} = 0.81\pm0.04$. The degeneracy can be broken further by incorporating X-ray \fgas{} data for relaxed clusters, which robustly measure \Omegam{} but do not constrain $\sigma_8$; the width of the confidence region therefore remains the same, $\sigma_8(\Omegam/0.3)^{0.21} = 0.81\pm0.04$. Adding X-ray mass proxies from \Chandra{} or ROSAT follow-up (XMP) refines constraints on the key X-ray luminosity--mass relation and its scatter and provides more precise mass estimates for individual clusters, shrinking the constraints to $\sigma_8(\Omegam/0.3)^{0.17} = 0.81\pm0.03$.

As we discuss in the next section, significant further improvements in cosmological constraints can be obtained by improving the mass calibration through additional lensing data. Nevertheless, the ability of X-ray and SZ mass proxies to provide more precise mass estimates for individual clusters, and their availability at the highest and lowest redshifts, where lensing observations are very challenging/expensive, underscore their utility for cosmology.

\subsection{The Benefits of Improved Weak Lensing Data} \label{sec:newlens}

\figref{}s~\ref{fig:mnus8} and \ref{fig:s8w}  emphasize that, in the context of non-minimal cosmological models, the cluster constraint on $\sigma_8$ can often break key parameter degeneracies (even when many cosmological probes are combined). With relatively modest improvements in lensing systematics (see \citealt{Applegate1208.0605}) and larger samples of clusters with high-quality weak lensing data, constraints on the cluster mass scale at the 5 per cent level are plausible in the near term. Given also a factor of $\sim2$ improvement in predictions of the halo mass function (compared with the 10 per cent uncertainty adopted here), doubling of the number of clusters with weak lensing data would then translate to a reduction in the uncertainty on $\sigma_8$ (at fixed \Omegam{}) from 4 per cent currently to $\sim2$ per cent from clusters alone.\footnote{The size of the lensing sample could be straightforwardly increased (approximately doubled) by incorporating data already present in the archives of SuprimeCam and MegaPrime/MegaCam, such as those gathered for the Local Cluster Substructure Survey \citep{Okabe0903.1103, Okabe1302.2728} and the Canadian Cluster Comparison Project \citep{Hoekstra0705.0358, Hoekstra1208.0606}. However, this would require the application of a consistent, rigorously tested reduction and analysis pipeline across the entire data set, and likely the gathering of additional data to ensure that a significant fraction of the clusters are observed in at least five well chosen bands (enabling robust estimates of photometric redshifts for individual lensed galaxies; \citealt{Applegate1208.0605}). The lack of such 5-band observations is currently the most serious limitation to exploiting these archival data.} At the same time, the new data could provide a $\sim5$ per cent precision constraint on \Omegam{} through the \fgas{} test (\citealt{Allen1307.8152}; \mamrakls), leading to a factor of four improvement in the joint $\Omegam$--$\sigma_8$ constraint.

We have importance sampled our results from \secref~\ref{sec:cosmores} to simulate the effect that such an improved $\sigma_8$ constraint would have, all other things being equal. For concreteness, we assume that the more precise cluster constraint is centered on the current best-fitting value from the combination of cosmological probes (keeping the WMAP and \Planck+WP cases separate) for constant-$w$ models with minimal neutrino mass (\secref~\ref{sec:demodels}). For constant-$w$ models, we find that, due to the degeneracy breaking shown in \figref~\ref{fig:s8w}, constraints on $w$ would improve by 28 (25) per cent for the combination using WMAP (\Planck+WP) data. Applying the same procedure to the \LCDM+\Mnu{} model, we would expect 95.4 per cent confidence intervals of $\Mnu=0.09^{+0.14}_{-0.09}$ ($0^{+0.15}_{-0.00}$)~eV from the WMAP (\Planck+WP) combination ($\sim60$ and 30 percent reductions in the upper limits). With \Mnu{} and $w$ both free, the upper limits on \Mnu{} would be reduced by 15--20 per cent. These estimates likely underestimate the true impact of additional lensing data, which may improve the cluster constraints on $w$, depending on the redshift range spanned by the expanded data set. Note also that the full \Planck{} data set (including polarization) should be significantly more powerful than the 1-year \Planck{} data, supplemented by WMAP polarization measurements, used here.

In the particular case of \Mnu{}, accurate and precise constraints on $\sigma_8$ are clearly an important step towards obtaining a robust cosmological detection of non-zero neutrino mass. However, breaking the $\sigma_8$--\Mnu{} degeneracy can only achieve so much, as \figref~\ref{fig:mnus8} makes clear. Tight constraints on other cosmological parameters, especially dark energy parameters, will also be required to fully exploit the power of a precise $\sigma_8$ determination to measure neutrino mass. Farther ahead, direct detection of the time-dependent effects of neutrino mass on the growth of structure may be possible, although such a measurement will be challenging.

\subsection{The Route to Improved Dark Energy Constraints}

While the addition of further high-quality weak lensing data for X-ray selected clusters at low-to-intermediate redshifts should lead to significant near-term benefits in the constraints on \Omegam{}, $\sigma_8$ and the neutrino mass, the route to obtaining improved knowledge of dark energy, gravity and non-Gaussianity from clusters lies primarily in extending the redshift range of the analysis. In this regard, the combination of X-ray and SZ-selected cluster surveys holds significant potential. Using simple Fisher matrix-based projections \citep{Wu0907.2690},\footnote{\url{http://risa.stanford.edu/cluster/}} we estimate that extending the redshift lever arm of our cluster growth measurements out to $z \approx 1.5$ by combining the RASS X-ray survey with an SZ survey with similar area and depth to the 2500 square degree SPT survey \citep{Bleem1409.0850}, and including available X-ray and lensing follow-up data, should improve the dark energy constraints shown in \figref~\ref{fig:wevol} by a factor of two or more, placing cluster measurements firmly in the vanguard of dark energy studies. Similar improvements can be expected for the constraints on modified gravity models, enabling us to move beyond the simple $\gamma$-parameterization shown in \figref~\ref{fig:gamma}, while even larger improvements are expected for inflation studies, which are exponentially sensitive to the presence of unusually massive clusters at high redshifts (relative to the evolved baseline population measured at low-$z$). As the field progresses, there will also be a need for increasingly sophisticated theoretical predictions -- for example mass functions calibrated to a few per cent precision spanning the full range of interest in mass and redshift, and an appropriate range of baryonic physics, dark energy and fundamental physics models.

\section{Conclusions} \label{sec:conclusions}

Earlier papers in the \wtg{} series have focussed on providing the most well characterized and unbiased constraints on the absolute cluster mass calibration possible, using measurements of weak gravitational lensing. Here we incorporate those data into a cosmological analysis that uses the number density of massive clusters as a function of time to probe the growth of cosmic structure. In addition to the \wtg{} lensing data, our analysis uses an X-ray selected cluster sample culled from the ROSAT All-Sky Survey, spanning redshifts $0<z<0.5$, along with  follow-up  X-ray  data to supply additional mass proxies. We additionally take advantage of cluster gas mass fraction data, which also benefit substantially from the lensing mass calibration, to provide an independent measurement of the cosmic expansion and tight constraints on \Omegam{}, breaking the main degeneracy (with $\sigma_8$) present in the analysis of cluster-counts data.

Our data provide marginalized constraints on the mean matter density and the amplitude of matter fluctuations, $\Omegam = 0.26 \pm 0.03$ and $\sigma_8 = 0.83 \pm 0.04$. These constraints are essentially identical for \LCDM{} models with and without curvature, as well as constant- and evolving-$w$ models of dark energy, and models with a free neutrino mass. The width of the confidence region in the \Omegam--$\sigma_8$ plane, which retains some degeneracy, is given by $\sigma_8(\Omegam/0.3)^{0.17} = 0.81 \pm 0.03$ (including all systematic uncertainties). These results are in good agreement with constraints from both WMAP and \Planck+WP CMB data, even under the restrictive assumption of a spatially flat \LCDM{} model, and also with our previous results using the same cluster catalogs (but without the lensing data). Our constraints are broadly similar to other recent results from clusters, although the agreement is not formally good within the quoted uncertainties, especially considering that the cluster samples used to provide the mass calibration often overlap to a large degree. This serves to underline the need for an unbiased mass calibration, as well as a robust characterization of the uncertainties in that calibration (as performed in the \wtg{} analysis).

Combining our cluster data with CMB, supernova and BAO data, we find no preference for non-zero neutrino mass, in contrast to some recent work. As measurements of $\sigma_8$ become even more precise, it will be critical to maintain good accuracy and control of systematic uncertainties affecting the cluster mass calibration, to obtain the most accurate constraints on neutrino properties.

The dark energy constraints available from cluster data remain highly competitive with the best available cosmological probes. From cluster data alone (including the survey, follow-up lensing and X-ray observations and \fgas{} data), we find $w=-0.98\pm0.15$ for flat, constant-$w$ models. The cluster data also constrain evolving-$w$ models: we find $w_0=-1.0^{+1.5}_{-1.4}$ and $\wet=-1.4^{+0.8}_{-1.1}$ for a flat, evolving model, marginalizing over the transition redshift of $w(z)$. Combining with CMB, supernova and BAO data, we continue to find consistency with flat \LCDM{}, even when global curvature and evolving dark energy are simultaneously included in the model.

The prospects for further improvements in the constraints on cosmology and fundamental physics from observations of galaxy clusters are substantial. A suite of major new surveys across the electromagnetic spectrum have or will soon come on line (e.g.\ DES, SPT-3G, Advanced ACT-Pol, eROSITA, LSST, WFIRST-AFTA, {\it Euclid}). Optimally leveraging the data from these surveys, as well as follow-up facilities, to produce robust cluster catalogs (with well understood purity and completeness), accurate absolute mass calibration (from weak lensing) and sufficient, low-scatter mass proxy information (from X-ray and SZ follow-up) will be critical to obtaining the tightest and most robust constraints possible.

In the near term, the path toward further reducing systematic uncertainties in the absolute mass calibration of low-redshift cluster samples using weak lensing methods seems clear (e.g.\ \citealt{Applegate1208.0605}), with important work already underway within the \citet{LSSTDESC1211.0310} and elsewhere. The most immediate and straightforward aspect of this would be an expansion of the weak lensing data set to $2$--$4\times$ more clusters, maintaining comparable data quality to the \wtg{} study. With this, the prospects for, e.g., quickly halving the statistical-plus-systematic uncertainty on $\sigma_8$ from clusters, and determining (in combination with new CMB measurements) improved constraints on neutrino properties, are strong. Likewise, for dark energy studies, the prospects for improved constraints by utilizing optimally the full mass and redshift lever arm of new and existing X-ray, optical and SZ-selected cluster samples are excellent.

\section*{Acknowledgments}

We thank Alastair Edge for sharing his list of likely AGN-dominated BCS and REFLEX clusters, as well as Risa Wechsler and Sam Skillman for insightful discussions. We thank the Dark Cosmology Centre for hosting collaboration meetings during the development of this paper. We also thank the referee for providing prompt, thorough, and very useful comments. Calculations for this work utilized the Coma, Orange and Bullet compute clusters at the SLAC National Accelerator Laboratory, and the HPC facility at the University of Copenhagen. AM was supported by National Science Foundation grants AST-0838187 and AST-1140019. DA acknowledges funding from the German Federal Ministry of Economics and Technology (BMWi) under project 50 OR 1210. S. Adhikari and S. Shandera are supported by the National Aeronautics and Space Administration (NASA) under Grant No.\ NNX12AC99G issued through the Astrophysics Theory Program. We acknowledge support from the U.S. Department of Energy under contract number DE-AC02-76SF00515; from NASA through Chandra Award Numbers GO8-9118X and TM1-12010X, issued by the Chandra X-ray Observatory Center, which is operated by the Smithsonian Astrophysical Observatory for and on behalf of NASA under contract NAS8-03060; as well as through program HST-AR-12654.01-A, provided by NASA through a grant from the Space Telescope Science Institute, which is operated by the Association of Universities for Research in Astronomy, Inc., under NASA contract NAS 5-26555. The Dark Cosmology Centre is funded by the Danish National Research Foundation.

Based in part on data collected at Subaru Telescope (University of Tokyo) and obtained from the SMOKA, which is operated by the Astronomy Data Center, National Astronomical Observatory of Japan.  Based on observations obtained with MegaPrime/MegaCam, a joint project of CFHT and CEA/DAPNIA, at the Canada-France-Hawaii Telescope (CFHT) which is operated by the National Research Council (NRC) of Canada, the Institute National des Sciences de l'Univers of the Centre National de la Recherche Scientifique of France, and the University of Hawaii.

\def \araa {ARA\&A}
\def \aj {AJ}
\def \aar {A\&AR}
\def \apj {ApJ}
\def \apjl {ApJL}
\def \apjs {ApJS}
\def \asl {Adv. Sci. Lett.} 
\def \mnras {MNRAS}
\def \nat {Nat}
\def \pasj {PASJ}
\def \pasp {PASP}
\def \science {Sci}
\def \gca {Geochim.\ Cosmochim.\ Acta}
\def \npa {Nucl.\ Phys.\ A}
\def \plb {Phys.\ Lett.\ B}
\def \prc {Phys.\ Rev.\ C}
\def \prd {Phys.\ Rev.\ D}
\def \prl {Phys.\ Rev.\ Lett.}
\def \jcap {J. Cosmology Astropart. Phys.} 
\def \physrep {Phys. Rep.} 
\def \aap {A\&A} 
\def \ijmpd {Int.\ J.\ Mod.\ Phys.\ D} 

\appendix

\section{Constraints on Scaling Relations} \label{sec:scaling}

Obtaining cosmological constraints from the mass function of clusters necessarily involves simultaneously constraining the scaling of cluster observables with mass (\secref~\ref{sec:model}). In this work, the X-ray follow-up observables we use are those derived in \maerd{}. For the cosmological results reported here, the impact of more recent updates to the Chandra calibration is expected to be very small. As noted in \secref~\ref{sec:data}, luminosity and gas mass measurements agree between \maerd{} and measurements using recent calibration updates at the per cent level (and luminosity is in any case cross-calibrated to the ROSAT standard in the cosmological analysis). Temperature measurements do differ; however, the primary impact that measured temperatures have on the present cosmological analysis is through the conversion of X-ray luminosity to flux (the K-correction), which is an exceedingly weak dependence for massive clusters at the relevant redshifts. In \emten{}, we explicitly tested the effect of assuming a fixed temperature of 5\,keV for every cluster and found this to produce identical cosmological constraints to fitting for the temperature--mass relation, a result which still applies to the current analysis. Thus, the temperature--mass relation can be considered a pure output of the combined cosmology/scaling analysis for our data set.

On the other hand, any astrophysical interpretation of the X-ray scaling relations should ideally be based on the latest \Chandra{} calibration. For this reason, we defer such discussion to an upcoming paper, WtG\,V, which will employ updated X-ray measurements. For completeness, the scaling relation constraints obtained from the current analysis (using the same X-ray calibration as \emten{}) are provided below.

In implementing the general scaling model introduced in \secref~\ref{sec:scalingmod}, we have assumed several of the off-diagonal terms of the intrinsic covariance matrix to be zero, specifically the terms linking X-ray luminosity or temperature to gas mass or lensing mass. (The luminosity--temperature correlation is free in our analysis.) While this simplification is required for computational reasons, it is also well motivated according to our best understanding of the observables involved. The marginal scatter in X-ray luminosity at fixed mass is dominated by the presence or absence of compact, bright cores found at the centers of some clusters (\maerd{}). Since our temperature measurements exclude cluster centers ($r<0.15\,r_{500}$; see \maerd), and because X-ray luminosity in the 0.1--2.4\,keV{} band is approximately temperature-independent for the $kT\gtsim 4\keV$ clusters in our sample, the $\ell$--$t$ covariance is expected (and measured) to be small. The marginal luminosity scatter is both large ($\sim40$ per cent) and physically different in origin from the scatters in \Mgas{} and \Mlens{}, both of which are most sensitive to larger spatial scales ($\sim r_{500}$ compared with $\ltsim 0.05\,r_{500}$; \citealt{Mantz2009PhDT........18M}). Covariances among the temperature, gas mass and lensing mass measured for a given cluster, due to e.g.\ asphericity or dynamical state, are similarly thought to be small (see e.g.\ calculations by \citealt{Gavazzi0503696} and \citealt{Buote2012MNRAS.421.1399B}, hydrodynamical simulations by \citealt{Stanek0910.1599}, and further discussion in WtG\,V). We therefore expect the impact of neglecting these cross-terms to be small compared with the overall level of systematic allowances in our analysis (e.g.\ \citealt{Allen1103.4829}).

The constraints on scaling relations from the present work appear in \tabref~\ref{tab:scaling}. The constraints on the luminosity--mass and temperature--mass relations are similar to those of \maerd{}, though note the change from base-10 to natural logarithms in the definition of the scaling relations (relevant to the normalizations and scatters). The largest shift, though still within errors, is in the $\ell$--$m$ normalization, which has degeneracies with cosmological parameters due to its role in the sample selection. Astrophysical interpretation of the scaling relation constraints, using an up-to-date \Chandra{} calibration, will be presented in WtG\,V. In the context of cosmological constraints, however, it is interesting to note that the constraint on intrinsic scatter in the $\mlens$--$m$ relation, $0.18\pm0.05$, is considerably tighter than the prior ($0.2\pm0.1$) and in good agreement with simulation predictions \citep{Becker1011.1681}. This exemplifies the complementary nature of lensing and other mass-proxy follow-up data. Namely, while lensing excels at providing an unbiased mean mass, the intrinsic scatter is relatively large. Mass proxies with smaller scatter, once calibrated in the mean by lensing data, can provide more precise mass estimates for individual clusters, as well as directly calibrate the size of the lensing intrinsic scatter.

\begin{table}
  \begin{center}
    \caption{
      Best-fitting values and 68.3 per cent confidence intervals for scaling relation parameters. The scaling relation model is introduced in \secref~\ref{sec:scalingmod}; for comparing normalizations and scatters to \maerd{}, note the change from base-10 to natural logarithms in our definition of the scaling relation parameters. This set of constraints results from an analysis of the cluster data alone, marginalizing over flat \LCDM{} cosmological models. Parameters which are only constrained by the prior, namely the $\mgas$--$m$ scatter and $\mlens$--$m$ normalization, are not listed.
    }
    \label{tab:scaling}
    \vspace{1ex}
    \begin{tabular}{lc}
      \hline
      Parameter & Constraint \\
      \hline
      $\ell$--$m$ normalization & \phmin$1.71 \pm 0.17$ \\
      $\ell$--$m$ slope & \phmin$1.34 \pm 0.07$ \\
      $\ell$--$m$ scatter & \phmin$0.42 \pm 0.05$ \\
      $t$--$m$ normalization & \phmin$2.04 \pm 0.06$ \\
      $t$--$m$ slope & \phmin$0.47 \pm 0.04$ \\
      $t$--$m$ scatter & \phmin$0.13 \pm 0.02$ \\
      $\ell$,$t$ correlation & \phmin$0.11 \pm 0.19$ \\
      $\mgas$--$m$ normalization & $-2.18 \pm 0.08$ \\
     $\mgas$--$m$ slope & \phmin$0.99 \pm 0.01$ \\
      $\mlens$--$m$ scatter & \phmin$0.18 \pm 0.05$ \\
      \hline
   \end{tabular}
  \end{center}
\end{table}

\section{Figures Using \Planck{} Data} \label{sec:planckfigs}

\figref{}~\ref{fig:planckfigs} shows results equivalent to \figref{}s~\ref{fig:mnu}b and \ref{fig:s8w}--\ref{fig:gamma}, with the substitution of \Planck{} 1-year data (plus WMAP polarization; \citealt{Planck1303.5076}) for WMAP 9-year data \citep{Hinshaw1212.5226}.

\begin{figure*}
  \centering
  \includegraphics[scale=\figscale]{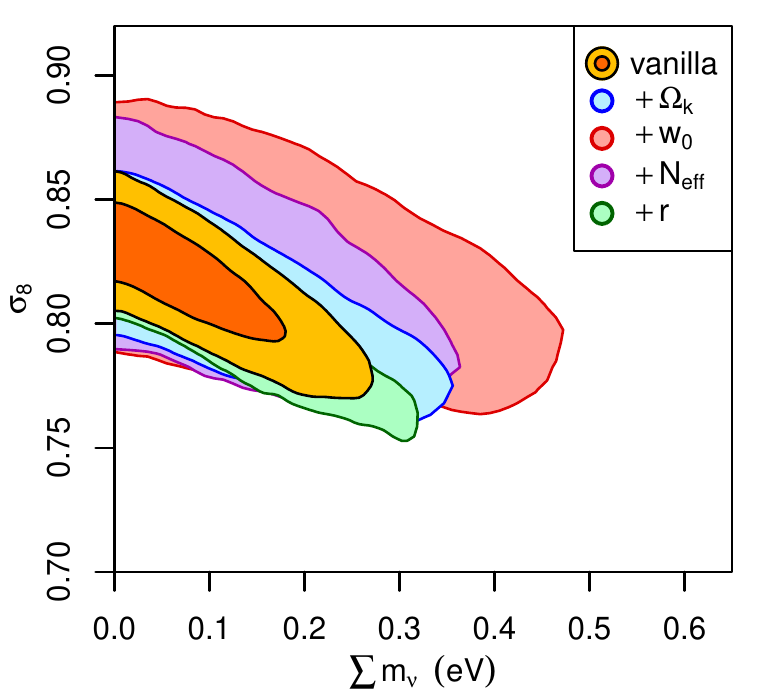}
  \hspace{1cm}
  \includegraphics[scale=\figscale]{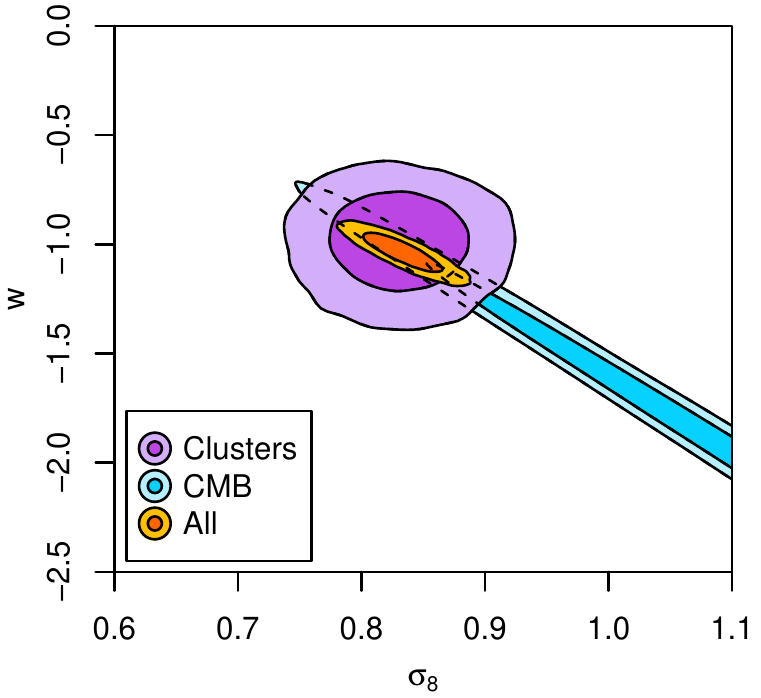}
  \includegraphics[scale=\figscale]{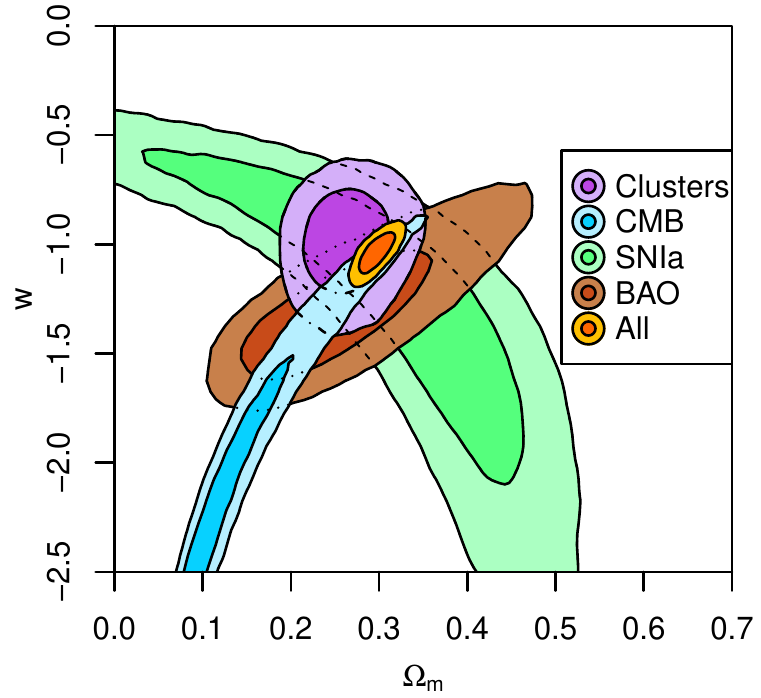}
  \hspace{1cm}
   \includegraphics[scale=\figscale]{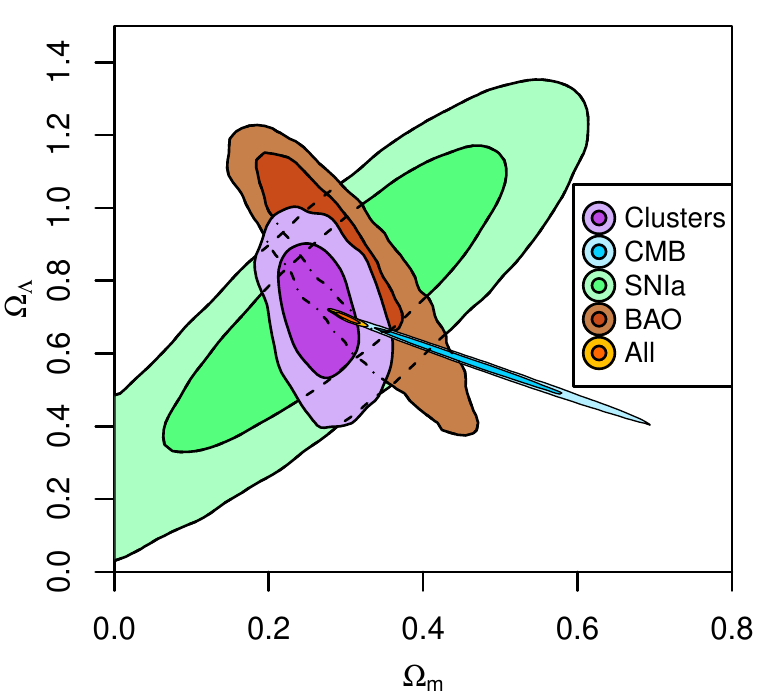}
   \includegraphics[scale=\figscale]{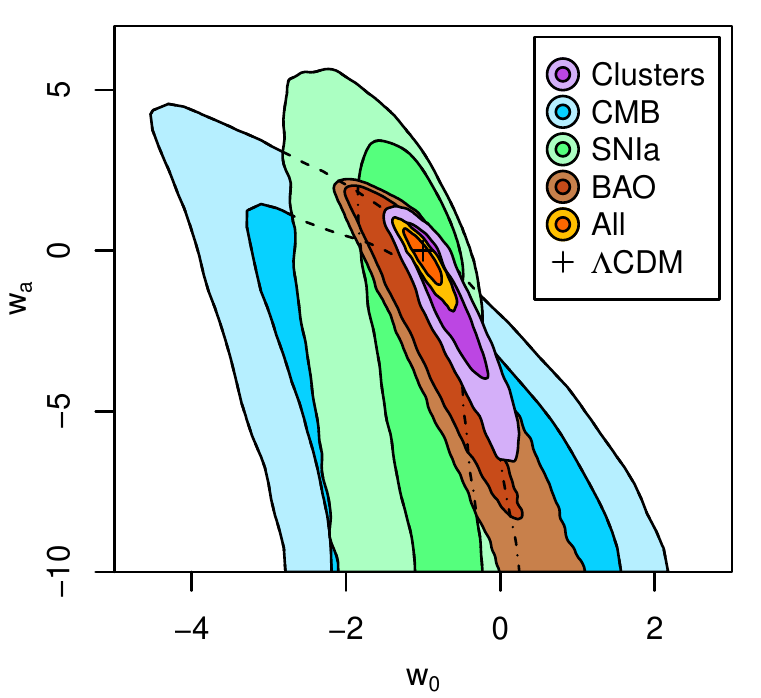}
   \hspace{1cm}
   \includegraphics[scale=\figscale]{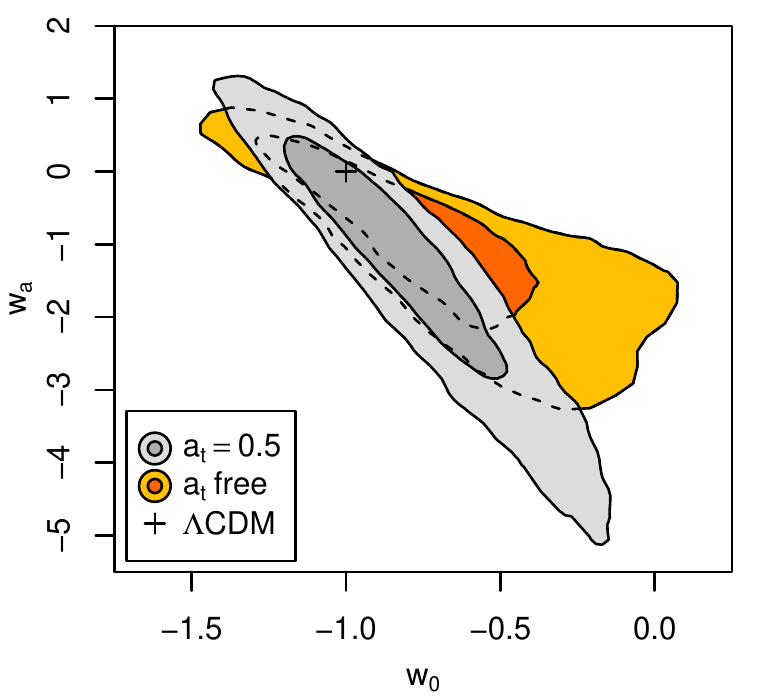}
  \caption{
    Constraints on cosmological models from the cluster data set, CMB data from \Planck+WP, ACT and SPT (\citealt{Keisler1105.3182, Reichardt1111.0932, Story1210.7231, Das1301.1037,Planck1303.5076}), type Ia supernovae (\citealt{Suzuki1105.3470}), baryon acoustic oscillations (\citealt{Beutler1106.3366, Padmanabhan1202.0090, Anderson1303.4666}), and their combination. These figures are identical to the equivalent ones in \secref~\ref{sec:cosmores} apart from the substitution of \Planck{} 1-year data (plus WMAP polarization) for WMAP 9-year data. Left to right and top to bottom, the panels correspond to \figref{}s~\ref{fig:mnu}b, \ref{fig:s8w}a, \ref{fig:s8w}b, \ref{fig:lcdm}, \ref{fig:clwevol}a and \ref{fig:wevol}b (this page) and \ref{fig:gamma} (second page).
  }
  \label{fig:planckfigs}
\end{figure*}

\begin{figure*}
  \centering
  \includegraphics[scale=\figscale]{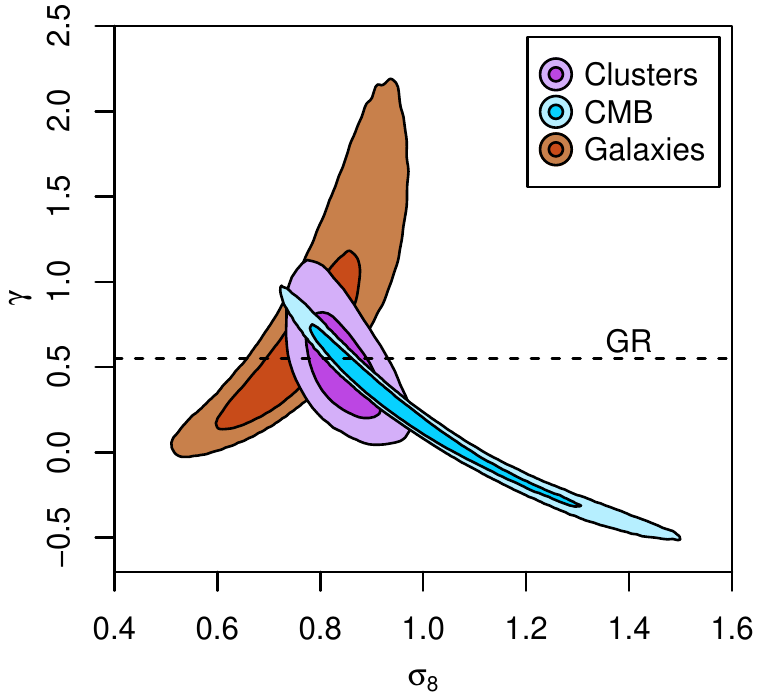}
  \hspace{1cm}
  \includegraphics[scale=\figscale]{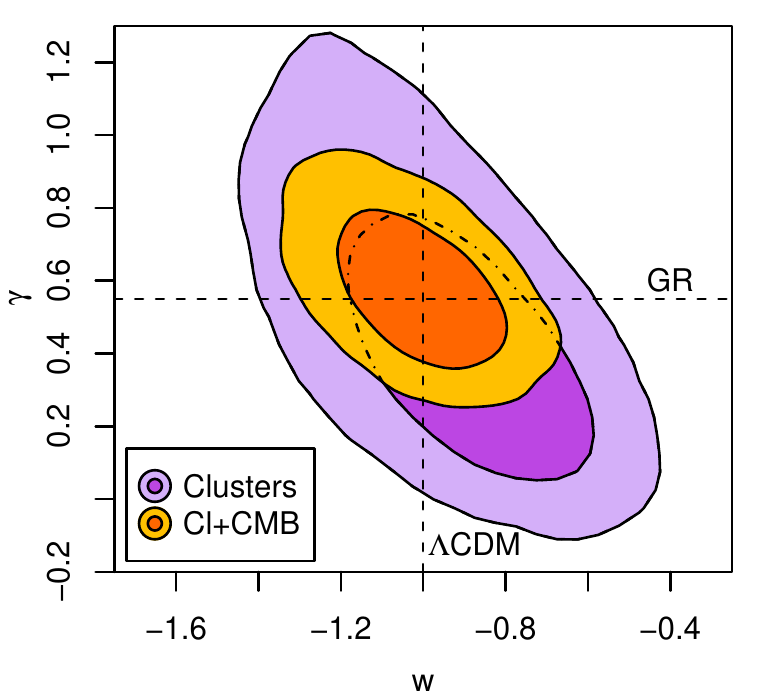}
  \contcaption{}
\end{figure*}

\section{The ISW Effect in Free Growth-Index Models} \label{sec:isw}

In our study of the growth index of cosmic structure (\secref~\ref{sec:growthindex}), as in our previous analyses, we obtain the contribution of the ISW effect to the anisotropy power spectrum of the CMB temperature fluctuations through an integral over time of the variation of the gravitational potential with respect to conformal time, $\dot{\phi}$ \citep{Weller0307104}. For the latter, we take the derivative of the gauge invariant Poisson equation
\begin{equation}
  k^2\phi = -4\pi G a^2\rho\Delta,
\end{equation}
where $\rho\Delta \equiv \sum_i \rho_i\delta_i+3\mathcal{H} \sum_i(\rho_i+P_i)\theta_i$, with $\mathcal{H}$ being the Hubble parameter in conformal time, $\rho_i$ and $P_i$ the densities and pressures for each species $i$, and each of the sums the mass-averaged density contrast, $\delta_i$, and velocity divergence, $\theta_i$, for a given gauge \citep{Bardeen1980PhRvD..22.1882B}. In synchronous gauge, we have $\rho_\mathrm{c}\Delta_\mathrm{c} \equiv \rho_\mathrm{c}\delta_\mathrm{c}$, $\rho_\mathrm{b}\Delta_\mathrm{b} \equiv \rho_\mathrm{b}\delta_\mathrm{b} + 3\mathcal{H}\rho_\mathrm{b} \theta_\mathrm{b} k^{-2}$, and $\rho_\mathrm{r}\Delta_\mathrm{r} \equiv \rho_\mathrm{r}\delta_\mathrm{r} + 4\mathcal{H}\rho_\mathrm{r} \theta_\mathrm{r} k^{-2}$ for CDM (c), baryons (b), and radiation (r, including massless neutrinos and photons), respectively \citep{Ma9506072}. Note that after recombination the baryon velocity fluctuations evolve as $\dot{\theta}_\mathrm{b} = -\mathcal{H}\theta_\mathrm{b} + c_\mathrm{s}^2 k^2 \delta_\mathrm{b}$, where $c_\mathrm{s}$ is the baryonic sound speed, which after baryon-photon decoupling is rapidly driven to zero by adiabatic cooling. This implies that at late times baryonic perturbations will, like those for CDM, follow mainly metric perturbations, $\dot{\delta}_\mathrm{b} \approx \dot{\delta}_\mathrm{c} = -\dot{h}/2$. For radiation, we have $\dot{\delta}_\mathrm{r} = -4\theta_\mathrm{r}/3 -2\dot{h}/3$ and $\dot{\theta}_\mathrm{r} = k^2 (\delta_\mathrm{r}/4-\sigma_\mathrm{r})$, where the anisotropic stress perturbation is defined as $(\rho +P)\sigma \equiv -(\hat{k}_i\hat{k}_j - \delta_{ij}/3)\Sigma_i^j$, with $i$ and $j$ denoting the indexes for the spatial components, and $\Sigma_i^j$ the shear stress, which is negligible after recombination. Therefore, in the matter-dominated era we have oscillating solutions for both $\delta_\mathrm{r}$ and $\theta_\mathrm{r}$, and both terms of $\rho_\mathrm{r} \Delta_\mathrm{r}$ (see above) become strongly suppressed by $\rho_\mathrm{r} \propto a^{-4}$. We finally obtain $k^2 \phi \approx -4\pi G a^2 \delta\rho_\mathrm{m}$, and thus
\begin{equation}
\dot{\phi} \approx 4\pi G (a^2/k^2)
\mathcal{H}\delta\rho_\mathrm{m}\left[ 1- \Omega_\mathrm{m}(a)^\gamma
  \right]. 
\end{equation}
More discussion can be found in \citet{Rapetti0812.2259, Rapetti0911.1787, Rapetti1205.4679}.

\label{lastpage}


\begin{thebibliography}{}

\bibitem[\protect\citeauthoryear{{Adhikari}, {Shandera} \& {Dalal}}{{Adhikari}
  et~al.}{2014}]{Adhikari1402.2336}
{Adhikari} S., {Shandera} S.,  {Dalal} N., 2014,
  \href{http://adsabs.harvard.edu/abs/2014JCAP...06..052A}{\textcolor{blue}{\jcap,
  6, 52}}

\bibitem[\protect\citeauthoryear{{Allen}, {Schmidt} \& {Fabian}}{{Allen}
  et~al.}{2002}]{Allen0205007}
{Allen} S., {Schmidt} R.,  {Fabian} A., 2002,
  \href{http://adsabs.harvard.edu/abs/2002MNRAS.334L..11A}{\textcolor{blue}{\mnras,
  334, L11}}

\bibitem[\protect\citeauthoryear{{Allen}, {Evrard} \& {Mantz}}{{Allen}
  et~al.}{2011}]{Allen1103.4829}
{Allen} S.~W., {Evrard} A.~E.,  {Mantz} A.~B., 2011,
  \href{http://adsabs.harvard.edu/abs/2011ARA%26A..49..409A}{\textcolor{blue}{\araa,
  49, 409}}

\bibitem[\protect\citeauthoryear{{Allen} et~al.}{{Allen}
  et~al.}{2013}]{Allen1307.8152}
{Allen} S.~W., {Mantz} A.~B., {Morris} R.~G., {Applegate} D.~E., {Kelly} P.~L.,
  {von der Linden} A., {Rapetti} D.~A.,  {Schmidt} R.~W., 2013,
  \href{http://arxiv.org/abs/1307.8152}{\textcolor{blue}{arXiv:1307.8152}}

\bibitem[\protect\citeauthoryear{{Allen} et~al.}{{Allen}
  et~al.}{2008}]{Allen0706.0033}
{Allen} S.~W., {Rapetti} D.~A., {Schmidt} R.~W., {Ebeling} H., {Morris} R.~G.,
  {Fabian} A.~C., 2008,
  \href{http://adsabs.harvard.edu/abs/2008MNRAS.383..879A}{\textcolor{blue}{\mnras,
  383, 879}}

\bibitem[\protect\citeauthoryear{{Allen}, {Schmidt} \& {Bridle}}{{Allen}
  et~al.}{2003}]{Allen0306386}
{Allen} S.~W., {Schmidt} R.~W.,  {Bridle} S.~L., 2003,
  \href{http://adsabs.harvard.edu/abs/2003MNRAS.346..593A}{\textcolor{blue}{\mnras,
  346, 593}}

\bibitem[\protect\citeauthoryear{{Allen} et~al.}{{Allen}
  et~al.}{2004}]{Allen0405340}
{Allen} S.~W., {Schmidt} R.~W., {Ebeling} H., {Fabian} A.~C.,  {van Speybroeck}
  L., 2004,
  \href{http://adsabs.harvard.edu/abs/2004MNRAS.353..457A}{\textcolor{blue}{\mnras,
  353, 457}}

\bibitem[\protect\citeauthoryear{{Allen} et~al.}{{Allen}
  et~al.}{2003}]{Allen0208394}
{Allen} S.~W., {Schmidt} R.~W., {Fabian} A.~C.,  {Ebeling} H., 2003,
  \href{http://adsabs.harvard.edu/cgi-bin/nph-bib_query?bibcode=2003MNRAS.342..287A&db_key=AST}{\textcolor{blue}{\mnras,
  342, 287}}

\bibitem[\protect\citeauthoryear{{Anderson} et~al.}{{Anderson}
  et~al.}{2014}]{Anderson1303.4666}
{Anderson} L. et~al., 2014,
  \href{http://adsabs.harvard.edu/abs/2014MNRAS.439...83A}{\textcolor{blue}{\mnras,
  439, 83}}

\bibitem[\protect\citeauthoryear{{Applegate} et~al.}{{Applegate}
  et~al.}{2014}]{Applegate1208.0605}
{Applegate} D.~E. et~al., 2014,
  \href{http://adsabs.harvard.edu/abs/2014MNRAS.439...48A}{\textcolor{blue}{\mnras,
  439, 48}} (WtG\,III)

\bibitem[\protect\citeauthoryear{{Bardeen}}{{Bardeen}}{1980}]{Bardeen1980PhRvD..22.1882B}
{Bardeen} J.~M., 1980,
  \href{http://adsabs.harvard.edu/abs/1980PhRvD..22.1882B}{\textcolor{blue}{\prd,
  22, 1882}}

\bibitem[\protect\citeauthoryear{{Barnaby} \& {Shandera}}{{Barnaby} \&
  {Shandera}}{2012}]{Barnaby1109.2985}
{Barnaby} N.,  {Shandera} S., 2012,
  \href{http://adsabs.harvard.edu/abs/2012JCAP...01..034B}{\textcolor{blue}{\jcap,
  1, 34}}

\bibitem[\protect\citeauthoryear{{Bartolo} et~al.}{{Bartolo}
  et~al.}{2004}]{Bartolo0406398}
{Bartolo} N., {Komatsu} E., {Matarrese} S.,  {Riotto} A., 2004,
  \href{http://adsabs.harvard.edu/abs/2004PhR...402..103B}{\textcolor{blue}{\physrep,
  402, 103}}

\bibitem[\protect\citeauthoryear{{Battaglia} et~al.}{{Battaglia}
  et~al.}{2013}]{Battaglia1209.4082}
{Battaglia} N., {Bond} J.~R., {Pfrommer} C.,  {Sievers} J.~L., 2013,
  \href{http://adsabs.harvard.edu/abs/2013ApJ...777..123B}{\textcolor{blue}{\apj,
  777, 123}}

\bibitem[\protect\citeauthoryear{{Becker} \& {Kravtsov}}{{Becker} \&
  {Kravtsov}}{2011}]{Becker1011.1681}
{Becker} M.~R.,  {Kravtsov} A.~V., 2011,
  \href{http://adsabs.harvard.edu/abs/2011ApJ...740...25B}{\textcolor{blue}{\apj,
  740, 25}}

\bibitem[\protect\citeauthoryear{{Bennett} et~al.}{{Bennett}
  et~al.}{2013}]{Bennett1212.5225}
{Bennett} C.~L. et~al., 2013,
  \href{http://adsabs.harvard.edu/abs/2013ApJS..208...20B}{\textcolor{blue}{\apjs,
  208, 20}}

\bibitem[\protect\citeauthoryear{{Benson} et~al.}{{Benson}
  et~al.}{2013}]{Benson1112.5435}
{Benson} B.~A. et~al., 2013,
  \href{http://adsabs.harvard.edu/abs/2013ApJ...763..147B}{\textcolor{blue}{\apj,
  763, 147}}

\bibitem[\protect\citeauthoryear{{Beutler} et~al.}{{Beutler}
  et~al.}{2011}]{Beutler1106.3366}
{Beutler} F. et~al., 2011,
  \href{http://adsabs.harvard.edu/abs/2011MNRAS.416.3017B}{\textcolor{blue}{\mnras,
  416, 3017}}

\bibitem[\protect\citeauthoryear{{Beutler} et~al.}{{Beutler}
  et~al.}{2012}]{Beutler1204.4725}
{Beutler} F. et~al., 2012,
  \href{http://adsabs.harvard.edu/abs/2012MNRAS.423.3430B}{\textcolor{blue}{\mnras,
  423, 3430}}

\bibitem[\protect\citeauthoryear{{Beutler} et~al.}{{Beutler}
  et~al.}{2014a}]{Beutler1403.4599}
{Beutler} F. et~al., 2014a,
  \href{http://adsabs.harvard.edu/abs/2014MNRAS.444.3501B}{\textcolor{blue}{\mnras,
  444, 3501}}

\bibitem[\protect\citeauthoryear{{Beutler} et~al.}{{Beutler}
  et~al.}{2014b}]{Beutler1312.4611}
{Beutler} F. et~al., 2014b,
  \href{http://adsabs.harvard.edu/abs/2014MNRAS.443.1065B}{\textcolor{blue}{\mnras,
  443, 1065}}

\bibitem[\protect\citeauthoryear{{BICEP2 Collaboration}}{{BICEP2
  Collaboration}}{2014}]{BICEP2-Collaboration1403.3985}
{BICEP2 Collaboration}, 2014,
  \href{http://adsabs.harvard.edu/abs/2014PhRvL.112x1101A}{\textcolor{blue}{\prl,
  112, 241101}}

\bibitem[\protect\citeauthoryear{{Blake} et~al.}{{Blake}
  et~al.}{2011}]{Blake1108.2637}
{Blake} C. et~al., 2011,
  \href{http://adsabs.harvard.edu/abs/2011MNRAS.418.1725B}{\textcolor{blue}{\mnras,
  418, 1725}}

\bibitem[\protect\citeauthoryear{{Bleem} et~al.}{{Bleem}
  et~al.}{2014}]{Bleem1409.0850}
{Bleem} L.~E. et~al., 2014,
  \href{http://arxiv.org/abs/1409.0850}{\textcolor{blue}{arXiv:1409.0850}}

\bibitem[\protect\citeauthoryear{{B{\"o}hringer} et~al.}{{B{\"o}hringer}
  et~al.}{2004}]{Bohringer0405546}
{B{\"o}hringer} H. et~al., 2004,
  \href{http://adsabs.harvard.edu/cgi-bin/nph-bib_query?bibcode=2004A%26A...425..367B&db_key=AST}{\textcolor{blue}{\aap,
  425, 367}}

\bibitem[\protect\citeauthoryear{{Borgani} et~al.}{{Borgani}
  et~al.}{2001}]{Borgani2001ApJ...561...13B}
{Borgani} S. et~al., 2001,
  \href{http://adsabs.harvard.edu/cgi-bin/nph-bib_query?bibcode=2001ApJ...561...13B&db_key=AST}{\textcolor{blue}{\apj,
  561, 13}}

\bibitem[\protect\citeauthoryear{{Buote} \& {Humphrey}}{{Buote} \&
  {Humphrey}}{2012}]{Buote2012MNRAS.421.1399B}
{Buote} D.~A.,  {Humphrey} P.~J., 2012,
  \href{http://adsabs.harvard.edu/abs/2012MNRAS.421.1399B}{\textcolor{blue}{\mnras,
  421, 1399}}

\bibitem[\protect\citeauthoryear{{Burenin}}{{Burenin}}{2013}]{Burenin1301.4791}
{Burenin} R.~A., 2013,
  \href{http://adsabs.harvard.edu/abs/2013AstL...39..357B}{\textcolor{blue}{Astronomy
  Letters, 39, 357}}

\bibitem[\protect\citeauthoryear{{Burenin} et~al.}{{Burenin}
  et~al.}{2007}]{Burenin0610739}
{Burenin} R.~A., {Vikhlinin} A., {Hornstrup} A., {Ebeling} H., {Quintana} H.,
  {Mescheryakov} A., 2007,
  \href{http://adsabs.harvard.edu/abs/2007ApJS..172..561B}{\textcolor{blue}{\apjs,
  172, 561}}

\bibitem[\protect\citeauthoryear{{Burenin} \& {Vikhlinin}}{{Burenin} \&
  {Vikhlinin}}{2012}]{Burenin1202.2889}
{Burenin} R.~A.,  {Vikhlinin} A.~A., 2012,
  \href{http://adsabs.harvard.edu/abs/2012AstL...38..347B}{\textcolor{blue}{Astronomy
  Letters, 38, 347}}

\bibitem[\protect\citeauthoryear{{Chevallier} \& {Polarski}}{{Chevallier} \&
  {Polarski}}{2001}]{Chevallier0009008}
{Chevallier} M.,  {Polarski} D., 2001,
  \href{http://adsabs.harvard.edu/abs/2001IJMPD..10..213C}{\textcolor{blue}{\ijmpd,
  10, 213}}

\bibitem[\protect\citeauthoryear{{Clifton} et~al.}{{Clifton}
  et~al.}{2012}]{Clifton1106.2476}
{Clifton} T., {Ferreira} P.~G., {Padilla} A.,  {Skordis} C., 2012,
  \href{http://adsabs.harvard.edu/abs/2012PhR...513....1C}{\textcolor{blue}{\physrep,
  513, 1}}

\bibitem[\protect\citeauthoryear{{Cooke} et~al.}{{Cooke}
  et~al.}{2014}]{Cooke1308.3240}
{Cooke} R.~J., {Pettini} M., {Jorgenson} R.~A., {Murphy} M.~T.,  {Steidel}
  C.~C., 2014,
  \href{http://adsabs.harvard.edu/abs/2014ApJ...781...31C}{\textcolor{blue}{\apj,
  781, 31}}

\bibitem[\protect\citeauthoryear{{Costanzi} et~al.}{{Costanzi}
  et~al.}{2013}]{Costanzi1311.1514}
{Costanzi} M., {Villaescusa-Navarro} F., {Viel} M., {Xia} J.-Q., {Borgani} S.,
  {Castorina} E.,  {Sefusatti} E., 2013,
  \href{http://adsabs.harvard.edu/abs/2013JCAP...12..012C}{\textcolor{blue}{\jcap,
  12, 12}}

\bibitem[\protect\citeauthoryear{{Dahle}}{{Dahle}}{2006}]{Dahle0608480}
{Dahle} H., 2006,
  \href{http://adsabs.harvard.edu/abs/2006ApJ...653..954D}{\textcolor{blue}{\apj,
  653, 954}}

\bibitem[\protect\citeauthoryear{{Das} et~al.}{{Das}
  et~al.}{2014}]{Das1301.1037}
{Das} S. et~al., 2014,
  \href{http://adsabs.harvard.edu/abs/2014JCAP...04..014D}{\textcolor{blue}{\jcap,
  4, 14}}

\bibitem[\protect\citeauthoryear{{di Porto} \& {Amendola}}{{di Porto} \&
  {Amendola}}{2008}]{di-Porto0707.2686}
{di Porto} C.,  {Amendola} L., 2008,
  \href{http://adsabs.harvard.edu/abs/2008PhRvD..77h3508D}{\textcolor{blue}{\prd,
  77, 083508}}

\bibitem[\protect\citeauthoryear{{Donahue} \& {Voit}}{{Donahue} \&
  {Voit}}{1999}]{Donahue9907333}
{Donahue} M.,  {Voit} G.~M., 1999,
  \href{http://adsabs.harvard.edu/abs/1999ApJ...523L.137D}{\textcolor{blue}{\apjl,
  523, L137}}

\bibitem[\protect\citeauthoryear{{Dvorkin} et~al.}{{Dvorkin}
  et~al.}{2014}]{Dvorkin1403.8049}
{Dvorkin} C., {Wyman} M., {Rudd} D.~H.,  {Hu} W., 2014,
  \href{http://adsabs.harvard.edu/abs/2014PhRvD..90h3503D}{\textcolor{blue}{\prd,
  90, 083503}}

\bibitem[\protect\citeauthoryear{{Ebeling} et~al.}{{Ebeling}
  et~al.}{1998}]{Ebeling1998MNRAS.301..881E}
{Ebeling} H., {Edge} A.~C., {Bohringer} H., {Allen} S.~W., {Crawford} C.~S.,
  {Fabian} A.~C., {Voges} W.,  {Huchra} J.~P., 1998,
  \href{http://adsabs.harvard.edu/cgi-bin/nph-bib_query?bibcode=1998MNRAS.301..881E&db_key=AST}{\textcolor{blue}{\mnras,
  301, 881}}

\bibitem[\protect\citeauthoryear{{Ebeling} et~al.}{{Ebeling}
  et~al.}{2010}]{Ebeling1004.4683}
{Ebeling} H., {Edge} A.~C., {Mantz} A., {Barrett} E., {Henry} J.~P., {Ma}
  C.~J.,  {van Speybroeck} L., 2010,
  \href{http://adsabs.harvard.edu/abs/2010MNRAS.407...83E}{\textcolor{blue}{MNRAS,
  407, 83}}

\bibitem[\protect\citeauthoryear{{Eke} et~al.}{{Eke} et~al.}{1998}]{Eke9802350}
{Eke} V.~R., {Cole} S., {Frenk} C.~S.,  {Henry} J.~P., 1998,
  \href{http://adsabs.harvard.edu/abs/1998MNRAS.298.1145E}{\textcolor{blue}{\mnras,
  298, 1145}}

\bibitem[\protect\citeauthoryear{{Ettori} et~al.}{{Ettori}
  et~al.}{2009}]{Ettori0904.2740}
{Ettori} S., {Morandi} A., {Tozzi} P., {Balestra} I., {Borgani} S., {Rosati}
  P., {Lovisari} L.,  {Terenziani} F., 2009,
  \href{http://adsabs.harvard.edu/abs/2009A%26A...501...61E}{\textcolor{blue}{\aap,
  501, 61}}

\bibitem[\protect\citeauthoryear{{Ettori}, {Tozzi} \& {Rosati}}{{Ettori}
  et~al.}{2003}]{Ettori0211335}
{Ettori} S., {Tozzi} P.,  {Rosati} P., 2003,
  \href{http://adsabs.harvard.edu/abs/2003A%26A...398..879E}{\textcolor{blue}{\aap,
  398, 879}}

\bibitem[\protect\citeauthoryear{{Flauger}, {Hill} \& {Spergel}}{{Flauger}
  et~al.}{2014}]{Flauger1405.7351}
{Flauger} R., {Hill} J.~C.,  {Spergel} D.~N., 2014,
  \href{http://adsabs.harvard.edu/abs/2014JCAP...08..039F}{\textcolor{blue}{\jcap,
  8, 39}}

\bibitem[\protect\citeauthoryear{{Frieman}, {Turner} \& {Huterer}}{{Frieman}
  et~al.}{2008}]{Frieman0803.0982}
{Frieman} J.~A., {Turner} M.~S.,  {Huterer} D., 2008,
  \href{http://adsabs.harvard.edu/abs/2008ARA%26A..46..385F}{\textcolor{blue}{\araa,
  46, 385}}

\bibitem[\protect\citeauthoryear{{Gavazzi}}{{Gavazzi}}{2005}]{Gavazzi0503696}
{Gavazzi} R., 2005,
  \href{http://adsabs.harvard.edu/abs/2005A%26A...443..793G}{\textcolor{blue}{\aap,
  443, 793}}

\bibitem[\protect\citeauthoryear{{Hasselfield} et~al.}{{Hasselfield}
  et~al.}{2013}]{Hasselfield1301.0816}
{Hasselfield} M. et~al., 2013,
  \href{http://adsabs.harvard.edu/abs/2013JCAP...07..008H}{\textcolor{blue}{\jcap,
  7, 8}}

\bibitem[\protect\citeauthoryear{{Henry}}{{Henry}}{2000}]{Henry0002365}
{Henry} J.~P., 2000,
  \href{http://adsabs.harvard.edu/abs/2000ApJ...534..565H}{\textcolor{blue}{\apj,
  534, 565}}

\bibitem[\protect\citeauthoryear{{Henry}}{{Henry}}{2004}]{Henry0404142}
{Henry} J.~P., 2004,
  \href{http://adsabs.harvard.edu/abs/2004ApJ...609..603H}{\textcolor{blue}{\apj,
  609, 603}}

\bibitem[\protect\citeauthoryear{{Henry} et~al.}{{Henry}
  et~al.}{2009}]{Henry0809.3832}
{Henry} J.~P., {Evrard} A.~E., {Hoekstra} H., {Babul} A.,  {Mahdavi} A., 2009,
  \href{http://adsabs.harvard.edu/abs/2009ApJ...691.1307H}{\textcolor{blue}{\apj,
  691, 1307}}

\bibitem[\protect\citeauthoryear{{Heymans} et~al.}{{Heymans}
  et~al.}{2013}]{Heymans1303.1808}
{Heymans} C. et~al., 2013,
  \href{http://adsabs.harvard.edu/abs/2013MNRAS.432.2433H}{\textcolor{blue}{\mnras,
  432, 2433}}

\bibitem[\protect\citeauthoryear{{Hinshaw} et~al.}{{Hinshaw}
  et~al.}{2013}]{Hinshaw1212.5226}
{Hinshaw} G. et~al., 2013,
  \href{http://adsabs.harvard.edu/abs/2013ApJS..208...19H}{\textcolor{blue}{\apjs,
  208, 19}}

\bibitem[\protect\citeauthoryear{{Hoekstra}}{{Hoekstra}}{2007}]{Hoekstra0705.0358}
{Hoekstra} H., 2007,
  \href{http://adsabs.harvard.edu/abs/2007MNRAS.379..317H}{\textcolor{blue}{\mnras,
  379, 317}}

\bibitem[\protect\citeauthoryear{{Hoekstra} et~al.}{{Hoekstra}
  et~al.}{2012}]{Hoekstra1208.0606}
{Hoekstra} H., {Mahdavi} A., {Babul} A.,  {Bildfell} C., 2012,
  \href{http://adsabs.harvard.edu/abs/2012MNRAS.427.1298H}{\textcolor{blue}{\mnras,
  427, 1298}}

\bibitem[\protect\citeauthoryear{{Joyce} et~al.}{{Joyce}
  et~al.}{2014}]{Joyce1407.0059}
{Joyce} A., {Jain} B., {Khoury} J.,  {Trodden} M., 2014,
  \href{http://arxiv.org/abs/1407.0059}{\textcolor{blue}{arXiv:1407.0059}}

\bibitem[\protect\citeauthoryear{{Keisler} et~al.}{{Keisler}
  et~al.}{2011}]{Keisler1105.3182}
{Keisler} R. et~al., 2011,
  \href{http://adsabs.harvard.edu/abs/2011ApJ...743...28K}{\textcolor{blue}{\apj,
  743, 28}}

\bibitem[\protect\citeauthoryear{{Kelly} et~al.}{{Kelly}
  et~al.}{2014}]{Kelly1208.0602}
{Kelly} P.~L. et~al., 2014,
  \href{http://adsabs.harvard.edu/abs/2014MNRAS.439...28K}{\textcolor{blue}{\mnras,
  439, 28}} (WtG\,II)

\bibitem[\protect\citeauthoryear{{Kilbinger} et~al.}{{Kilbinger}
  et~al.}{2013}]{Kilbinger1212.3338}
{Kilbinger} M. et~al., 2013,
  \href{http://adsabs.harvard.edu/abs/2013MNRAS.430.2200K}{\textcolor{blue}{\mnras,
  430, 2200}}

\bibitem[\protect\citeauthoryear{{Koester} et~al.}{{Koester}
  et~al.}{2007}]{Koester0701265}
{Koester} B.~P. et~al., 2007,
  \href{http://adsabs.harvard.edu/abs/2007ApJ...660..239K}{\textcolor{blue}{\apj,
  660, 239}}

\bibitem[\protect\citeauthoryear{{Lesgourgues} \& {Pastor}}{{Lesgourgues} \&
  {Pastor}}{2006}]{Lesgourgues0603494}
{Lesgourgues} J.,  {Pastor} S., 2006,
  \href{http://adsabs.harvard.edu/abs/2006PhR...429..307L}{\textcolor{blue}{\physrep,
  429, 307}}

\bibitem[\protect\citeauthoryear{{Lewis} \& {Bridle}}{{Lewis} \&
  {Bridle}}{2002}]{Lewis0205436}
{Lewis} A.,  {Bridle} S., 2002,
  \href{http://adsabs.harvard.edu/abs/2002PhRvD..66j3511L}{\textcolor{blue}{\prd,
  66, 103511}}

\bibitem[\protect\citeauthoryear{{Lewis}, {Challinor} \& {Lasenby}}{{Lewis}
  et~al.}{2000}]{Lewis9911177}
{Lewis} A., {Challinor} A.,  {Lasenby} A., 2000,
  \href{http://adsabs.harvard.edu/abs/2000ApJ...538..473L}{\textcolor{blue}{\apj,
  538, 473}}

\bibitem[\protect\citeauthoryear{{Linder}}{{Linder}}{2003}]{Linder0208512}
{Linder} E.~V., 2003,
  \href{http://adsabs.harvard.edu/abs/2003PhRvL..90i1301L}{\textcolor{blue}{\prl,
  90, 091301}}

\bibitem[\protect\citeauthoryear{{Linder}}{{Linder}}{2005}]{Linder0507263}
{Linder} E.~V., 2005,
  \href{http://adsabs.harvard.edu/abs/2005PhRvD..72d3529L}{\textcolor{blue}{\prd,
  72, 043529}}

\bibitem[\protect\citeauthoryear{{Lo Verde} et~al.}{{Lo Verde}
  et~al.}{2008}]{LoVerde0711.4126}
{Lo Verde} M., {Miller} A., {Shandera} S.,  {Verde} L., 2008,
  \href{http://adsabs.harvard.edu/abs/2008JCAP...04..014L}{\textcolor{blue}{\jcap,
  4, 14}}

\bibitem[\protect\citeauthoryear{{LoVerde}}{{LoVerde}}{2014}]{LoVerde1405.4858}
{LoVerde} M., 2014,
  \href{http://arxiv.org/abs/1405.4858}{\textcolor{blue}{arXiv:1405.4858}}

\bibitem[\protect\citeauthoryear{{LSST Dark Energy Science
  Collaboration}}{{LSST Dark Energy Science
  Collaboration}}{2012}]{LSSTDESC1211.0310}
{LSST Dark Energy Science Collaboration} 2012,
  \href{http://arxiv.org/abs/1211.0310}{\textcolor{blue}{arXiv:1211.0310}}

\bibitem[\protect\citeauthoryear{{Ma} \& {Bertschinger}}{{Ma} \&
  {Bertschinger}}{1995}]{Ma9506072}
{Ma} C.-P.,  {Bertschinger} E., 1995,
  \href{http://adsabs.harvard.edu/abs/1995ApJ...455....7M}{\textcolor{blue}{\apj,
  455, 7}}

\bibitem[\protect\citeauthoryear{{Majumdar} \& {Mohr}}{{Majumdar} \&
  {Mohr}}{2004}]{Majumdar0305341}
{Majumdar} S.,  {Mohr} J.~J., 2004,
  \href{http://adsabs.harvard.edu/abs/2004ApJ...613...41M}{\textcolor{blue}{\apj,
  613, 41}}

\bibitem[\protect\citeauthoryear{{Mana} et~al.}{{Mana}
  et~al.}{2013}]{Mana1303.0287}
{Mana} A., {Giannantonio} T., {Weller} J., {Hoyle} B., {H{\"u}tsi} G.,
  {Sartoris} B., 2013,
  \href{http://adsabs.harvard.edu/abs/2013MNRAS.434..684M}{\textcolor{blue}{\mnras,
  434, 684}}

\bibitem[\protect\citeauthoryear{{Mantz}}{{Mantz}}{2009}]{Mantz2009PhDT........18M}
{Mantz} A., 2009, Ph.D. thesis, Stanford University

\bibitem[\protect\citeauthoryear{{Mantz} et~al.}{{Mantz}
  et~al.}{2008}]{Mantz0709.4294}
{Mantz} A., {Allen} S.~W., {Ebeling} H.,  {Rapetti} D., 2008,
  \href{http://adsabs.harvard.edu/abs/2008MNRAS.387.1179M}{\textcolor{blue}{\mnras,
  387, 1179}}

\bibitem[\protect\citeauthoryear{{Mantz} et~al.}{{Mantz}
  et~al.}{2010a}]{Mantz0909.3098}
{Mantz} A., {Allen} S.~W., {Rapetti} D.,  {Ebeling} H., 2010a,
  \href{http://adsabs.harvard.edu/abs/2010MNRAS.406.1759M}{\textcolor{blue}{MNRAS,
  406, 1759}} (M10a)

\bibitem[\protect\citeauthoryear{{Mantz} et~al.}{{Mantz}
  et~al.}{2010b}]{Mantz0909.3099}
\hypertarget{MtenBht}{}
{Mantz} A., {Allen} S.~W., {Ebeling} H., {Rapetti} D.,  {Drlica-Wagner} A.,
  2010b,
  \href{http://adsabs.harvard.edu/abs/2010MNRAS.406.1773M}{\textcolor{blue}{MNRAS,
  406, 1773}} (M10b)

\bibitem[\protect\citeauthoryear{{Mantz}, {Allen} \& {Rapetti}}{{Mantz}
  et~al.}{2010}]{Mantz0911.1788}
{Mantz} A., {Allen} S.~W.,  {Rapetti} D., 2010,
  \href{http://adsabs.harvard.edu/abs/2010MNRAS.406.1805M}{\textcolor{blue}{MNRAS,
  406, 1805}}

\bibitem[\protect\citeauthoryear{{Mantz} et~al.}{{Mantz}
  et~al.}{2014}]{Mantz1402.6212}
{Mantz} A.~B., {Allen} S.~W., {Morris} R.~G., {Rapetti} D.~A., {Applegate}
  D.~E., {Kelly} P.~L., {von der Linden} A.,  {Schmidt} R.~W., 2014,
  \href{http://adsabs.harvard.edu/abs/2014MNRAS.440.2077M}{\textcolor{blue}{\mnras,
  440, 2077}} (M14)

\bibitem[\protect\citeauthoryear{{Mortonson} \& {Seljak}}{{Mortonson} \&
  {Seljak}}{2014}]{Mortonson1405.5857}
{Mortonson} M.~J.,  {Seljak} U., 2014,
  \href{http://adsabs.harvard.edu/abs/2014JCAP...10..035M}{\textcolor{blue}{\jcap,
  10, 35}}

\bibitem[\protect\citeauthoryear{{Navarro}, {Frenk} \& {White}}{{Navarro}
  et~al.}{1997}]{Navarro9611107}
{Navarro} J.~F., {Frenk} C.~S.,  {White} S.~D.~M., 1997,
  \href{http://adsabs.harvard.edu/abs/1997ApJ...490..493N}{\textcolor{blue}{\apj,
  490, 493}}

\bibitem[\protect\citeauthoryear{{Nesseris} \& {Perivolaropoulos}}{{Nesseris}
  \& {Perivolaropoulos}}{2008}]{Nesseris0710.1092}
{Nesseris} S.,  {Perivolaropoulos} L., 2008,
  \href{http://adsabs.harvard.edu/abs/2008PhRvD..77b3504N}{\textcolor{blue}{\prd,
  77, 023504}}

\bibitem[\protect\citeauthoryear{{Neto} et~al.}{{Neto}
  et~al.}{2007}]{Neto0706.2919}
{Neto} A.~F. et~al., 2007,
  \href{http://adsabs.harvard.edu/abs/2007MNRAS.381.1450N}{\textcolor{blue}{\mnras,
  381, 1450}}

\bibitem[\protect\citeauthoryear{{Okabe} et~al.}{{Okabe}
  et~al.}{2013}]{Okabe1302.2728}
{Okabe} N., {Smith} G.~P., {Umetsu} K., {Takada} M.,  {Futamase} T., 2013,
  \href{http://adsabs.harvard.edu/abs/2013ApJ...769L..35O}{\textcolor{blue}{\apjl,
  769, L35}}

\bibitem[\protect\citeauthoryear{{Okabe} et~al.}{{Okabe}
  et~al.}{2010}]{Okabe0903.1103}
{Okabe} N., {Takada} M., {Umetsu} K., {Futamase} T.,  {Smith} G.~P., 2010,
  \href{http://adsabs.harvard.edu/abs/2010PASJ...62..811O}{\textcolor{blue}{\pasj,
  62, 811}}

\bibitem[\protect\citeauthoryear{{Padmanabhan} et~al.}{{Padmanabhan}
  et~al.}{2012}]{Padmanabhan1202.0090}
{Padmanabhan} N., {Xu} X., {Eisenstein} D.~J., {Scalzo} R., {Cuesta} A.~J.,
  {Mehta} K.~T.,  {Kazin} E., 2012,
  \href{http://adsabs.harvard.edu/abs/2012MNRAS.427.2132P}{\textcolor{blue}{\mnras,
  427, 2132}}

\bibitem[\protect\citeauthoryear{{Pen}}{{Pen}}{1997}]{Pen9610090}
{Pen} U., 1997,
  \href{http://adsabs.harvard.edu/abs/1997NewA....2..309P}{\textcolor{blue}{NewA,
  2, 309}}

\bibitem[\protect\citeauthoryear{{Pierpaoli} et~al.}{{Pierpaoli}
  et~al.}{2003}]{Pierpaoli0210567}
{Pierpaoli} E., {Borgani} S., {Scott} D.,  {White} M., 2003,
  \href{http://adsabs.harvard.edu/abs/2003MNRAS.342..163P}{\textcolor{blue}{\mnras,
  342, 163}}

\bibitem[\protect\citeauthoryear{{Planck Collaboration}}{{Planck
  Collaboration}}{2013a}]{Planck1303.5089}
{Planck Collaboration}, 2013a,
  \href{http://arxiv.org/abs/1303.5089}{\textcolor{blue}{arXiv:1303.5089}}

\bibitem[\protect\citeauthoryear{{Planck Collaboration}}{{Planck
  Collaboration}}{2013b}]{Planck1303.5076}
{Planck Collaboration}, 2013b,
  \href{http://arxiv.org/abs/1303.5076}{\textcolor{blue}{arXiv:1303.5076}}

\bibitem[\protect\citeauthoryear{{Planck Collaboration}}{{Planck
  Collaboration}}{2013c}]{Planck1303.5075}
{Planck Collaboration}, 2013c,
  \href{http://arxiv.org/abs/1303.5075}{\textcolor{blue}{arXiv:1303.5075}}

\bibitem[\protect\citeauthoryear{{Planck Collaboration}}{{Planck
  Collaboration}}{2013d}]{Planck1303.5080}
{Planck Collaboration}, 2013d,
  \href{http://arxiv.org/abs/1303.5080}{\textcolor{blue}{arXiv:1303.5080}}

\bibitem[\protect\citeauthoryear{{Planck Collaboration}}{{Planck
  Collaboration}}{2013e}]{Planck1303.5084}
{Planck Collaboration}, 2013e,
  \href{http://arxiv.org/abs/1303.5084}{\textcolor{blue}{arXiv:1303.5084}}

\bibitem[\protect\citeauthoryear{{Planelles} et~al.}{{Planelles}
  et~al.}{2013}]{Planelles1209.5058}
{Planelles} S., {Borgani} S., {Dolag} K., {Ettori} S., {Fabjan} D., {Murante}
  G.,  {Tornatore} L., 2013,
  \href{http://adsabs.harvard.edu/abs/2013MNRAS.431.1487P}{\textcolor{blue}{\mnras,
  431, 1487}}

\bibitem[\protect\citeauthoryear{{Polarski} \& {Gannouji}}{{Polarski} \&
  {Gannouji}}{2008}]{Polarski0710.1510}
{Polarski} D.,  {Gannouji} R., 2008,
  \href{http://adsabs.harvard.edu/abs/2008PhLB..660..439P}{\textcolor{blue}{Physics
  Letters B, 660, 439}}

\bibitem[\protect\citeauthoryear{{Rapetti} et~al.}{{Rapetti}
  et~al.}{2009}]{Rapetti0812.2259}
{Rapetti} D., {Allen} S.~W., {Mantz} A.,  {Ebeling} H., 2009,
  \href{http://adsabs.harvard.edu/abs/2009MNRAS.400..699R}{\textcolor{blue}{MNRAS,
  400, 699}}

\bibitem[\protect\citeauthoryear{{Rapetti} et~al.}{{Rapetti}
  et~al.}{2010}]{Rapetti0911.1787}
{Rapetti} D., {Allen} S.~W., {Mantz} A.,  {Ebeling} H., 2010,
  \href{http://adsabs.harvard.edu/abs/2010MNRAS.406.1796R}{\textcolor{blue}{MNRAS,
  406, 1796}}

\bibitem[\protect\citeauthoryear{{Rapetti}, {Allen} \& {Weller}}{{Rapetti}
  et~al.}{2005}]{Rapetti0409574}
{Rapetti} D., {Allen} S.~W.,  {Weller} J., 2005,
  \href{http://adsabs.harvard.edu/abs/2005MNRAS.360..555R}{\textcolor{blue}{\mnras,
  360, 555}}

\bibitem[\protect\citeauthoryear{{Rapetti} et~al.}{{Rapetti}
  et~al.}{2013}]{Rapetti1205.4679}
{Rapetti} D., {Blake} C., {Allen} S.~W., {Mantz} A., {Parkinson} D.,  {Beutler}
  F., 2013,
  \href{http://adsabs.harvard.edu/abs/2013MNRAS.432..973R}{\textcolor{blue}{\mnras,
  432, 973}}

\bibitem[\protect\citeauthoryear{{Reichardt} et~al.}{{Reichardt}
  et~al.}{2012}]{Reichardt1111.0932}
{Reichardt} C.~L. et~al., 2012,
  \href{http://adsabs.harvard.edu/abs/2012ApJ...755...70R}{\textcolor{blue}{\apj,
  755, 70}}

\bibitem[\protect\citeauthoryear{{Reichardt} et~al.}{{Reichardt}
  et~al.}{2013}]{Reichardt1203.5775}
{Reichardt} C.~L. et~al., 2013,
  \href{http://adsabs.harvard.edu/abs/2013ApJ...763..127R}{\textcolor{blue}{\apj,
  763, 127}}

\bibitem[\protect\citeauthoryear{{Reid} et~al.}{{Reid}
  et~al.}{2012}]{Reid1203.6641}
{Reid} B.~A. et~al., 2012,
  \href{http://adsabs.harvard.edu/abs/2012MNRAS.426.2719R}{\textcolor{blue}{\mnras,
  426, 2719}}

\bibitem[\protect\citeauthoryear{{Reid} et~al.}{{Reid}
  et~al.}{2010}]{Reid0910.0008}
{Reid} B.~A., {Verde} L., {Jimenez} R.,  {Mena} O., 2010,
  \href{http://adsabs.harvard.edu/abs/2010JCAP...01..003R}{\textcolor{blue}{\jcap,
  1, 3}}

\bibitem[\protect\citeauthoryear{{Reiprich} \& {B{\"o}hringer}}{{Reiprich} \&
  {B{\"o}hringer}}{2002}]{Reiprich0111285}
{Reiprich} T.~H.,  {B{\"o}hringer} H., 2002,
  \href{http://adsabs.harvard.edu/cgi-bin/nph-bib_query?bibcode=2002ApJ...567..716R&db_key=AST}{\textcolor{blue}{\apj,
  567, 716}}

\bibitem[\protect\citeauthoryear{{Riemer-Sorensen} et~al.}{{Riemer-Sorensen}
  et~al.}{2012}]{Riemer-Sorensen1112.4940}
{Riemer-Sorensen} S. et~al., 2012,
  \href{http://adsabs.harvard.edu/abs/2012PhRvD..85h1101R}{\textcolor{blue}{\prd,
  85, 081101}}

\bibitem[\protect\citeauthoryear{{Riess} et~al.}{{Riess}
  et~al.}{2011}]{Riess1103.2976}
{Riess} A.~G. et~al., 2011,
  \href{http://adsabs.harvard.edu/abs/2011ApJ...730..119R}{\textcolor{blue}{\apj,
  730, 119}}

\bibitem[\protect\citeauthoryear{{Rozo} et~al.}{{Rozo}
  et~al.}{2010}]{Rozo0902.3702}
{Rozo} E. et~al., 2010,
  \href{http://adsabs.harvard.edu/abs/2010ApJ...708..645R}{\textcolor{blue}{\apj,
  708, 645}}

\bibitem[\protect\citeauthoryear{{Rykoff} et~al.}{{Rykoff}
  et~al.}{2014}]{Rykoff1303.3562}
{Rykoff} E.~S. et~al., 2014,
  \href{http://adsabs.harvard.edu/abs/2014ApJ...785..104R}{\textcolor{blue}{\apj,
  785, 104}}

\bibitem[\protect\citeauthoryear{{Samushia} et~al.}{{Samushia}
  et~al.}{2013}]{Samushia1206.5309}
{Samushia} L. et~al., 2013,
  \href{http://adsabs.harvard.edu/abs/2013MNRAS.429.1514S}{\textcolor{blue}{\mnras,
  429, 1514}}

\bibitem[\protect\citeauthoryear{{Samushia} et~al.}{{Samushia}
  et~al.}{2014}]{Samushia1312.4899}
{Samushia} L. et~al., 2014,
  \href{http://adsabs.harvard.edu/abs/2014MNRAS.439.3504S}{\textcolor{blue}{\mnras,
  439, 3504}}

\bibitem[\protect\citeauthoryear{{Sasaki}}{{Sasaki}}{1996}]{Sasaki9611033}
{Sasaki} S., 1996,
  \href{http://adsabs.harvard.edu/abs/1996PASJ...48L.119S}{\textcolor{blue}{\pasj,
  48, L119}}

\bibitem[\protect\citeauthoryear{{Schmidt}, {Vikhlinin} \& {Hu}}{{Schmidt}
  et~al.}{2009}]{Schmidt0908.2457}
{Schmidt} F., {Vikhlinin} A.,  {Hu} W., 2009,
  \href{http://adsabs.harvard.edu/abs/2009PhRvD..80h3505S}{\textcolor{blue}{\prd,
  80, 083505}}

\bibitem[\protect\citeauthoryear{{Schrabback} et~al.}{{Schrabback}
  et~al.}{2010}]{Schrabback0911.0053}
{Schrabback} T. et~al., 2010,
  \href{http://adsabs.harvard.edu/abs/2010A%26A...516A..63S}{\textcolor{blue}{\aap,
  516, A63}}

\bibitem[\protect\citeauthoryear{{Schuecker} et~al.}{{Schuecker}
  et~al.}{2003}]{Schuecker0208251}
{Schuecker} P., {B{\"o}hringer} H., {Collins} C.~A.,  {Guzzo} L., 2003,
  \href{http://adsabs.harvard.edu/abs/2003A%26A...398..867S}{\textcolor{blue}{\aap,
  398, 867}}

\bibitem[\protect\citeauthoryear{{Sehgal} et~al.}{{Sehgal}
  et~al.}{2011}]{Sehgal1010.1025}
{Sehgal} N. et~al., 2011,
  \href{http://adsabs.harvard.edu/abs/2011ApJ...732...44S}{\textcolor{blue}{\apj,
  732, 44}}

\bibitem[\protect\citeauthoryear{{Seljak}}{{Seljak}}{2002}]{Seljak0111362}
{Seljak} U., 2002,
  \href{http://adsabs.harvard.edu/abs/2002MNRAS.337..769S}{\textcolor{blue}{\mnras,
  337, 769}}

\bibitem[\protect\citeauthoryear{{Shandera} et~al.}{{Shandera}
  et~al.}{2013a}]{Shandera1211.7361}
{Shandera} S., {Erickcek} A.~L., {Scott} P.,  {Galarza} J.~Y., 2013a,
  \href{http://adsabs.harvard.edu/abs/2013PhRvD..88j3506S}{\textcolor{blue}{\prd,
  88, 103506}}

\bibitem[\protect\citeauthoryear{{Shandera} et~al.}{{Shandera}
  et~al.}{2013b}]{Shandera1304.1216}
{Shandera} S., {Mantz} A., {Rapetti} D.,  {Allen} S.~W., 2013b,
  \href{http://adsabs.harvard.edu/abs/2013JCAP...08..004S}{\textcolor{blue}{\jcap,
  8, 4}}

\bibitem[\protect\citeauthoryear{{Sif{\'o}n} et~al.}{{Sif{\'o}n}
  et~al.}{2013}]{Sifon1201.0991}
{Sif{\'o}n} C. et~al., 2013,
  \href{http://adsabs.harvard.edu/abs/2013ApJ...772...25S}{\textcolor{blue}{\apj,
  772, 25}}

\bibitem[\protect\citeauthoryear{{Stanek} et~al.}{{Stanek}
  et~al.}{2010}]{Stanek0910.1599}
{Stanek} R., {Rasia} E., {Evrard} A.~E., {Pearce} F.,  {Gazzola} L., 2010,
  \href{http://adsabs.harvard.edu/abs/2010ApJ...715.1508S}{\textcolor{blue}{\apj,
  715, 1508}}

\bibitem[\protect\citeauthoryear{{Story} et~al.}{{Story}
  et~al.}{2013}]{Story1210.7231}
{Story} K.~T. et~al., 2013,
  \href{http://adsabs.harvard.edu/abs/2013ApJ...779...86S}{\textcolor{blue}{\apj,
  779, 86}}

\bibitem[\protect\citeauthoryear{{Suzuki} et~al.}{{Suzuki}
  et~al.}{2012}]{Suzuki1105.3470}
{Suzuki} N. et~al., 2012,
  \href{http://adsabs.harvard.edu/abs/2012ApJ...746...85S}{\textcolor{blue}{\apj,
  746, 85}}

\bibitem[\protect\citeauthoryear{{Tegmark} et~al.}{{Tegmark}
  et~al.}{2004}]{Tegmark0310723}
{Tegmark} M. et~al., 2004,
  \href{http://adsabs.harvard.edu/abs/2004PhRvD..69j3501T}{\textcolor{blue}{\prd,
  69, 103501}}

\bibitem[\protect\citeauthoryear{{Tereno} et~al.}{{Tereno}
  et~al.}{2009}]{Tereno0810.0555}
{Tereno} I., {Schimd} C., {Uzan} J.-P., {Kilbinger} M., {Vincent} F.~H.,  {Fu}
  L., 2009,
  \href{http://adsabs.harvard.edu/abs/2009A%26A...500..657T}{\textcolor{blue}{\aap,
  500, 657}}

\bibitem[\protect\citeauthoryear{{Thomas}, {Abdalla} \& {Lahav}}{{Thomas}
  et~al.}{2010}]{Thomas0911.5291}
{Thomas} S.~A., {Abdalla} F.~B.,  {Lahav} O., 2010,
  \href{http://adsabs.harvard.edu/abs/2010PhRvL.105c1301T}{\textcolor{blue}{Physical
  Review Letters, 105, 031301}}

\bibitem[\protect\citeauthoryear{{Tinker} et~al.}{{Tinker}
  et~al.}{2008}]{Tinker0803.2706}
{Tinker} J., {Kravtsov} A.~V., {Klypin} A., {Abazajian} K., {Warren} M.,
  {Yepes} G., {Gottl{\"o}ber} S.,  {Holz} D.~E., 2008,
  \href{http://adsabs.harvard.edu/abs/2008ApJ...688..709T}{\textcolor{blue}{\apj,
  688, 709}}

\bibitem[\protect\citeauthoryear{{Tr\"umper}}{{Tr\"umper}}{1993}]{Truemper1993Sci...260.1769T}
{Tr\"umper} J., 1993,
  \href{http://adsabs.harvard.edu/abs/1993Sci...260.1769T}{\textcolor{blue}{Science,
  260, 1769}}

\bibitem[\protect\citeauthoryear{{Viana}, {Nichol} \& {Liddle}}{{Viana}
  et~al.}{2002}]{Viana0111394}
{Viana} P.~T.~P., {Nichol} R.~C.,  {Liddle} A.~R., 2002,
  \href{http://adsabs.harvard.edu/abs/2002ApJ...569L..75V}{\textcolor{blue}{\apjl,
  569, L75}}

\bibitem[\protect\citeauthoryear{{Vikhlinin} et~al.}{{Vikhlinin}
  et~al.}{2009}]{Vikhlinin0812.2720}
{Vikhlinin} A. et~al., 2009,
  \href{http://adsabs.harvard.edu/abs/2009ApJ...692.1060V}{\textcolor{blue}{\apj,
  692, 1060}}

\bibitem[\protect\citeauthoryear{{Vikhlinin} et~al.}{{Vikhlinin}
  et~al.}{2003}]{Vikhlinin0212075}
{Vikhlinin} A. et~al., 2003,
  \href{http://adsabs.harvard.edu/abs/2003ApJ...590...15V}{\textcolor{blue}{\apj,
  590, 15}}

\bibitem[\protect\citeauthoryear{{Voevodkin} \& {Vikhlinin}}{{Voevodkin} \&
  {Vikhlinin}}{2004}]{Voevodkin0305549}
{Voevodkin} A.,  {Vikhlinin} A., 2004,
  \href{http://adsabs.harvard.edu/abs/2004ApJ...601..610V}{\textcolor{blue}{\apj,
  601, 610}}

\bibitem[\protect\citeauthoryear{von~der {Linden} et~al.}{von~der {Linden}
  et~al.}{2014a}]{von-der-Linden1208.0597}
von~der {Linden} A. et~al., 2014a,
  \href{http://adsabs.harvard.edu/abs/2014MNRAS.439....2V}{\textcolor{blue}{\mnras,
  439, 2}} (WtG\,I)

\bibitem[\protect\citeauthoryear{von~der {Linden} et~al.}{von~der {Linden}
  et~al.}{2014b}]{von-der-Linden1402.2670}
von~der {Linden} A. et~al., 2014b,
  \href{http://adsabs.harvard.edu/abs/2014MNRAS.443.1973V}{\textcolor{blue}{\mnras,
  443, 1973}}

\bibitem[\protect\citeauthoryear{{Weinberg} et~al.}{{Weinberg}
  et~al.}{2013}]{Weinberg1201.2434}
{Weinberg} D.~H., {Mortonson} M.~J., {Eisenstein} D.~J., {Hirata} C., {Riess}
  A.~G.,  {Rozo} E., 2013,
  \href{http://adsabs.harvard.edu/abs/2013PhR...530...87W}{\textcolor{blue}{\physrep,
  530, 87}}

\bibitem[\protect\citeauthoryear{{Weller} \& {Lewis}}{{Weller} \&
  {Lewis}}{2003}]{Weller0307104}
{Weller} J.,  {Lewis} A.~M., 2003,
  \href{http://adsabs.harvard.edu/abs/2003MNRAS.346..987W}{\textcolor{blue}{\mnras,
  346, 987}}

\bibitem[\protect\citeauthoryear{{Williamson} et~al.}{{Williamson}
  et~al.}{2011}]{Williamson1101.1290}
{Williamson} R. et~al., 2011,
  \href{http://adsabs.harvard.edu/abs/2011ApJ...738..139W}{\textcolor{blue}{\apj,
  738, 139}}

\bibitem[\protect\citeauthoryear{{Wright} \& {Brainerd}}{{Wright} \&
  {Brainerd}}{2000}]{Wright2000ApJ...534...34W}
{Wright} C.~O.,  {Brainerd} T.~G., 2000,
  \href{http://adsabs.harvard.edu/abs/2000ApJ...534...34W}{\textcolor{blue}{\apj,
  534, 34}}

\bibitem[\protect\citeauthoryear{{Wu}, {Rozo} \& {Wechsler}}{{Wu}
  et~al.}{2010}]{Wu0907.2690}
{Wu} H., {Rozo} E.,  {Wechsler} R.~H., 2010,
  \href{http://adsabs.harvard.edu/abs/2010ApJ...713.1207W}{\textcolor{blue}{\apj,
  713, 1207}}

\bibitem[\protect\citeauthoryear{{Zhang} et~al.}{{Zhang}
  et~al.}{2008}]{Zhang0802.0770}
{Zhang} Y.-Y., {Finoguenov} A., {B{\"o}hringer} H., {Kneib} J.-P., {Smith}
  G.~P., {Kneissl} R., {Okabe} N.,  {Dahle} H., 2008,
  \href{http://adsabs.harvard.edu/abs/2008A%26A...482..451Z}{\textcolor{blue}{\aap,
  482, 451}}

\end{thebibliography}
\end{document}